\documentclass{SciPost}
\hypersetup{
    colorlinks,
    linkcolor={red!50!black},
    citecolor={blue!50!black},
    urlcolor={blue!80!black}
}
\usepackage[bitstream-charter]{mathdesign}

\usepackage{subfig}
\usepackage[normalem]{ulem}
\DeclareSymbolFont{usualmathcal}{OMS}{cmsy}{m}{n}
\DeclareSymbolFontAlphabet{\mathcal}{usualmathcal}

\fancypagestyle{SPstyle}{
\fancyhf{}
\lhead{\colorbox{scipostblue}{\bf \color{white} ~SciPost Physics }}
\rhead{{\bf \color{scipostdeepblue} ~Submission }}

\fancyfoot[C]{\textbf{\thepage}}
}

\usepackage{graphicx} 
\usepackage{xcolor} 
\definecolor{dgreen}{rgb}{0.0,0.5,0.0}
\definecolor{pink}{rgb}{1,0,0.9}

\begin{document}
\pagestyle{SPstyle}

\begin{center}{\Large \textbf{\color{scipostdeepblue}{
CTMRG as a tensor-network message-passing algorithm\\
}}}\end{center}

\begin{center}\textbf{
Illia Lukin\textsuperscript{1,2$\star$} and
Andrii Sotnikov\textsuperscript{1,3$\dagger$}
}\end{center}

\begin{center}
{\bf 1} National Science Center "Kharkiv Institute of Physics and Technology", Akademichna str. 1, 61108 Kharkiv, Ukraine
\\
{\bf 2} Haiqu, Inc., 95 Third Street, San Francisco, CA 94103, USA
\\
{\bf 3} V.N. Karazin Kharkiv National University,  Svobody Square 4, 61022 Kharkiv, Ukraine
\\[\baselineskip]
$\star$ \href{mailto:illya.lukin11@gmail.com}{\small illya.lukin11@gmail.com}\,,\quad
$\dagger$ \href{mailto:a_sotnikov@kipt.kharkov.ua}{\small a\_sotnikov@kipt.kharkov.ua}
\end{center}

\section*{\color{scipostdeepblue}{Abstract}}
\textbf{\boldmath{%
We formulate the corner transfer matrix equations, and the corner transfer matrix renormalization group built on them, as a set of local equations on general planar graphs. These equations depend only on the local subgraphs of the planar graph and are glued together consistently to describe the whole graph. Solving them yields message tensors and corner matrices, which can be used to construct an MPS environment for any connected subgraph. We test the proposed equations on the ferromagnetic Ising model, on a variety of infinite lattices (including hyperbolic ones) and on finite graphs, and find systematic improvement over belief propagation.
}}

\vspace{\baselineskip}

\noindent\textcolor{white!90!black}{%
\fbox{\parbox{0.975\linewidth}{%
\textcolor{white!40!black}{\begin{tabular}{lr}%
  \begin{minipage}{0.6\textwidth}%
    {\small Copyright attribution to authors. \newline
    This work is a submission to SciPost Physics. \newline
    License information to appear upon publication. \newline
    Publication information to appear upon publication.}
  \end{minipage} & \begin{minipage}{0.4\textwidth}
    {\small Received Date \newline Accepted Date \newline Published Date}%
  \end{minipage}
\end{tabular}}
}}
}


\vspace{10pt}
\noindent\rule{\textwidth}{1pt}
\tableofcontents
\noindent\rule{\textwidth}{1pt}
\vspace{10pt}

\section{Introduction}

The central problem addressed in this paper is the contraction of tensor networks~\cite{Orus2019natrev, Cirac2021rmp, Banuls2023routemap, Okunishi2022review, Ran2020lnp} defined on two-dimensional (2d) planar graphs. This problem is of direct physical significance: both local 2d statistical-mechanics problems on planar graphs and the calculation of observables for 2d tensor network states on planar graphs reduce to it. In most cases, the contraction cannot be solved exactly, since exact algorithms have memory and computation costs that scale exponentially with the size of the underlying graph. For this reason, approximate algorithms with controlled error scaling are of great value. The most widely used ones fall into three broad classes: those based on tensor renormalization group methods~\cite{Levin2007trg, Xie2009srg, Xie2012hotrg, Gu2009tefr, Evenbly2015tnr, Hauru2018gilt}, those based on auxiliary matrix product states~\cite{Jordan2008ipeps, Verstraete2004peps}, and approaches related to belief propagation.

In recent years, tensor-network algorithms have been substantially enriched by methods from the belief propagation literature~\cite{Tindall2023gauging, Alkabetz2021bp}. In particular, belief propagation is now routinely used, in various forms~\cite{Park2025ifbp}, to simulate time dynamics with tensor networks, including in recent IBM QPU experiments~\cite{Tindall2024eagle, Tindall2024confinement, Tindall2026disordered, Kshetrimayum2026advantage, Leviatan2026prethermal, Luchnikov2024annealing, Patra2024ibm, Bermejo2026echoes}. It is also routinely employed to approximately compute observables with tensor network states. This influence has driven the development of methods to systematically correct belief-propagation-based tensor network calculations: loop~\cite{Chertkov2006loopcalculus, Chertkov2006loopseries, Evenbly2026loopseries} and cluster expansions~\cite{Gray2026loopcluster, Midha2026beyondbp, Midha2026locality, Midha2026rigorous, Sim2026stochasticloop} on top of belief propagation, block belief propagation~\cite{Guo2023blockbp}, and generalized belief propagation~\cite{Tindall2026gbp}. In principle, these methods allow strongly correlated 2d and 3d quantum systems to be investigated systematically using belief-propagation-based methods alone~\cite{Mello2026ruby, Lukin2026su4hyperhoneycomb}, without relying on boundary MPS methods~\cite{Verstraete2004peps, Murg2007hardcore, Jordan2008ipeps, Orus2008itebd} or the corner transfer matrix renormalization group (CTMRG)~\cite{Nishino1996ctmrg, Nishino1997ctmrg, Nishino2001tpvs, Orus2009ctm, Corboz2011stripes}.

At the same time, the CTMRG algorithm has also seen substantial progress. Recent developments include generalizations of CTMRG to triangular, kagome, and other Archimedean lattices~\cite{Lukin2024archimedean}, building on earlier work on the honeycomb lattice~\cite{Lukin2023honeycomb, Nyckees2023c3ctmrg} and on hyperbolic-lattice CTMRG~\cite{Ueda2007hyperbolic, Krcmar2008hyperbolic, Krcmar2008tricritical, Iharagi2010hyperbolic, Gendiar2008clock, Gendiar2012triangular, Gendiar2014meanfield, Gendiar2020arealaw, Lee2016boundary, Genzor2017review, Daniska2015pentagonal, Daniska2016quantum, Serina2016freeenergy, Mosko2025vertex}; a new 3d CTMRG for the hyperbolic lattice, along with experiments on the cubic lattice~\cite{Mosko2026dodecahedral}; a formulation of CTMRG for the reflection-symmetric square lattice as a bilevel variational problem~\cite{Liu2022vctmrg}; novel methods for finding CTMRG projectors based on QR decomposition~\cite{Zhang2026qrctmrg, Yang2025qrhoneycomb}; Grassmann CTMRG~\cite{Kong2026grassmannctmrg, Kong2026gnw}; and the application of CTMRG in PEPS tensor-network excitation-ansatz optimization~\cite{Ponsioen2020excitations, Ponsioen2022ad, Ponsioen2023summations, Tu2021generating} and in implicit differentiation for PEPS variational optimization~\cite{Burgelman2026implicit}. These developments have also led to practical applications, such as those in Refs.~\cite{Naumann2026triangular, Zhang2025kitaev, Zhang2025preconditioning}.

In this study, we aim to reconcile these two branches of research. Specifically, we show that CTMRG can be formulated as a message-update algorithm, similar to belief propagation, on general planar graphs. In our formulation, CTMRG consists of local equations that depend only on small neighborhoods of the full planar graph; these local equations are then glued together consistently to describe the whole graph. This mirrors belief propagation, whose equations likewise depend only on the individual vertices of the graph. A further similarity is that, in this formulation, CTMRG tensors can be used to describe the environment of an arbitrary connected subgraph of the underlying graph -- much as in belief propagation, where the environment of an arbitrary subgraph is approximated as a product of messages on the subgraph's external legs. In CTMRG the situation is analogous, except that the environment is approximated as a generalized MPS on the subgraph's external legs.

This work is a natural continuation of our previous study~\cite{Lukin2024archimedean}. There, we attempted to construct CTMRG on Archimedean lattices and succeeded for most examples, but our constructions were lengthy and not very systematic, requiring a case-by-case approach for every lattice. Moreover, we were unable to extend our method of CTMRG derivation to hyperbolic lattices, even though a CTMRG algorithm is known to exist for them. In addition, the tensors, corner matrices, and projectors in that earlier approach showed non-trivial relations upon convergence that were neither assumed nor predicted by the derivation. This pointed to a more general hidden structure underlying CTMRG -- one that could be used to formulate it on essentially any planar lattice while reducing the number of independent variables and equations. This hidden structure is precisely the message-passing formulation we present in this work.

\section{Vertex models}\label{sec:vertex_models}

\subsection{Tensor networks and MPS messages}
We introduce the general framework of our study as follows. Let us define a graph~$G$ consisting of a set of vertices~$V$ and a set of edges~$E$. On every vertex $v$ of the graph we place a tensor $T_{v}$ whose rank equals the valency $z_{v}$ of the vertex, where each index of dimension $D$ corresponds to one of the edges $e$ incident to $v$. Let us also introduce an edge orientation, denoting by $\bar{e}$ the same edge as $e$ but traversed in the opposite direction. The tensor network $Z$ is then formed as the product of all the tensors $T_{v}$, with the indices of every edge contracted pairwise between the two tensors sharing that edge. We additionally consider only vertices with valency $z_{v} > 2$ (except for the boundary) and without contractable tree parts (the latter can be always contracted using belief propagation). Another requirement for the graph $G$ is that the graph needs to be a planar one --- it should be possible to place it on the plane without edge intersections. 

\begin{figure}
\centering
\includegraphics[width= 0.5\linewidth]{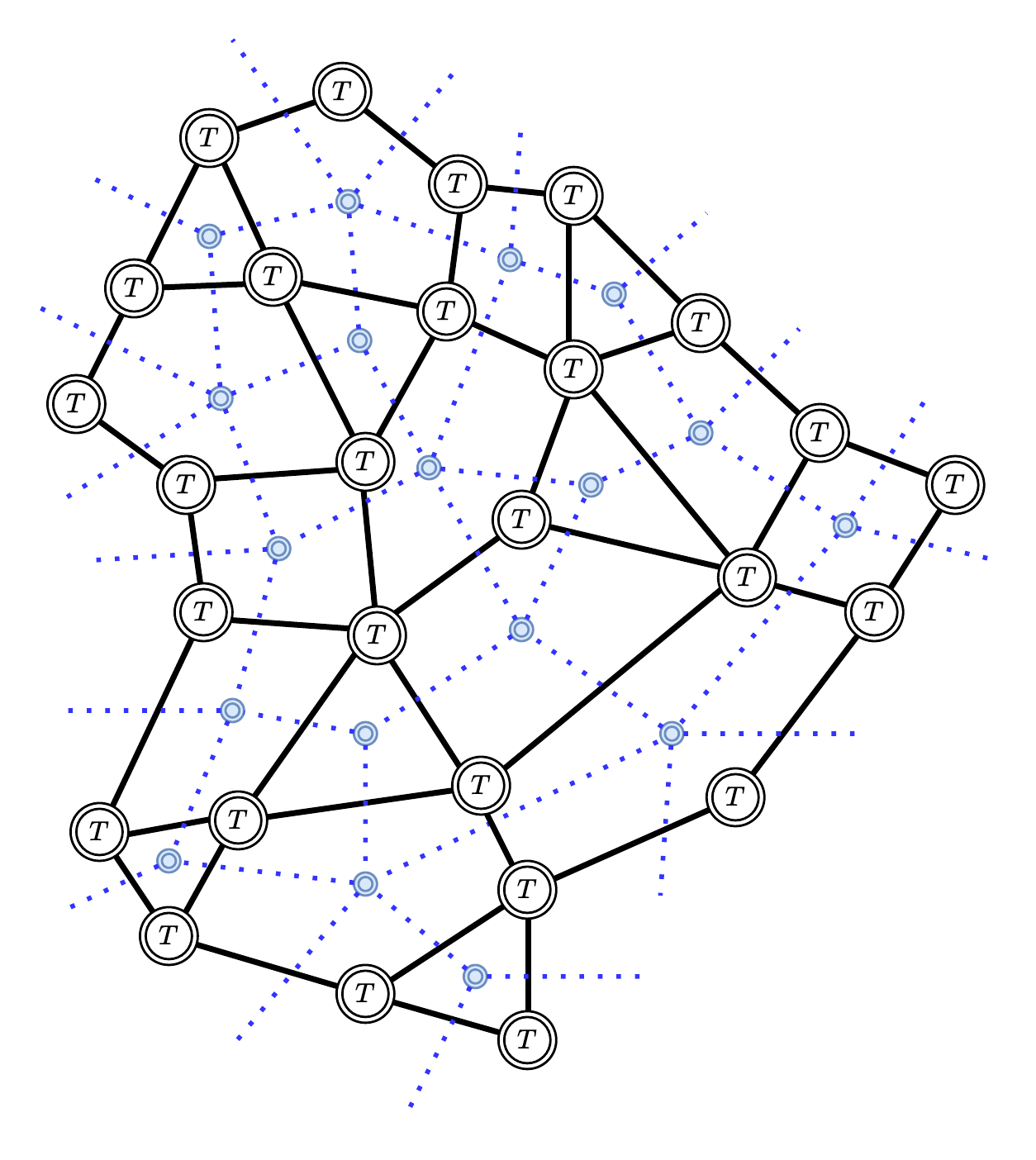} 
 \caption{\label{fig:dual_graph}%
       Tensors $T$ are placed on the vertices of the graph $G$, and contracted indices correspond to graph edges in black lines. The dual graph is constructed by placing dual vertices (in blue) on the faces of the original graph and connecting the dual vertices by the edge (dotted blue lines) if the original faces share the edge. }
\end{figure}

The graph $G$ can be either finite or infinite; in the latter case we additionally need to impose periodicity on the tensors $T$. In most of what follows, we deal with periodic infinite planar or hyperbolic lattices. An example of such a tensor network is shown in Fig.~\ref{fig:dual_graph}. 

Note that every graph $G$ also admits a dual graph~$G^{*}$, whose vertices correspond to the faces of $G$ and whose faces correspond to the vertices of $G$. The construction of the dual graph is illustrated in Fig.~\ref{fig:dual_graph}. The planar graphs are also dual to cell decompositions of the surface, which makes the vertex type tensor network constructions on such graphs naturally dual to the face model constructions on dual graphs. 

\begin{figure}
\centering
\includegraphics[width= 0.9\linewidth]{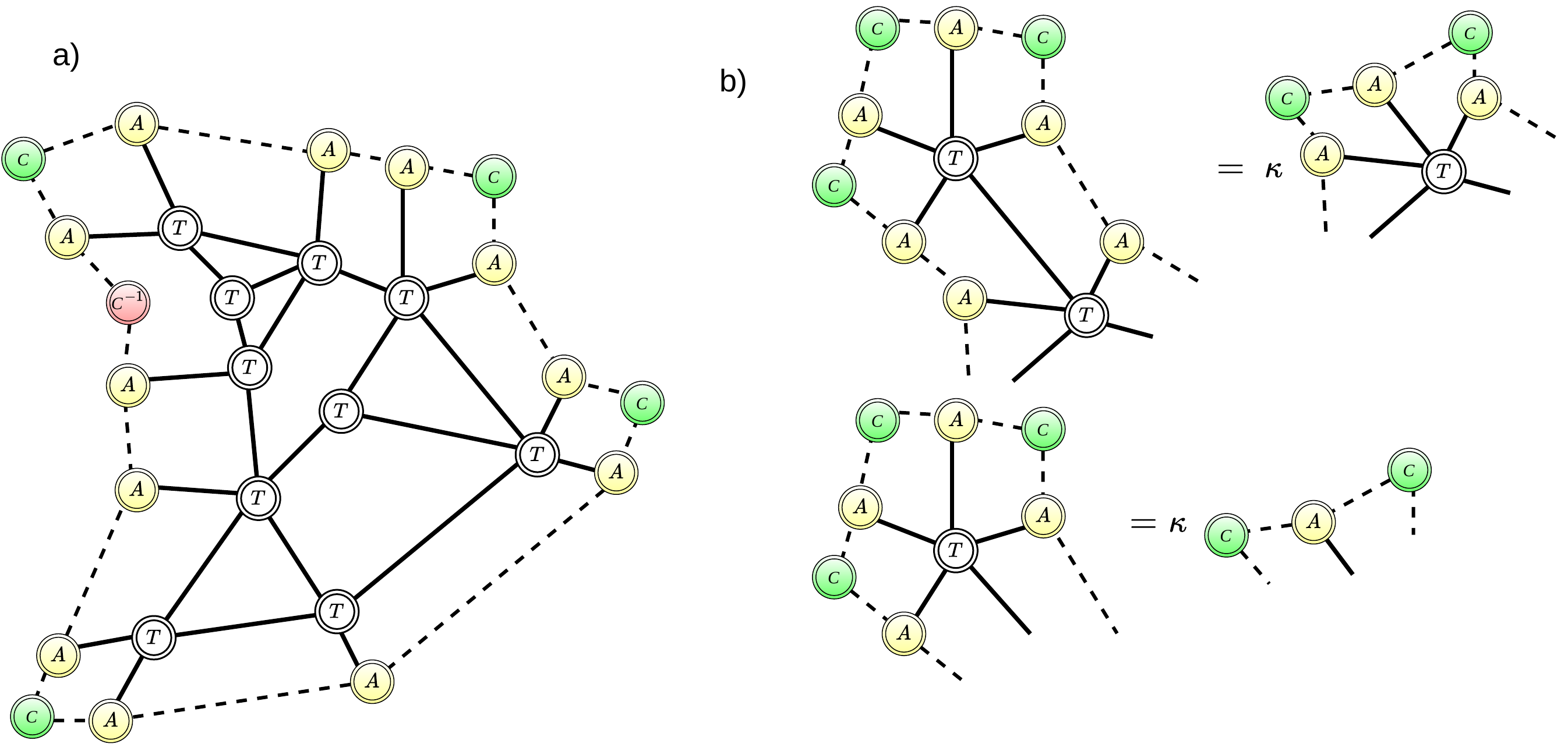} 
 \caption{\label{fig:mps_messages}%
   Definition of the MPS messages. (a)  A general subgraph consisting of tensors $T$ with varying connectivity, together with its boundary represented as an MPS; the MPS consists of the message tensors $A$ and various corner matrices $C, C^{-1}, \dots$. (b) Derivation of the self-consistency equation for the tensor $A$: a single bulk tensor $T$ can be fully isolated by an MPS ring (except for one bond), and this isolated contraction must equal the message tensor on the remaining bond together with the appropriate corner matrices. This yields the update equation for $A$.   }
\end{figure}

Our goal is now to contract the tensor network. To this end, we introduce matrix product state (MPS) messages, as illustrated in Fig.~\ref{fig:mps_messages}. The MPS messages live on the edges of the graph, with the two orientations of each edge carrying two different messages. The MPS message $A_{v_{1} \to v_{2}}$ associated with the edge $(v_{1}, v_{2})$ is a rank-$3$ tensor: one index is the edge index of the tensor network, while the other two are virtual indices of dimension~$\chi$ used to connect the message tensors on neighboring edges. These virtual indices effectively live on the faces of the graph, so that, in principle, one could assign a different bond dimension $\chi_{f}$ to every face $f$. To combine the edge messages into a single MPS, we could first attempt to form an MPS line on the dual graph, by directly contracting the virtual indices of the messages on edges sharing a face. This naive contraction, however, is not gauge covariant: changing the gauge on one virtual index of a message tensor would force a simultaneous gauge change on \emph{every} other message sharing that face, which is far too restrictive a gauge freedom and, moreover, does not lead to anything resembling CTMRG.

Instead, we use a more general construction that introduces ``connector'' or ``corner'' matrices $C_{e_{1} e_{2}}$, which connect two edge messages $A_{e_{1}}$ and $A_{e_{2}}$ on the virtual index associated with their shared face. Consider a particular face of the graph and orient all its edges counterclockwise, so that each edge also has an opposite, clockwise-pointing partner. The corner matrix $C_{e_{1}, e_{2}}$ exists only when one of $e_{1}, e_{2}$ points clockwise and the other counterclockwise; otherwise, this $C$ matrix cannot be used to form an MPS on the boundary. With this construction, every virtual index of the message tensors can be gauged independently, since the full MPS remains invariant once the corner matrices $C$ are transformed accordingly. More generally, the edge messages $A_{e}$ together with the connector matrices $C_{e_{1} e_{2}}$ can be used to construct an MPS along an arbitrary polygonal line on the dual graph.

Let us now introduce the preferred gauge condition used throughout this study. Consider a face (polygon) of the graph, and order its edges counterclockwise as $(e_{1}, e_{2}, \ldots, e_{n})$; each edge also has an oppositely oriented partner $\bar{e}_{1}, \bar{e}_{2}, \ldots, \bar{e}_{n}$. We fix the corner matrices $C$ on the edge pairs $(e_{1}, \bar{e}_{3})$, $(e_{2}, \bar{e}_{4})$, \ldots, $(e_{n-1}, \bar{e}_{1})$, $(e_{n}, \bar{e}_{2})$ to be identity matrices. This gauge is convenient for two reasons: (i) it exists for every polygon, including triangles, and (ii) on lattices such as the square lattice it reduces to the gauge conventionally used in standard CTMRG. 

This gauge always exists for arbitrary message tensors, but it can be incompatible with certain lattice symmetries: for instance, if we impose reflection symmetry on the message tensors $A$, the residual gauge freedom of the symmetric tensors may not be enough to eliminate all of these corner matrices (we return to this case later, for the triangular lattice). The gauge is illustrated in Fig.~\ref{fig:mps_messages}(a). Aside from the gauged matrices, there remain corner matrices of the form $C_{e_{k}, \bar{e}_{k+1}}$, connecting the two edges that share a vertex; we will usually denote these as $C_{k}$. These are the most important corner matrices in what follows, since every other corner matrix can be expressed in terms of them, and they are the ones that appear naturally in the equations.

\subsection{Self-consistency and the belief-propagation limit}

Next, let us discuss what we want from the message tensors. Our overall goal is to compute the contraction of the tensor network and its correlation functions (correlation functions are defined just as for the original tensor network, but with the tensors $T_{v}$ at several chosen vertices $v$ replaced by other ``impurity'' tensors $I_{v}$). To do this, we choose some small connected subgraph of the graph $G$ and replace the rest of the lattice by the message MPS defined on the polygonal line bounding the subgraph, as shown in Fig.~\ref{fig:mps_messages}(a). This boundary MPS can then be used to compute correlation functions, and, with a suitable normalization, the partition function $Z$ itself, as we describe below. In effect, the message MPS represents the information from outside the subgraph pouring into it --- the subgraph's environment.

When the virtual bond dimension is $\chi = 1$, the corner matrices $C$ become trivial scalars that can be absorbed into an overall normalization, and the message tensors reduce to ordinary belief propagation (BP) messages. The well-known BP equations can also be understood geometrically, as a self-consistency condition between the environments of different subgraphs: the message MPS of a smaller subgraph must agree with that of any larger subgraph containing it. This is exactly the principle we use to define our equations for the message tensors $A_{e}$ and the corner matrices $C$. In Fig.~\ref{fig:mps_messages}(b), for instance, we show two such subgraphs together with their message MPS environments; requiring the two descriptions to be consistent yields a set of self-consistency equations or update equations for the messages $A_{e}$.

Note that in this equation we label every message and corner matrix identically to keep the diagram simple; in general, these tensors and matrices differ from edge to edge and depend on their position in the graph. A few points are worth emphasizing. For every vertex $v$ and every outgoing edge $e$ there is one such equation; since this enumerates all edges in both directions, these equations --- together with the corresponding equations for the corner matrices $C$ --- fully determine the message tensors $A_{e}$. Each equation also introduces a normalization factor $\kappa$ (we postpone fixing the normalization of $A_{e}$ and $C$ until later), which corresponds to removing one site from the subgraph; the full partition function $Z$ is then given by the product of these factors $\kappa$ over all vertices, as we discuss more rigorously below.

\subsection{CTM equations for the corner matrices}

\begin{figure}
\centering
\includegraphics[width= 0.9\linewidth]{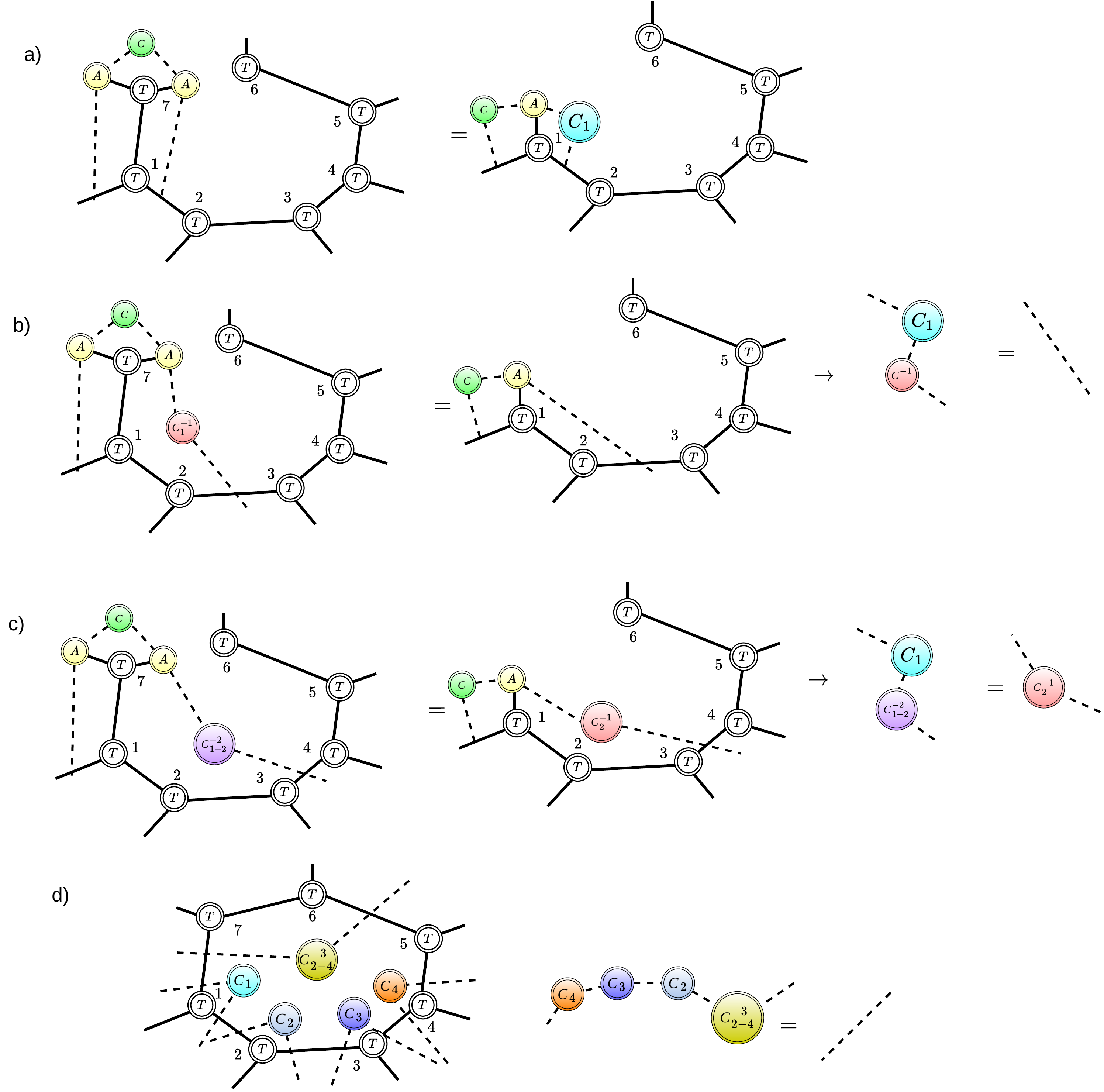} 
 \caption{\label{fig:corner_matrix_relations}%
   Derivation of the relations between different corner matrices: for a $7$-gon face with corner matrices $C_{1}, \dots, C_{7}$, an MPS message isolating a single vertex $7$ can be continued across different edges of the polygon, giving equivalent representations of the same environment --- with no corner matrix in the chosen gauge (a), with the inverse matrix $C_{1}^{-1}$ (red) (b), or with the two-index matrix $C_{1-2}^{-2} = C_{1}^{-1} C_{2}^{-1}$ (violet) (c). Comparing these equivalent representations fixes the relations between the different corner matrices. (d) An illustration of the general corner matrices, using $C^{-3}_{2-4}$ as an example: this is the corner matrix that lies on one side of vertices $2, 3, 4$, without touching them. It can in turn be represented as $C^{-1}_{2} C^{-1}_{3} C^{-1}_{4}$, where each matrix~$C_{n}$ corresponds to the pair of edges incoming to vertex $n$.}
\end{figure}

We now turn to deriving the equations for the corner matrices $C$. Before doing so, we first establish the relations between the different corner matrices, illustrated in Fig.~\ref{fig:corner_matrix_relations}(a-c). Consider a polygon with $7$ vertices and the $7$ corresponding principal matrices $C_{k}$, $k = 1, 2, \ldots, 7$. We can isolate one vertex of the polygon with an MPS message that crosses different edges inside the polygon, picking up different corner matrices along the way. Fig.~\ref{fig:corner_matrix_relations}(a-c) shows three such cases: the MPS message carries no corner matrix at all (the gauged case), the inverse matrix $C_{1}^{-1}$ (red), or the two-index matrix $C_{1-2}^{-2}$ (violet). In this notation, $C^{m-n-1}_{m-n}$ denotes the corner matrix connecting edges $m-1$ and $n+2$ (see also Fig.~\ref{fig:corner_matrix_relations}(d)); as we show below, this choice of notation directly reflects how the corner matrix is expressed in terms of the $C_{k}$.

These three cases must be mutually consistent, which constrains the corner matrices. Specifically, reducing the isolated vertex in each of the three cases using the equations for $A_{e}$ yields, in turn, an MPS message carrying the corner matrix $C_{1}$, one carrying the identity (no corner matrix), and one carrying $C_{2}^{-1}$. Consistency between the three cases and with the equations for $A_{e}$ then forces the corner matrix labeled $C_{1}^{-1}$ to coincide with the literal matrix inverse $(C_{1})^{-1}$ --- confirming that the notation was well chosen --- and further gives $C_{1-2}^{-2} = C_{1}^{-1} C_{2}^{-1}$, which explains the general notation. The same argument generalizes to $C^{m-n-1}_{m-n} = C_{m}^{-1} C_{m+1}^{-1} \cdots C_{n}^{-1}$. These relations show that every corner matrix can be expressed in terms of the simple matrices $C_{k}$, so that only $n$ independent equations need to be derived for each $n$-gon face of the graph.

\begin{figure}
\centering
\includegraphics[width= 0.8\linewidth]{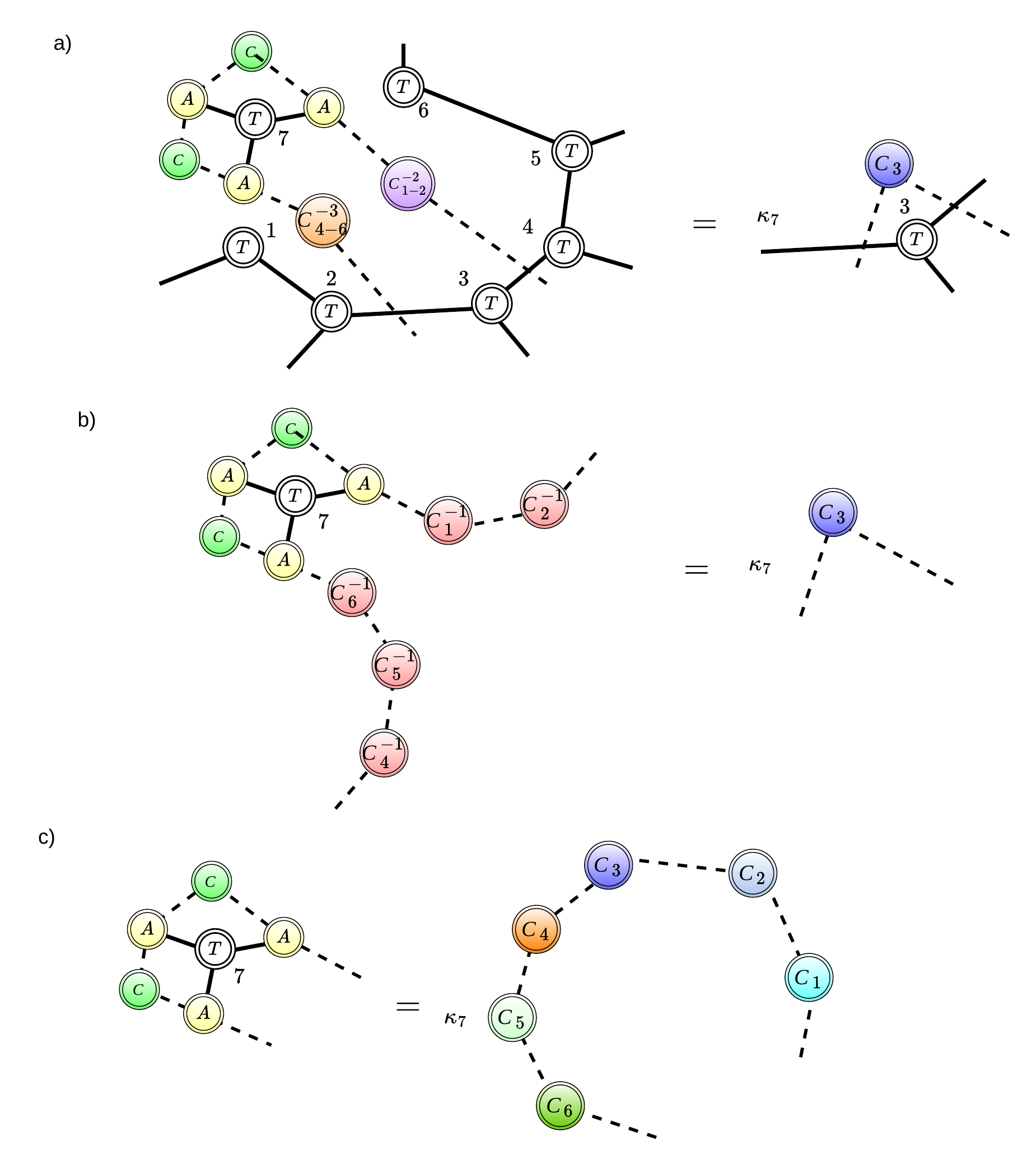} 
 \caption{\label{fig:CTM_cornerr}%
   The derivation of the CTM equations for the corner matrices. (a) Consider a face of the graph --- a polygon, here a $7$-gon --- with its vertices labeled for convenience. We isolate vertex $7$ with an MPS message and continue this MPS until it crosses bonds $2$--$3$ and $3$--$4$; consistency of the message environment then forces this isolated-vertex contraction to equal the $C_{3}$ corner matrix at vertex $3$. (b) Rewriting the matrices $C_{1-2}^{-2}$ and $C_{4-6}^{-3}$ in terms of $C_{1}, C_{2}, \ldots, C_{7}$ casts the equation purely in terms of the tensors $A$ and the matrices $C$. (c) The equivalent, most universal form of the equation: the MPS message isolating vertex $k$ equals the product of all the corner matrices around the $n$-gon except the one associated with (i.e., incoming to) vertex~$k$.}
\end{figure}
Figure~\ref{fig:CTM_cornerr} shows the derivation of the equations for the corner matrices $C_{k}$. In Fig.~\ref{fig:CTM_cornerr}(a), we again consider the $7$-gon face of the graph, with its vertices labeled. We isolate vertex $7$ with an MPS message and continue the MPS until it crosses the edges $(2,3)$ and $(3,4)$, inserting the relevant $C$ matrices for these edge pairs along the way. On the other hand, the whole isolated vertex should be replaceable by just the corner matrix $C_{3}$; this consistency requirement forces an equation relating the corner matrices and the message tensors. This equation is shown in Fig.~\ref{fig:CTM_cornerr}(b), where we have additionally rewritten the matrices $C_{1-2}^{-2}$ and $C_{4-6}^{-3}$ in terms of the $C_{k}$. Finally, Fig.~\ref{fig:CTM_cornerr}(c) shows the final, most universal form of the equation: the isolated vertex with its message MPS equals the product of all the simple corner matrices $C_{k}$ around the face, except for the one associated with the isolated vertex itself (here, $C_{7}$). Note that this derivation does not depend in any significant way on the fact that the faces are $7$-gons. The derivation can be repeated for any $n$-gon faces. 

Note that this equation does not single out $C_{3}$ in any way --- we could equally well have chosen the initial MPS message in Fig.~\ref{fig:CTM_cornerr}(a) to cross a different pair of edges, and the final equation in Fig.~\ref{fig:CTM_cornerr}(c) would be unchanged. This gives $n$ equations for every $n$-gon face of the graph, which, together with the equations for the message tensors $A_{e}$, fully determine the MPS message. We refer to these as the CTM (corner transfer matrix) equations, and we will show below that, for the square and honeycomb lattices, they reduce to Baxter's original CTM equations~\cite{Baxter1968dimers, Baxter1976ctm8v, Baxter1977ctm8vII, Baxter1978variational, Baxter1981ctm, Baxter1982book}. Although the CTM equations can in principle be solved directly, it is more convenient to introduce a few auxiliary tensors and recast them as a CTMRG algorithm.

\subsection{CTMRG projectors}

\begin{figure}
\centering
\includegraphics[width= \linewidth]{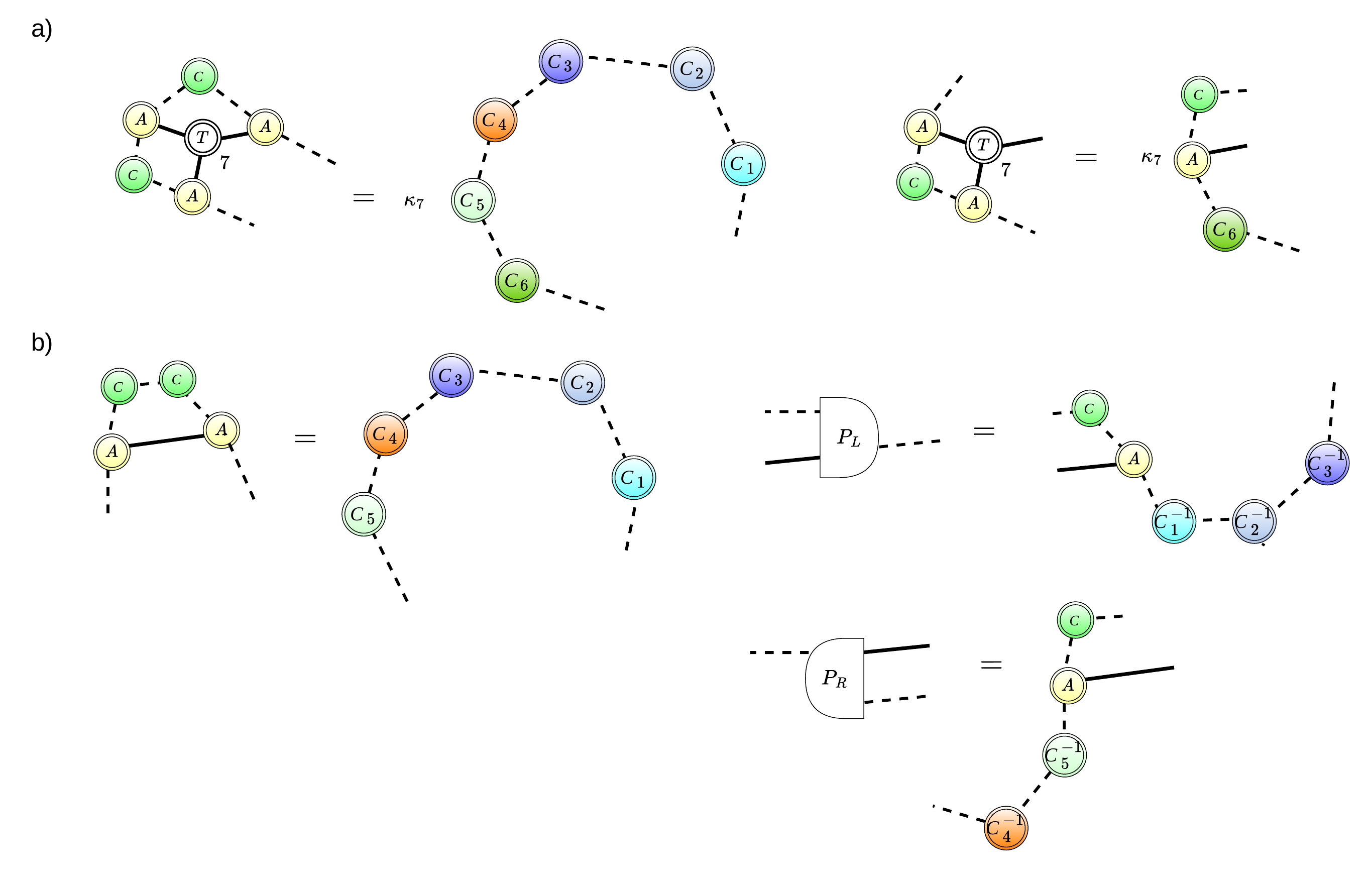} 
   \caption{\label{fig:Projectors}%
   The derivation of the CTMRG projectors. (a) The corner-matrix equation at vertex $7$, together with the update equation for the tensor $A$ at the same vertex. The diagram for the $A$ update is a subgraph of the diagram for the $C$ update, so the two equations can be combined. (b) Combining the two equations in (a) gives the biorthogonal relation shown here: with a suitable normalization, the vectors carried by two $A$ tensors on the same bond are mutually orthogonal. This relation lets us define the biorthogonal projectors $P_{L}, P_{R}$. Note that the two projectors carry different numbers of inverse $C$ matrices; this asymmetry is purely conventional --- we split the corner matrices evenly between $P_{L}$ and $P_{R}$, and when their number is odd we assign the extra matrix to $P_{L}$.}
\end{figure}
In addition to edge tensors and corner matrices, CTMRG usually operates with additional orthogonal or biorthogonal projectors $P$. Figure~\ref{fig:Projectors} shows how these projectors can be derived naturally from the CTM equations. In Fig.~\ref{fig:Projectors}(a) we show the corner-matrix equation and the message-tensor update equation, both taken at the same vertex of the graph. Since the diagram for the message equation is a subgraph of the diagram for the corner-matrix equation, the former can be substituted into the latter; the result is shown in Fig.~\ref{fig:Projectors}(b). We refer to this type of equation as a biorthogonal relation, since it relates a generalized inner product of the vectors carried by two message tensors $A_{e}$ to a product of $C$ matrices.

As shown in Fig.~\ref{fig:Projectors}(b), the biorthogonal relation allows us to define the biorthogonal projectors $P_{L}, P_{R}$. This definition is not unique --- we could, for instance, have inserted $C_{3}^{-1}$ into the definition of $P_{R}$ instead. Our convention is that when the number of corner matrices in the biorthogonal relation is even, we split them evenly between $P_{L}$ and $P_{R}$; when it is odd, we assign the extra matrix to $P_{L}$.

Unfortunately, this definition of the projectors is not very useful in practice, since it involves many inverse corner matrices, which can be ill-conditioned. In practice, CTMRG algorithms usually obtain the projectors from an auxiliary eigendecomposition problem instead. Our next goal is to show that the projectors defined above are in fact solutions of such an eigenproblem, which follows naturally from the corner-matrix update equations.

\begin{figure}
\centering
\includegraphics[width= 0.9\linewidth]{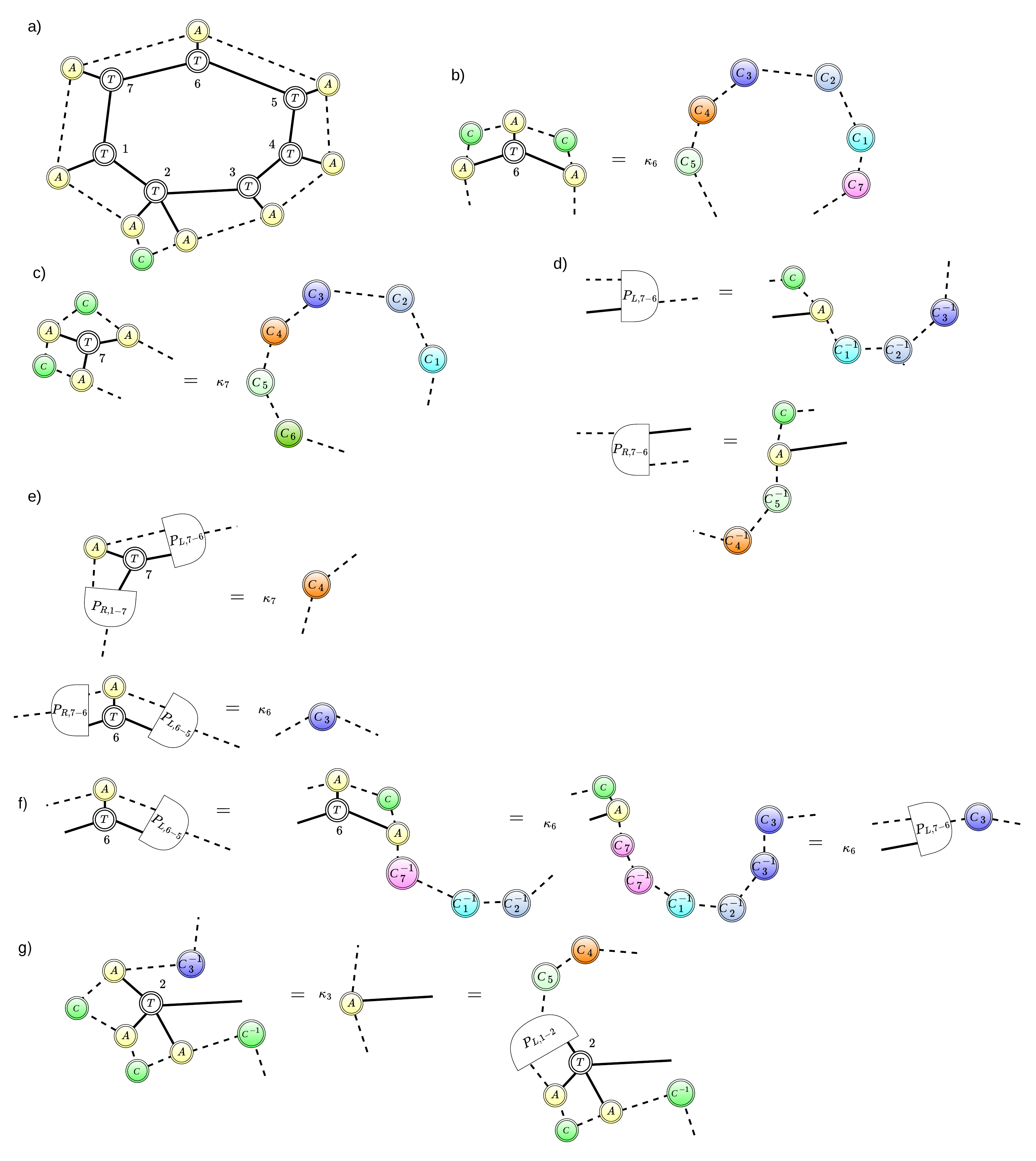} 
   \caption{\label{fig:CTM_eigenproblem}%
   The derivation of the CTM projectors via the auxiliary eigenproblem. (a) Consider again a polygon with an MPS message boundary. (b--c) For every vertex of the polygon one can form the corresponding corner equation; here we show these equations for vertices $6$ and $7$ as an example. On each polygon bond one can likewise define a pair of biorthogonal projectors; here we show the projectors on the bond connecting vertices $6$ and $7$. (d) In terms of these projectors, the update equations for the $C$ matrices can be reformulated: the new equation involves an ``enlarged'' corner, built from the $A$ messages incoming to the vertex along the outer bonds together with the vertex tensor $T$; this enlarged corner is then projected onto the polygon-bond indices using the newly defined projectors. (e)  The update equations for the corner matrices in terms of the newly defined projectors at vertices $6$ and $7$, giving the new corner matrices $C_{3}, C_{4}$. (f) This projection can also be shown to correspond to the truncated simultaneous diagonalization of all the enlarged corner matrices around the polygon: applying one projector to an enlarged corner yields the projector on the next bond, times the (truncated) corner matrix. This system of equations, coupled across all the enlarged corners of the polygon, constitutes the eigenvalue problem that defines the projectors. (g) The projectors can also be reinserted into the update equations for $A$.}
\end{figure}
Figure~\ref{fig:CTM_eigenproblem}(a) shows a face of the graph together with the corresponding MPS message on its boundary; the projectors can be constructed solely from the tensors shown in this panel. First, recall the update equations for the $C$ matrices, shown in Fig.~\ref{fig:CTM_eigenproblem}(b--c) for vertices $6$ and $7$ (the equations for the other vertices follow analogously). The same panels also show, again, the definition of the projectors on the bond connecting vertices $6$ and $7$; all other corner updates and bond projectors can be constructed completely analogously, by rotating these definitions around the polygon.

Now, in Fig.~\ref{fig:CTM_eigenproblem}(e) we rewrite the update equations for the corner matrices in terms of the newly defined projectors: the new matrices $C_{3}, C_{4}$ are constructed as projections of the tensors $T, A$ from both sides. There is one important point in this construction: even before projection, the contraction of these combinations of $T, A$ (and, in some cases, additional $C$ matrices from other polygons, as for vertex $2$) already reproduces the tensor network shown in Fig.~\ref{fig:CTM_eigenproblem}(a). We refer to these unprojected blocks as the ``enlarged corner matrices''; the face tensor network in Fig.~\ref{fig:CTM_eigenproblem}(a) is then simply the product of the enlarged matrices at all $7$ vertices. Hence, in general, the new corner matrices are obtained as projections of the enlarged ones.

Let us now show that the projectors indeed correspond to eigenvectors of these enlarged matrices; this is illustrated in Fig.~\ref{fig:CTM_eigenproblem}(f). Here we apply the projector $P_{L,6-5}$ from bond $6-5$ to the enlarged corner matrix at vertex $6$. Using the expression for $P_{L}$ in terms of $A, C$ together with the message update equation for $A$, one can show that this projector application equals the projection operator $P_{L,7-6}$ on bond $7-6$, times the matrix $C_{3}$ --- the same $C_{3}$ matrix obtained in Fig.~\ref{fig:CTM_eigenproblem}(e) from the projection of the enlarged corner matrix at vertex $6$. We can now apply $P_{L,7-6}$ again to the enlarged matrix at vertex $7$ and obtain the new projector on bond $1-7$. Repeating this procedure for all the enlarged corners in counterclockwise order, we eventually return to the original projector $P_{L,6-5}$, multiplied by the product of all the corner matrices $C_{k}$ around the face. If we additionally choose the gauge in which all the matrices $C_{k}$ are diagonal, this becomes the eigendecomposition of the product of all the enlarged corner matrices over the face: the projectors $P_{L}, P_{R}$ are the leading left and right eigenvectors of this product, and the corner matrices $C_{k}$ are the corresponding eigenvalues. This constitutes the auxiliary eigenproblem for every graph face, which can be used to obtain the projectors on the face bonds together with the face corner matrices.

Finally, as shown in Fig.~\ref{fig:CTM_eigenproblem}(g), the projectors can also be used to reformulate the update equations for the message tensors $A_{e}$; we discuss this further below, in the presentation of specific examples.

\subsection{The CTMRG algorithm}

As a result, the general theory for deriving the equations for the MPS messages on a particular graph or lattice runs as follows:
\begin{enumerate}
\item Fix the partial gauge to remove the corner matrices of the form $C_{e_{1}, \bar{e}_{3}}$.
\item Use the equations for the $C$ matrices to rewrite most of them in terms of the simplest ones, $C_{k}$.
\item Write down the self-consistency equations for all the message tensors $A_{e}$ and the remaining corner matrices.
\item For each vertex $v$, pair the corner-matrix equation with the message-update equation to derive the biorthogonal relation for each graph edge, and define the corresponding projectors.
\item Form the auxiliary eigenproblem for every graph face, and use it to find the projectors and corner matrices on that face.
\end{enumerate}

\begin{figure}
\centering
\includegraphics[width= 0.7\linewidth]{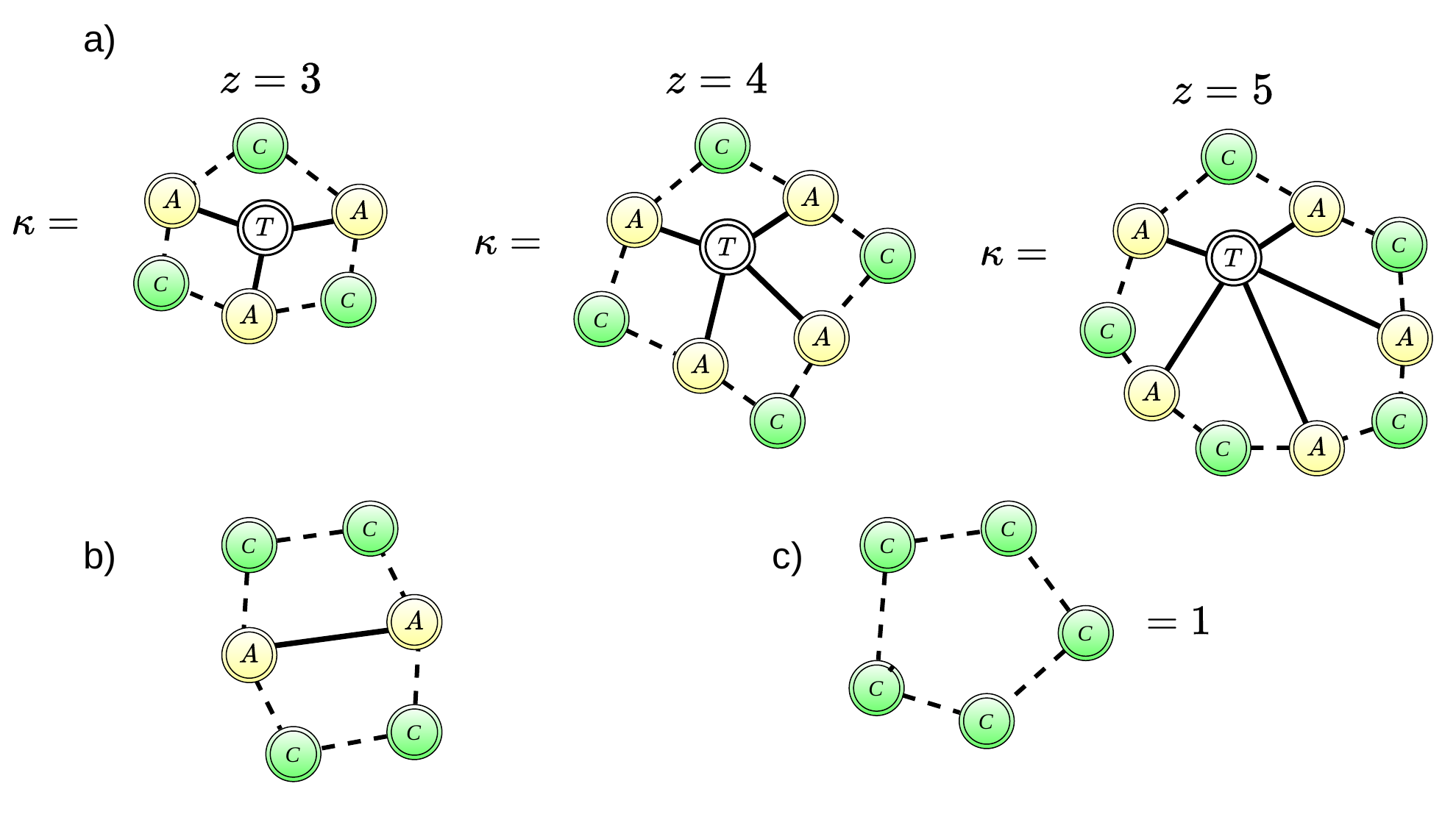} 
 \caption{\label{fig:CTM_normalization}%
      Calculation of the partition function per site $\kappa$ and of the normalization conventions for corner matrices $C$ and message tensors $A$. (a) Expression for $\kappa$ at different valency~$z$ of the vertex. (b) Normalization condition on every graph edge. (c) Normalization condition for the corner matrices around the face. Note that here we illustrate it using a pentagon as an example. For a general $p$-gon, the product of the $p$ corner matrices is equal to $1$.}
\end{figure}

Solving these equations in practice requires a few additional steps, which together make up the general CTMRG cycle:
\begin{enumerate}
\item Solve the eigenproblem for every graph face. Once solved, impose the additional gauge choice that all the matrices $C_{k}$ within the same face are equal; this fixes the new corner matrices and projectors. Note that the gauge of the projectors is not yet fully fixed at this point, as we discuss in detail in Appendix~\ref{app:residual_gauge}.
\item Use the projectors, together with both the new and old corner matrices, to update all the edge messages $A_{e}$ via the message equations. Because the gauge of the projectors is not fully fixed, the resulting message tensors may still differ in gauge from the previous iteration; we fix this residual gauge by requiring the new message tensors to be element-wise close to those from the previous step.
\item Normalize all the message tensors and corner matrices, as shown in Fig.~\ref{fig:CTM_normalization}, and compute the factors $\kappa$; the algorithm's convergence can then be monitored from the behavior of $\kappa$.
\end{enumerate}

Below, we illustrate this CTMRG construction on a variety of example lattices and graphs. Along the way, we discuss additional subtleties related to reflection and rotational symmetries and to trivalent vertices, and we show that the construction reduces to previously known CTMRG algorithms for the square, honeycomb, and triangular lattices.

\section{Examples}
\subsection{Square lattice}
\begin{figure}
\includegraphics[width= \linewidth]{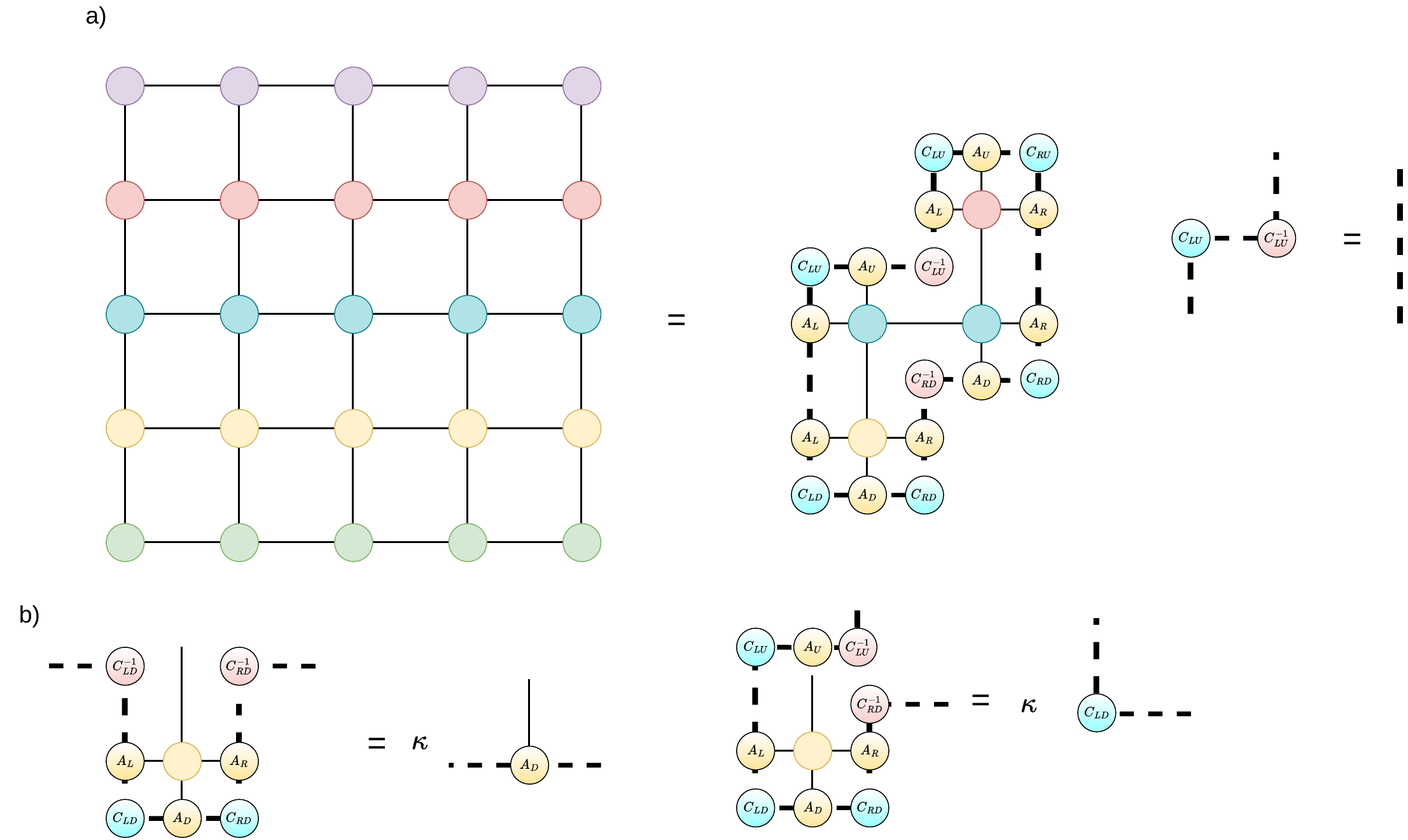} 
 \caption{\label{fig:CTM_square}%
   Derivation of the square-lattice CTM equations as a message-passing algorithm. (a) The finite $5 \times 5$ square graph and its subgraph with the MPS message environment; the environment consists of both conventional $C$ matrices and inverse $C^{-1}$ matrices. (b) Update equations for the message tensors $A$ and the corner matrices~$C$. }
\end{figure}

Let us first consider a finite $L \times L$ square lattice. There are vertical and horizontal edges, each with two directions, so we can introduce four different MPS messages, which we call left, right, up, and down according to the direction of the corresponding edge. The partial gauge on the square lattice is likewise natural: we fix the $C$ matrices connecting opposite edges of the square to Kronecker deltas, leaving the remaining $C$ matrices as our non-trivial degrees of freedom. The typical MPS message for a region of the square lattice is shown in Fig.~\ref{fig:CTM_square}(a); note that it also involves $C^{-1}$ matrices, shown in pink.

\begin{figure}
\includegraphics[width= \linewidth]{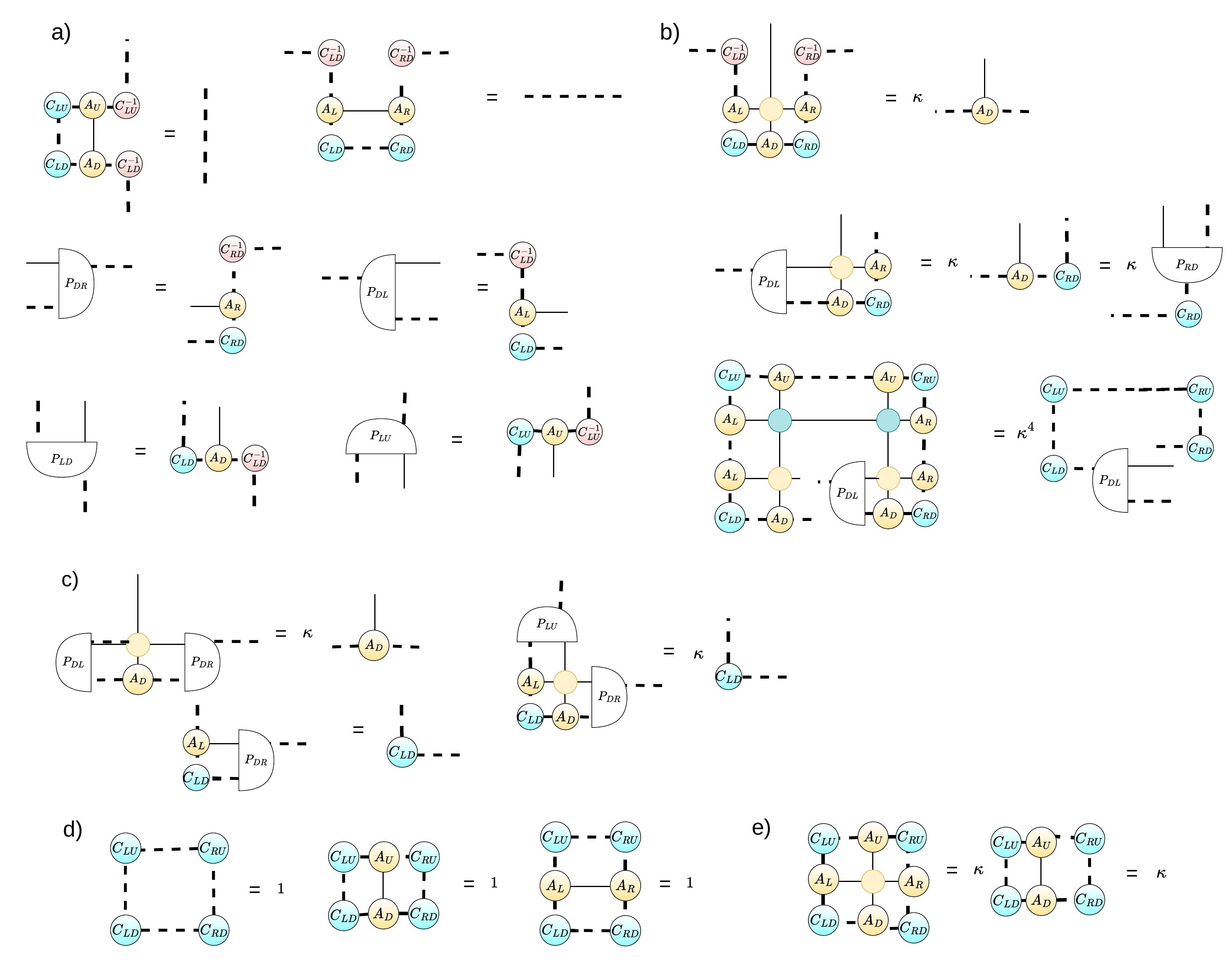} 
 \caption{\label{fig:CTMRG_square}%
   Derivation of the square-lattice CTMRG.  (a) Combining the equations in Fig.~\ref{fig:CTM_square}(b) makes it possible to construct the biorthogonal projectors. (b) Using the representation of the projectors in terms of $A, C$ together with the update equations for $A$, one can show that the projectors are matrices of eigenvectors of the enlarged corner matrices, obtainable from the eigenvalue problem for the enlarged corner matrix of the full square-lattice plaquette. (c) The newly obtained projectors are then used again in the update equations for $A, C$. (d) ``Canonical'' normalization conditions for $C, A$ can be derived that agree both with each other and with the projector definition. (e) With this canonical normalization, the partition function per site $\kappa$ can be computed from a simple MPS message contraction around the lattice site.}
\end{figure}

The next step is to construct the CTM equations for the messages $A$ and matrices $C$; these are shown in Fig.~\ref{fig:CTM_square}(b) and follow directly from the general construction of Sec.~\ref{sec:vertex_models}. The equations for $C$ can be reduced in two different ways using the equations for $A_{e}$, as shown in Fig.~\ref{fig:CTMRG_square}(a); together these yield the biorthogonal relations, which in turn define the biorthogonal projectors $P_{L}, P_{R}$. Note that the projectors are defined in terms of the message tensors $A_{e}$, the $C$ matrices, and their inverses. Because this expression involves inverses of $C$, it is not very numerically stable --- especially since the spectrum of $C$ typically decays exponentially in practice.

To avoid this, we instead treat the projectors as independent variables and derive equations for them directly from the equations for the message tensors $A_{e}$; an example of such an equation is shown in Fig.~\ref{fig:CTMRG_square}(b). These projector equations mean that the projectors project onto the dominant subspace of the enlarged corner matrices, so they can equivalently be defined as the dominant eigenspaces of the enlarged corner matrices. The projectors can be chosen so as to simultaneously diagonalize --- and equalize --- all four projected corner matrices on the same square-lattice face, which nearly fully fixes their gauge. We fix the remaining residual gauge freedom by comparing the new MPS tensors with their values from the previous iteration and adjusting the gauge so that the tensors converge element-wise.

Finally, the original equations for $C$ and $A_{e}$ can also be represented with the help of the projectors, as shown in Fig.~\ref{fig:CTMRG_square}(c); we also show there a version of the biorthogonal relation with a single projector, used in some variants of square-lattice CTMRG. These equations are identical to the standard CTMRG equations, differing only in how the projectors are defined.

\begin{figure}
\includegraphics[width= 0.8\linewidth]{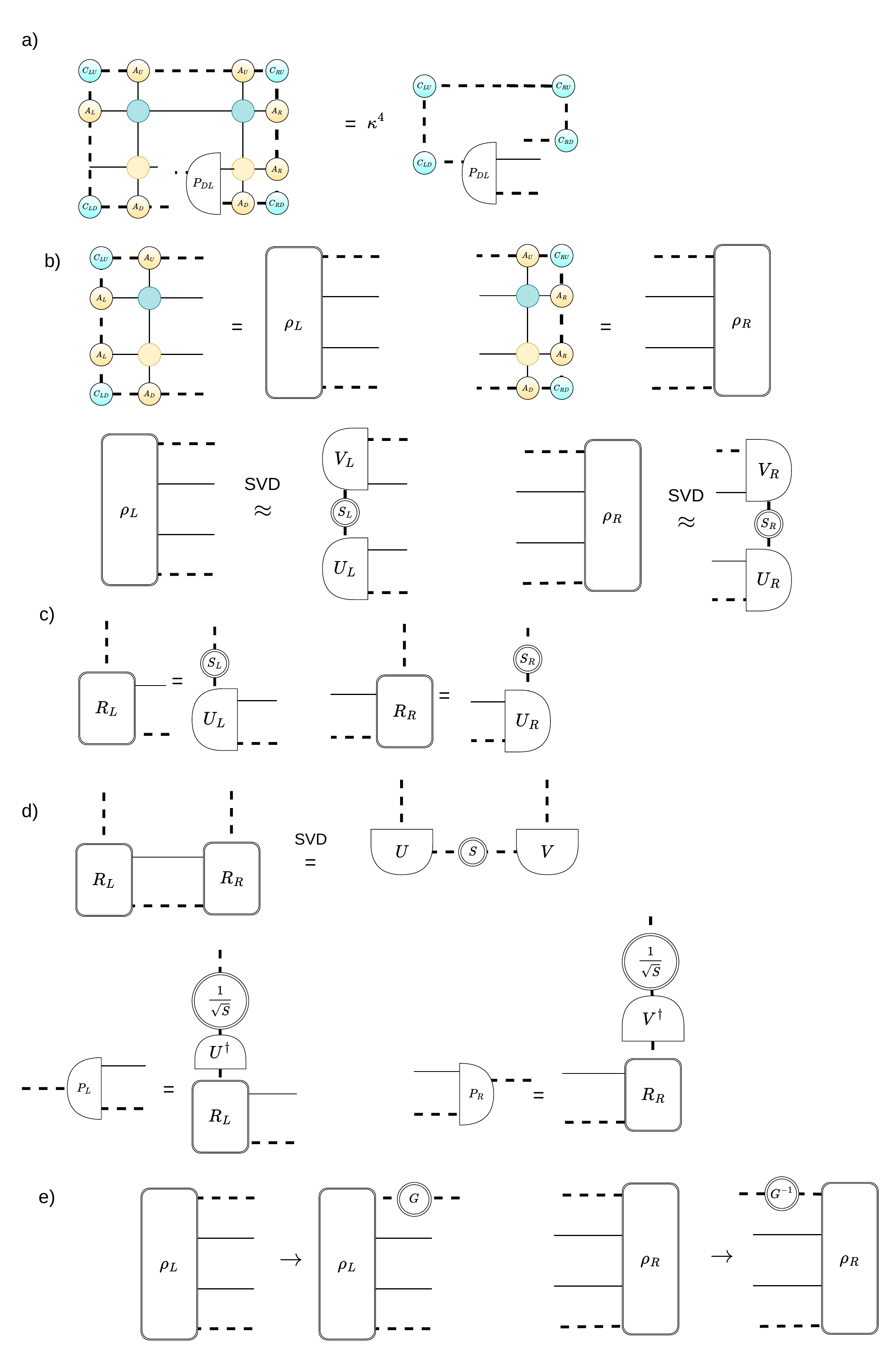} 
  \caption{\label{fig:Projectors_square}%
   Different definitions of the CTMRG projectors. (a) In our construction, the projectors are derived from the eigenvectors of the product of the enlarged corner matrices around the face (cf.\ Fig.~\ref{fig:CTM_eigenproblem}). (b) The most widely used definition, from Ref.~\cite{corboz2014competing}, instead first constructs the half-system tensors $\rho_{L}, \rho_{R}$, which are then decomposed via SVD, with possible truncation. (c) This decomposition defines new tensors $R_{L}, R_{R}$. (d) Finally, $R_{L}, R_{R}$ are biorthogonalized to form the projectors; during this step, the projected bond is truncated according to the spectrum $S$. (e) A gauge transformation $G$ on the virtual index of $\rho_{L}, \rho_{R}$, under which this alternative construction is not invariant --- unlike the eigenvector-based definition in (a).}
\end{figure}

While the eigendecomposition-based definition of the projectors is generally accepted in the rotationally symmetric, Hermitian case, the more general non-symmetric case usually instead uses a different definition~\cite{corboz2014competing}, shown in Fig.~\ref{fig:Projectors_square}(b--d). One could argue that our definition is theoretically preferable, since it is invariant under the gauge transformations shown in Fig.~\ref{fig:Projectors_square}(e), whereas the earlier approach is generally not.

In practice, however, both constructions project onto approximately the same space, because the eigenvalues of $\rho_{L}, \rho_{R}$ typically decay very quickly. To see this, consider matrices $\rho_{L}, \rho_{R}$ with rapidly decaying SVD spectra $S_{L}, S_{R}$. Such matrices are effectively low-rank, and hence so is their product. This low-rank property means that the eigenvectors of the combined matrix $\rho_{L} \times \rho_{R}$ and the matrices $U_{L}, U_{R}$ span the same linear space wherever these matrices have non-trivial rank --- and it is precisely this space that the projectors correspond to.

Let us now discuss the normalization of the message tensors and the corner matrices. The message tensors have a natural normalization condition inherited from their relation to the projectors, which fixes their overall scale (see Fig.~\ref{fig:CTMRG_square}(d)). For the corner matrices, we take the product of the four $C$ matrices around a graph face and its trace, and normalize this trace to unity. This is a natural choice, since such a contraction of corner matrices corresponds to the MPS message environment of the empty subgraph, and it is natural to assign it the value $1$, since the partition-function factors are expected to already be extracted in the form of the $\kappa$ associated with each vertex.

Finally, from the equations for the edge tensors $A_{e}$ it is possible to find the partition function per vertex $\kappa_{v}$. The overall partition function can then be computed as a product over the vertices: $\prod_{v} \kappa_{v}$. The definition of $\kappa_{v}$ is shown in Fig.~\ref{fig:CTMRG_square}(e). Note that this definition can be simplified in multiple ways using the equations for $A_{e}$, which simultaneously forces the $\kappa_{v}$ appearing in the equations for different $A_{e}$ at the same vertex $v$ to coincide.

\begin{figure}
\centering
\includegraphics[width= 0.4\linewidth]{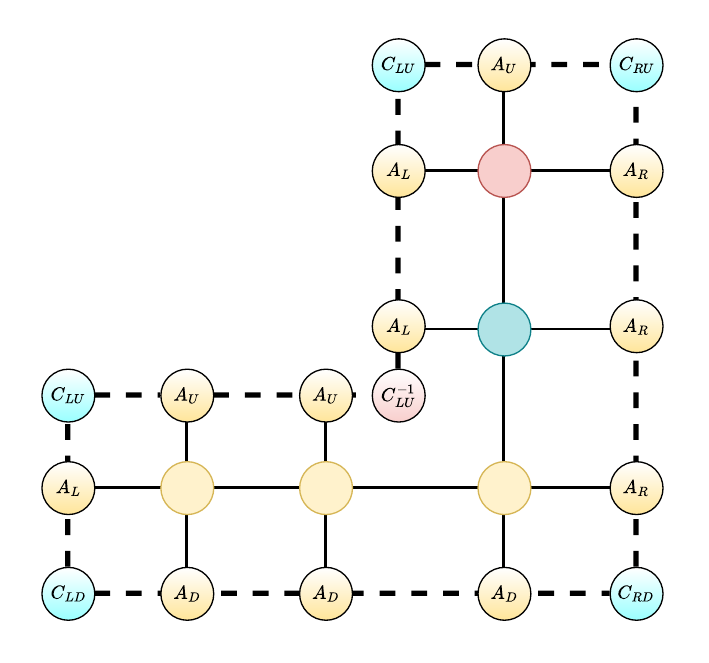} 
 \caption{\label{fig:LR_correlator}%
   MPS messages together with both $C$ and $C^{-1}$ corner matrices can be used to compute long-range correlation functions. Here the environment traces an L-shaped path: a horizontal MPS segment (bottom left, built from plain bulk tensors) is joined to a vertical MPS segment (right, containing the two colored ``impurity'' tensors marking the two correlator sites) via the inverse corner matrix $C_{LU}^{-1}$, which lets the boundary turn the corner between the two lattice directions. More generally, gluing straight MPS segments together with $C$ and $C^{-1}$ corner matrices at each bend lets the environment trace an arbitrary path on the lattice, giving access to the correlation function between any two sites --- not only those separated along a single row or column.}
\end{figure}

There is an additional point about this definition of MPS messages worth discussing. These tensors can be used to measure arbitrary two-site correlation functions, as shown in Fig.~\ref{fig:LR_correlator}. This construction closely resembles the gradient-computation methodology of Ref.~\cite{2016_Vanderstraeten_gradients}, which was one of the motivations for this paper. We discuss the connection between our approach and that of Ref.~\cite{2016_Vanderstraeten_gradients} further in Appendix~\ref{app:inverse_matr}.

\subsection{Hyperbolic $(4, 5)$ lattice}

\begin{figure}
\includegraphics[width= \linewidth]{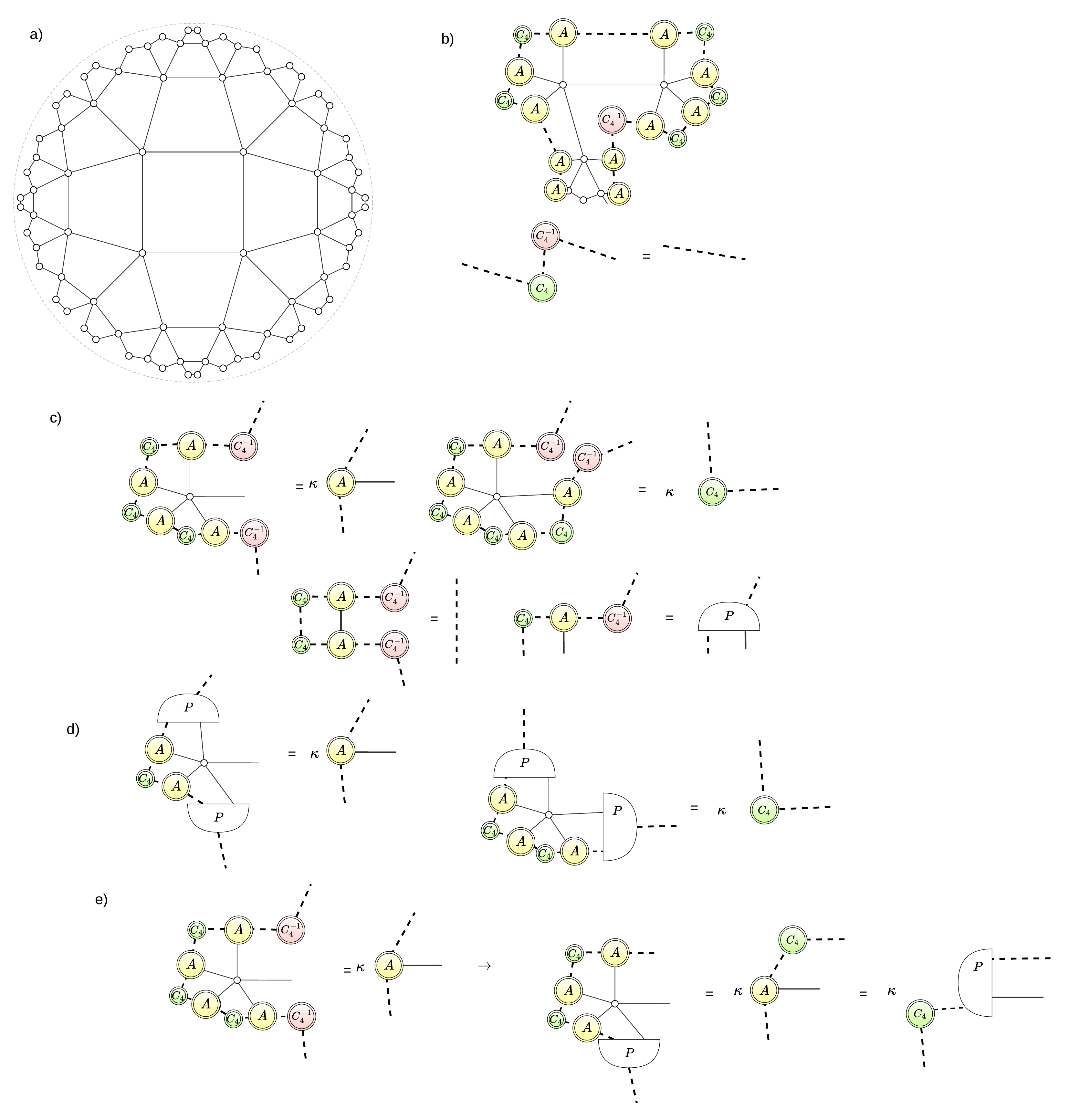} 
  \caption{\label{fig:CTM_hyperbolic_4_5}%
   Derivation of the CTM equations on the hyperbolic $(4, 5)$ lattice, assuming all lattice symmetries. (a) The hyperbolic $(4, 5)$ lattice. (b) The lattice subgraph with the corresponding MPS message environment, illustrating the connections between different $A$ tensors for different line configurations; the matrix $C^{-1}_{4}$ is the inverse of $C_{4}$. (c) The general CTM equation for the $A$ MPS messages and for the $C_{4}$ matrix; combining these equations, one can derive the projectors $P$, which take the same form as in the square-lattice case. (d) The projectors can be reinserted into the update equations for the message tensors $A$ and the $C_{4}$ matrix. (e) The projectors can also be derived from the diagonalization of the enlarged corner matrix.}
\end{figure}
The hyperbolic $(4, 5)$ lattice is shown in Fig.~\ref{fig:CTM_hyperbolic_4_5}(a). It consists of square faces meeting at valence-$5$ vertices. Let us assume for simplicity that all the tensor-network vertices are identical and invariant under rotations and reflections. In general, these assumptions can be relaxed, but imposing them here lets us derive a very simple system of just two equations, which can easily be solved numerically. Since the faces of the graph are identical (all squares), the equations for the corner matrices are evidently very similar to the square-lattice case. As a result, CTMRG on this lattice closely resembles the square-lattice case, but with a couple of interesting differences. The CTMRG on $(4, n)$ lattices with $n > 5$ can be constructed completely analogously.

The typical MPS message environment on the $(4, 5)$ lattice is shown in Fig.~\ref{fig:CTM_hyperbolic_4_5}(b). As in the square-lattice case, there are two relevant corner matrices, denoted $C_{4}$ and $C^{-1}_{4}$, where $C^{-1}_{4}$ is the inverse of $C_{4}$. The equations for the messages $A_{e}$ and the corner matrices are then shown in Fig.~\ref{fig:CTM_hyperbolic_4_5}(c). The main difference from the square-lattice case is that the equations now involve more $A$ tensors: for a $z$-valent vertex, the update equation for $A_{e}$ involves $z-1$ message tensors, while the update equation for the corner matrix involves $z$ of them.

Combining the equations for $A$ and $C$ yields the biorthogonal relations. In fact, these biorthogonal relations and the corresponding projectors are identical to those on the square lattice. This is, in fact, a general rule: the biorthogonal relations depend only on the number of boundary edges of the corresponding face, so any $(4, n)$ lattice has the same biorthogonal relations and the same definition of projectors, since these lattices are all formed from square faces. Fig.~\ref{fig:CTM_hyperbolic_4_5}(d) then shows how the projectors can be used to rewrite the update equations for $A_{e}$ and $C_{4}$, and Fig.~\ref{fig:CTM_hyperbolic_4_5}(e) further confirms that the projector diagonalizes the enlarged corner matrix. Together, these establish how to construct the projectors on this lattice and use them to update the corner matrices and message tensors.

Note that this algorithm precisely matches dualization of the face model CTMRG from Ref.~\cite{Ueda2007hyperbolic} on the dual $(5, 4)$ lattice.

\subsection{Hyperbolic $(5, 4)$ lattice}

\begin{figure}
\includegraphics[width= \linewidth]{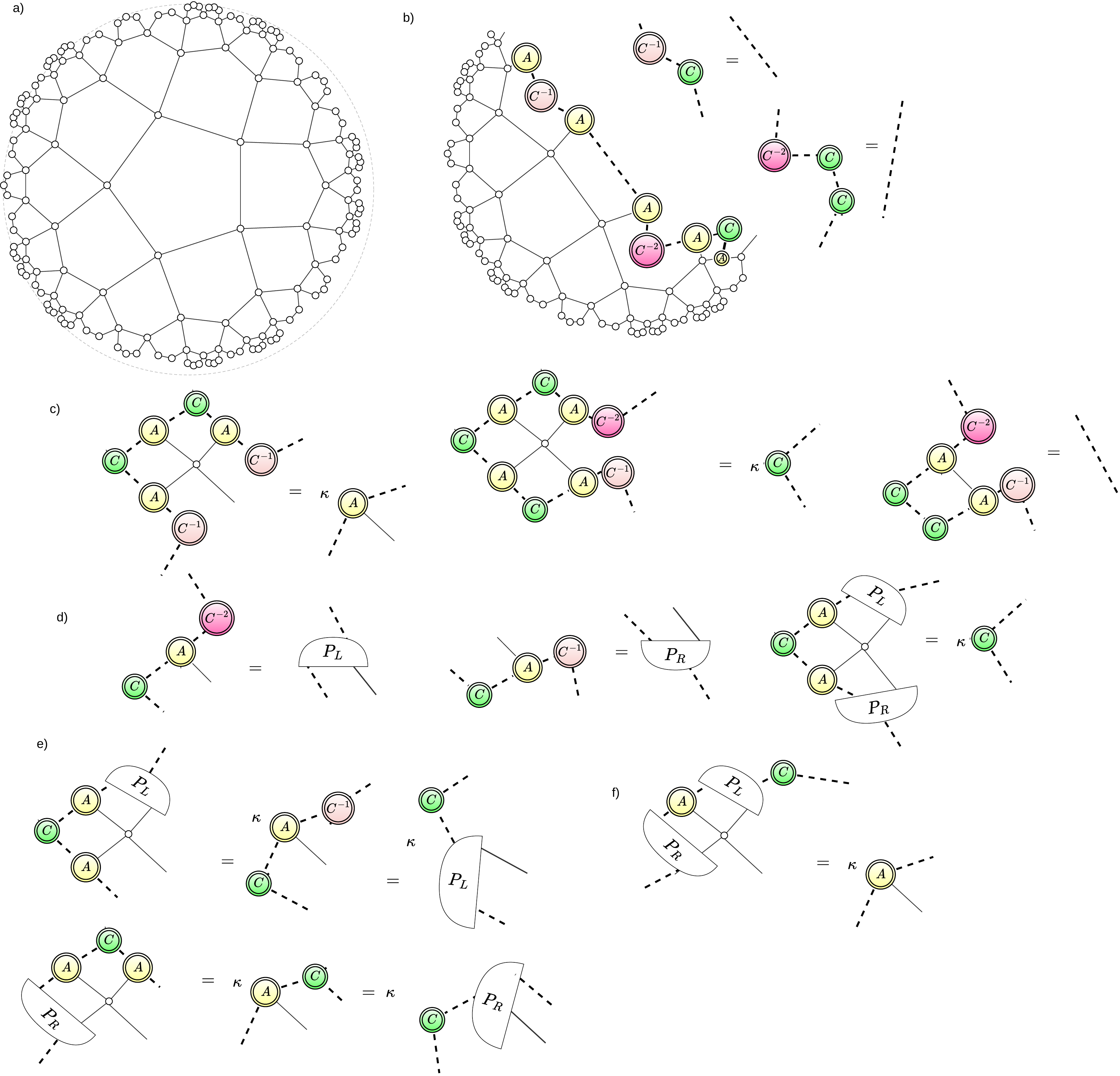} 
\caption{\label{fig:CTM_hyperbolic_5_4}%
   Derivation of the CTM equations on the hyperbolic $(5, 4)$ lattice, assuming all lattice symmetries. (a) The hyperbolic $(5, 4)$ lattice. (b) Part of the lattice together with its MPS message boundary, showing the different corner matrices $C, C^{-1}, C^{-2}$ and their relations. (c) The general CTM equations for the $A$ MPS messages and for the corner matrices $C$; combining these equations also makes it possible to derive the biorthogonal relation defining the projectors. (d) Definition of the biorthogonal projectors $P_{L}, P_{R}$ (note their asymmetric definitions) and the update equation for $C$ in terms of these projectors. (e) The eigenvalue equations defining the projectors $P_{L}, P_{R}$ via the enlarged corner matrices. (f) The update of the message tensor $A$ in terms of the new projectors.}
\end{figure}
The hyperbolic $(5, 4)$ lattice, and similar $(n, 4)$ lattices in general, consist of $n$-gons meeting at $4$-valent vertices. The lattice is shown in Fig.~\ref{fig:CTM_hyperbolic_5_4}(a). The equations for $A_{e}$ are quite similar to those on the square lattice (since the vertices are $4$-valent), while the structure of the corner matrices is quite different. Let us simplify things a bit and consider only tensor networks that are invariant under rotations and translations. We do not impose reflection symmetry here, since it is broken anyway by the asymmetric choice of projectors. The construction can be generalized further, but the resulting equations become considerably more cumbersome; our goal here is simply to illustrate a more general case of the projector construction and of the $C$-matrix relations.

The typical MPS message on this lattice is shown in Fig.~\ref{fig:CTM_hyperbolic_5_4}(b). There are several different $C$ matrices, corresponding to different combinations of connected edges; in particular, the figure shows the corner matrices $C$, $C^{-1}$, and $C^{-2}$, where $C^{-1}$ and $C^{-2}$ are simply the inverse powers of $C$.

The equations for $A_{e}$ and the $C$ matrices, together with the biorthogonal relation, are shown in Fig.~\ref{fig:CTM_hyperbolic_5_4}(c). There is, however, a problem with this biorthogonal relation: it involves asymmetric powers of the corner matrices. This prevents us from defining the left and right projectors $P_{L}, P_{R}$ symmetrically, since doing so would require a non-integer power $C^{3/2}$ of a corner matrix that need not be positive (as we will see below, for the triangular lattice it does make sense to define half-integer powers of the corner matrix, since positivity can be established there).

This means we are forced to introduce the projectors asymmetrically from the start, as shown in Fig.~\ref{fig:CTM_hyperbolic_5_4}(d); the same panel also shows the update of the corner matrices in terms of these projectors. Note that, in principle, we do not need to introduce the projectors at all: it is possible to simply solve the CTM equations for $A_{e}$ and $C$ directly. Such a procedure is fully correct in principle, but can be numerically unstable, since the defining equations involve matrix inverses.

With the projectors defined asymmetrically, we can write new equations for the matrices $C$ and the messages $A$, shown in Fig.~\ref{fig:CTM_hyperbolic_5_4}(d, f). Note that these equations are not reflection symmetric, as a consequence of the non-reflection-symmetric choice of projectors. Finally, the projectors themselves can be found from the diagonalization of the enlarged corner matrix, exactly as before, as shown in Fig.~\ref{fig:CTM_hyperbolic_5_4}(e). Note also that for $(n, 4)$ hyperbolic lattices with even $n$, it is, in fact, possible to define the projectors symmetrically.

The crucial point we want to emphasize about the $(5, 4)$-lattice equations is that the message-tensor update equation shown in Fig.~\ref{fig:CTM_hyperbolic_5_4}(f) contains one additional corner matrix $C$, applied after the projectors. This extra factor compensates for an asymmetry in the projector definitions: the CTM equations for $A$ involve only a single inverse power of $C$ on the left and right virtual indices, whereas the left projector is defined using the \emph{second} inverse power of the corner matrix. In the numerical implementation, this compensating $C$ must be the already-projected corner matrix, i.e., the one expressed in the same gauge as the projectors' outer index. More generally, for $n$-gons one needs to introduce $n-4$ such compensating corner matrices; for triangular faces in particular, this amounts to one additional $C^{-1}$ factor.

\subsection{Honeycomb lattice}

\begin{figure}
\includegraphics[width= 0.9\linewidth]{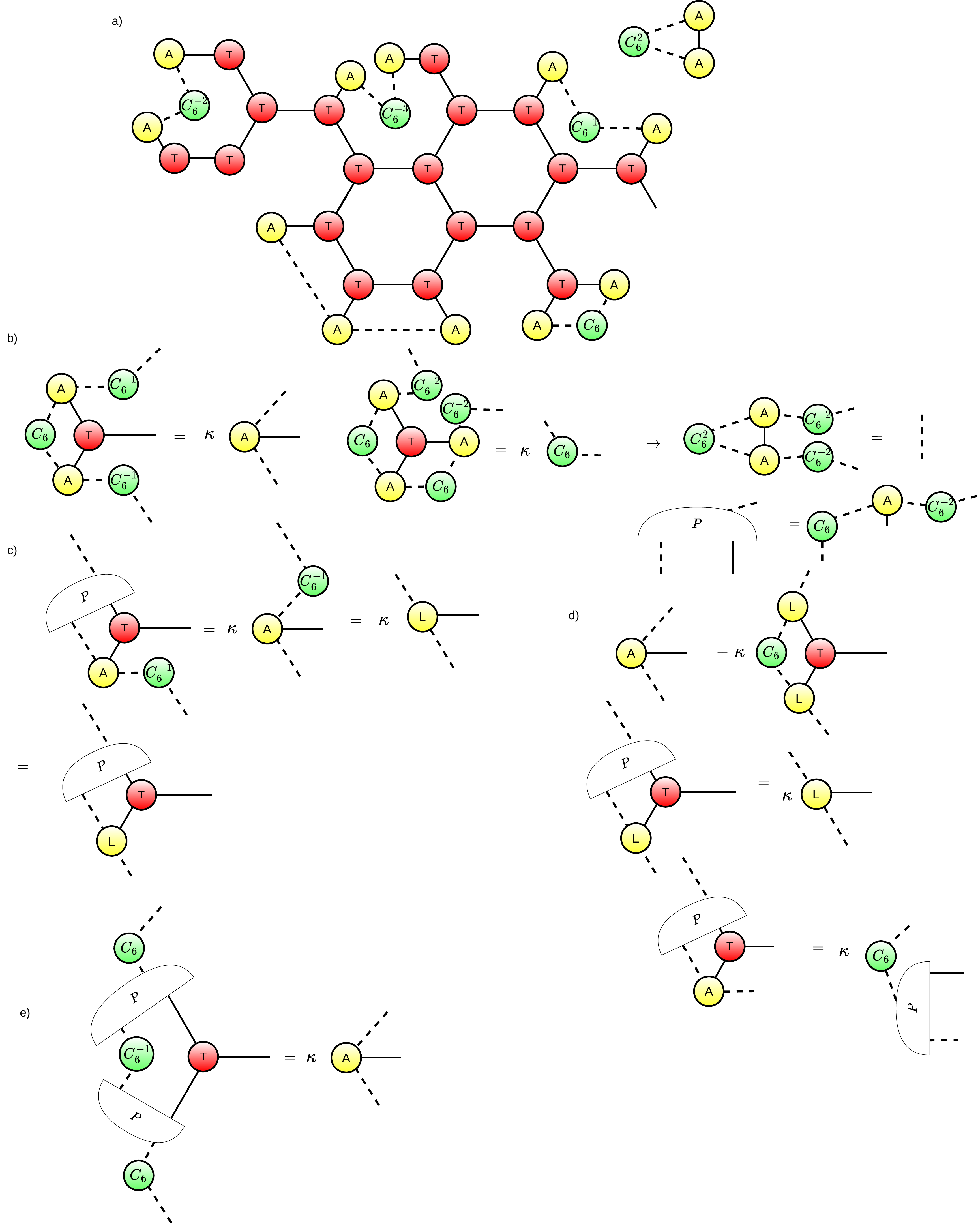} 
\caption{\label{fig:CTM_honeycomb}%
   CTM equations on the honeycomb lattice. (a) The honeycomb lattice, assumed translationally, rotationally, and reflection symmetric, showing the different combinations of message tensors $A$ and corner matrices appearing in the message environments. (b) The update equations for the message tensor $A$ and for the corner matrix $C_{6}$; combining these equations also yields the projectors $P$. (c) Since the honeycomb lattice has three-valent vertices, the usual formulation of the $A$ update with two projectors is singular. It is instead easier to define the update equation with a single projector $P$, and to combine $A$ with the inverse of $C_6$ into a new tensor $L$ (and its mirror reflection $R$); the update equation for $L$, shown here, uses one projector $P$ and involves no $C$-matrix inverses. (d) The full set of equations connecting $C, A, L, P$, constituting the CTMRG algorithm on the honeycomb lattice. (e) Alternative update of messages with two projectors and inverse matrix between them. Note the presence of compensating corner matrices after projectors.}
\end{figure}

In this subsection, we consider a translationally, rotationally, and reflection-invariant tensor network on the honeycomb lattice. The typical MPS message environments are shown in Fig.~\ref{fig:CTM_honeycomb}(a); the environment consists of the message tensors $A$, the simple corner matrix $C_{6}$, and its various inverse powers (up to $C_{6}^{-3}$).

This choice of tensors already appears in the original work of Baxter and Tsang on CTM for IRF models on the triangular lattice~\cite{Tsang1977triangular, Baxter1980hardhexagons} (recall that IRF CTMRG on the triangular lattice is dual to vertex CTMRG on the honeycomb lattice).

The defining equations for the corner matrix $C_{6}$ and for the tensors $A_{e}$ are shown in Fig.~\ref{fig:CTM_honeycomb}(b). Together, these equations also yield the biorthogonal relation and the definition of symmetric projectors. Note that these CTM equations already appeared in Refs.~\cite{Tsang1977triangular, Baxter1980hardhexagons}. The projectors can be reinserted into the equation for $C_{6}$, and can be obtained naturally from the truncated decomposition of the enlarged corner matrices.

It is somewhat more difficult, however, to express the equations for $A_{e}$ in terms of projectors, since the honeycomb lattice has $3$-valent vertices. For a $3$-valent vertex, the update equation for the new $A$ tensor involves only two old $A$ tensors, connected by a single $C_{6}$ matrix. If we tried to replace these old $A$ tensors with projectors, as in the square-lattice case, no old $A$ tensor would remain in the update equation --- and worse, the resulting equation would involve an inverse power of $C_{6}$, which is precisely what the introduction of projectors was meant to avoid. The resulting update equation is shown in Fig.~\ref{fig:CTM_honeycomb}(e). Note the inverse corner matrix between the projectors and two compensating corner matrices after the projectors. 

In the symmetric case, we can resolve this in two steps. First, we define the update equation for $A$ using only a single projector, as shown in Fig.~\ref{fig:CTM_honeycomb}(c); this still leaves some residual inverse powers of $C_{6}$. To remove them, we introduce an auxiliary tensor $L$, defined as the contraction of $A$ with the inverse of $C_{6}$ on one of its bonds. Together with $A$ and $C_{6}$, this new tensor $L$ admits a closed system of equations, shown in Fig.~\ref{fig:CTM_honeycomb}(d). This system is equivalent to the CTMRG presented in Refs.~\cite{Lukin2023honeycomb, Lukin2024archimedean}.

We can offer a brief critique of this approach to honeycomb-lattice CTMRG. While the construction works in this restricted (symmetric) case, it somewhat overcomplicates matters, introducing two tensors, $A$ and $L$, for the same edge and converging them simultaneously. Moreover, the update equation for $L$ uses only a single projector per step, which makes it harder to control the gauge ambiguity carefully, since only one virtual index of $L$ may change gauge during the update. For a general graph with $3$-valent vertices, it may be more convenient to forgo the auxiliary tensors altogether and simply accept the matrix inverses in the equations. In the benchmarks below we use the version with two projectors from Fig.~\ref{fig:CTM_honeycomb}(e), as it is more directly generalizable to arbitrary graphs.

\subsection{Hyperbolic $(8, 3)$ lattices}

\begin{figure}
\includegraphics[width= \linewidth]{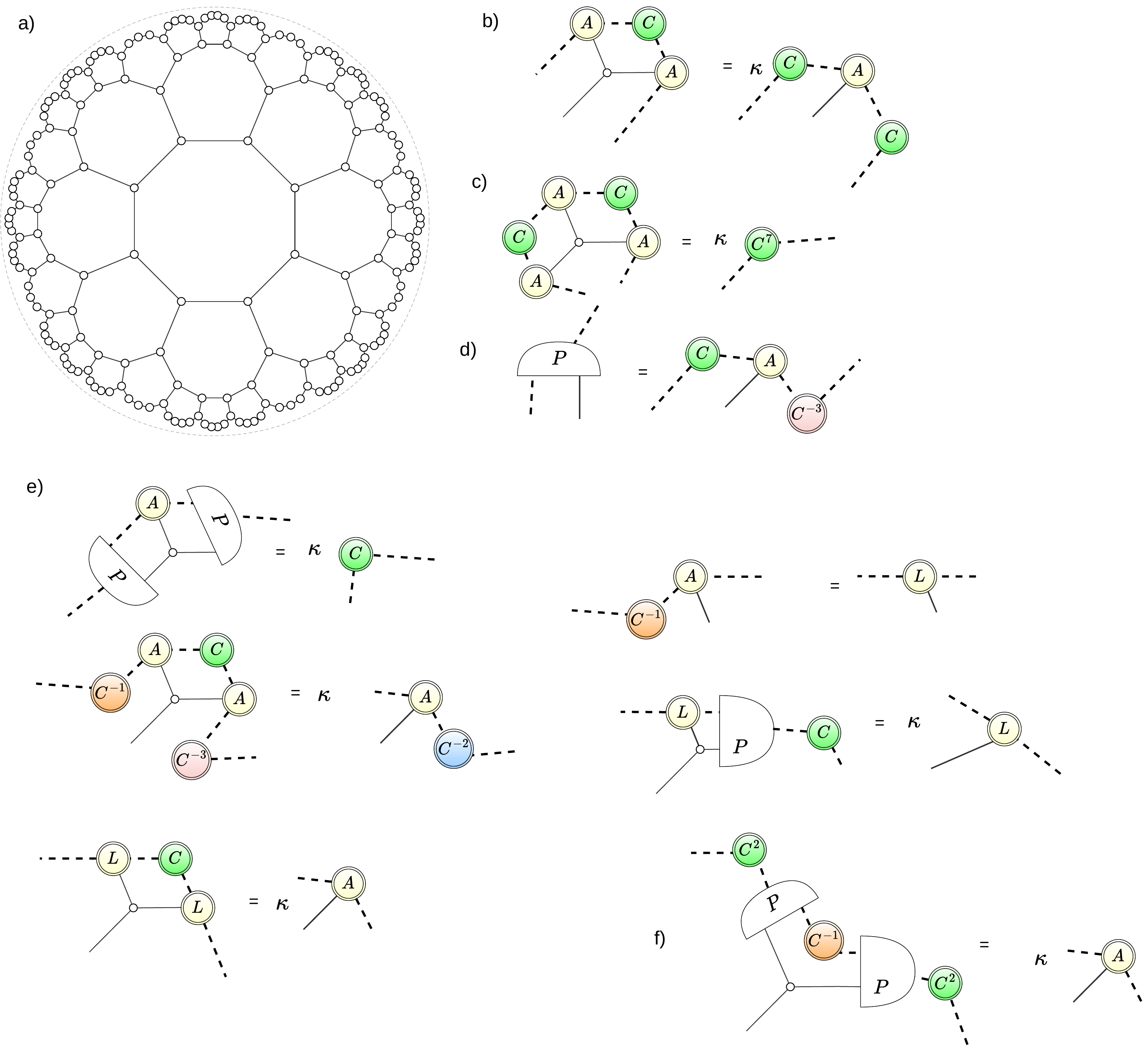} 
  \caption{\label{fig:CTM_hyperbolic_8_3}%
   Derivation of the CTM equations on the hyperbolic $(8, 3)$ lattice, assuming all lattice symmetries. (a) The hyperbolic $(8, 3)$ lattice. (b) The update equation for the message tensor $A$. (c) The update equation for the corner matrix $C$. (d) Definition of the projector $P$. (e) The update equation for the corner matrix $C$ in terms of the projectors, obtained as the diagonalization of the enlarged corner matrix; to define the update equation for $A$ in terms of projectors, we additionally introduce the auxiliary tensor $L$, exactly as on the honeycomb lattice, giving a closed system of equations for $A, L, C, P$. (f) Alternative variant of the message update with two projectors and inverse corner matrix. }
\end{figure}

The hyperbolic $(8, 3)$ lattice (see Fig.~\ref{fig:CTM_hyperbolic_8_3}(a)) is rather similar to the honeycomb lattice: it consists of octagons meeting at $3$-valent vertices. As before, we first assume rotational and translational invariance, so that all the messages are identical. The update equations for the message tensor $A_{e}$ and for the main corner matrix $C$ are then depicted in Fig.~\ref{fig:CTM_hyperbolic_8_3}(b, c). Combining these two equations makes it possible to find the projectors (Fig.~\ref{fig:CTM_hyperbolic_8_3}(d)) and to obtain equations for these projectors as the diagonalization of the enlarged corner matrices (Fig.~\ref{fig:CTM_hyperbolic_8_3}(e)).

The update of the $A_{e}$ tensors can likewise be rewritten in terms of projectors, but the resulting equations again contain matrix inverses $C^{-1}$, exactly as on the honeycomb lattice. We resolve this the same way as before: we introduce an auxiliary tensor $L = A C^{-1}$ as a new independent variable, and write the combined CTMRG equations for $L, A, C, P$, as shown in Fig.~\ref{fig:CTM_hyperbolic_8_3}(e). An alternative variant of the update equation for the message tensors is shown in Fig.~\ref{fig:CTM_hyperbolic_8_3}(f). This variant uses two projectors with inverse corner matrix between them, and four compensating corner matrices. This is the variant we employ below in numerical calculations. 

The CTMRG on the $(8, 3)$ lattice can also be obtained from dualization of the face model CTMRG on the dual $(3, 8)$ lattice, which was introduced in Ref.~\cite{Gendiar2012triangular}. Our construction is equivalent to~\cite{Gendiar2012triangular} after the redefinition that their tensors $L$ are equal to $C^{-1} L C^{-1}$ of our tensors $L$.

\subsection{Triangular lattice}

\begin{figure}
\centering
\includegraphics[width= 0.9\linewidth]{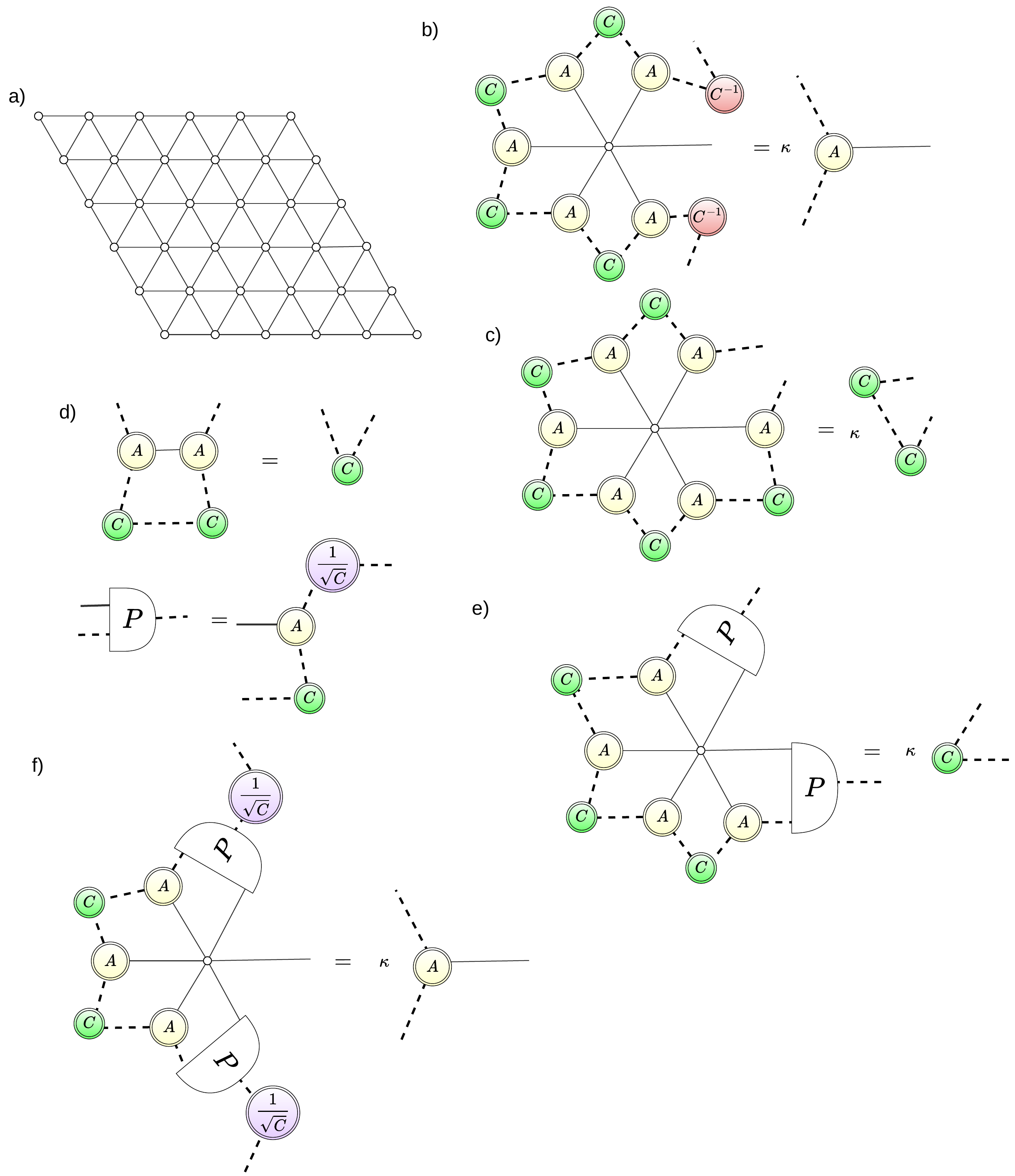} 
  \caption{\label{fig:CTM_triangular}%
   Derivation of the CTM equations on the triangular lattice, assuming all lattice symmetries. (a) The triangular lattice. (b) The general CTM equation for the $A$ MPS messages. (c) The general CTM equation for the corner matrices $C$. (d) Combining the equations in (b) and (c) yields the definition of the orthogonal projector~$P$. (e) With these projectors, the update of the corner matrix $C$ can be reformulated as an eigenvalue problem for the enlarged $C$ matrix. (f) The update of the tensor $A$ can likewise be rewritten in terms of the new projectors.}
\end{figure}

The triangular lattice is shown in Fig.~\ref{fig:CTM_triangular}(a). As before, we assume translational, rotational, and reflection invariance; as we will see below, we also need to assume certain positivity properties of the bulk tensors $T$. Under these assumptions, the equations for the messages are derived in Fig.~\ref{fig:CTM_triangular}(b), and the update equation for the corner matrices is shown in Fig.~\ref{fig:CTM_triangular}(c). From these equations, one can derive the biorthogonal relations shown in Fig.~\ref{fig:CTM_triangular}(d).

Note that these biorthogonal relations, together with the assumed reflection invariance of the $A$ messages, imply that the matrix $C$ is positive: the left-hand side of the biorthogonal relation is manifestly positive, and hence so is the right-hand side. Since $C$ is positive, its square root $\sqrt{C}$ is well defined, allowing us to define the projectors symmetrically using $1/\sqrt{C}$. With these symmetric projectors, the update equation for the corner matrix can be rewritten as shown in Fig.~\ref{fig:CTM_triangular}(e), and for the message tensors as shown in Fig.~\ref{fig:CTM_triangular}(f). Note that the update equation for the message tensor still retains some residual compensating factors of $1/\sqrt{C}$; unfortunately, these matrix inverses cannot be fully eliminated in this formulation.

Let us now address more carefully the bulk-tensor positivity assumption mentioned at the start of this subsection. Looking at Fig.~\ref{fig:CTM_triangular}(e), the right-hand side is manifestly positive, but the left-hand side is guaranteed to be positive only if the bulk tensor $T$ is itself positive (i.e., its eigendecomposition with respect to the symmetric grouping of its legs has only positive eigenvalues). If the bulk tensors are not positive, we must drop the assumption of reflection invariance for the $A$ tensors, along with the resulting conclusion that $C$ is positive. A CTMRG algorithm can still be formulated without reflection invariance, but it will be considerably less symmetric.

It is worth asking why fully symmetric bulk tensors can nonetheless force non-symmetric message tensors, given that positivity has no intrinsic connection to symmetry. The reason lies in our partial gauge fixing: we set the corner matrices of the form $C_{e_{1}, \bar{e}_{3}}$ to identity from the outset, which is always possible when the message tensors differ across edges, or are at least not reflection symmetric. Reflection, translational, and rotational symmetries, however, severely restrict the gauge group, since gauge transformations are not permitted to break these symmetries. If the matrices $C_{e_{1}, \bar{e}_{3}}$ are not positive, they cannot be turned into the identity without breaking the reflection symmetry of the $A_{e}$ tensors. This is the mechanism by which positivity interferes with symmetry.

\begin{figure}
\centering
\includegraphics[width= 0.9\linewidth]{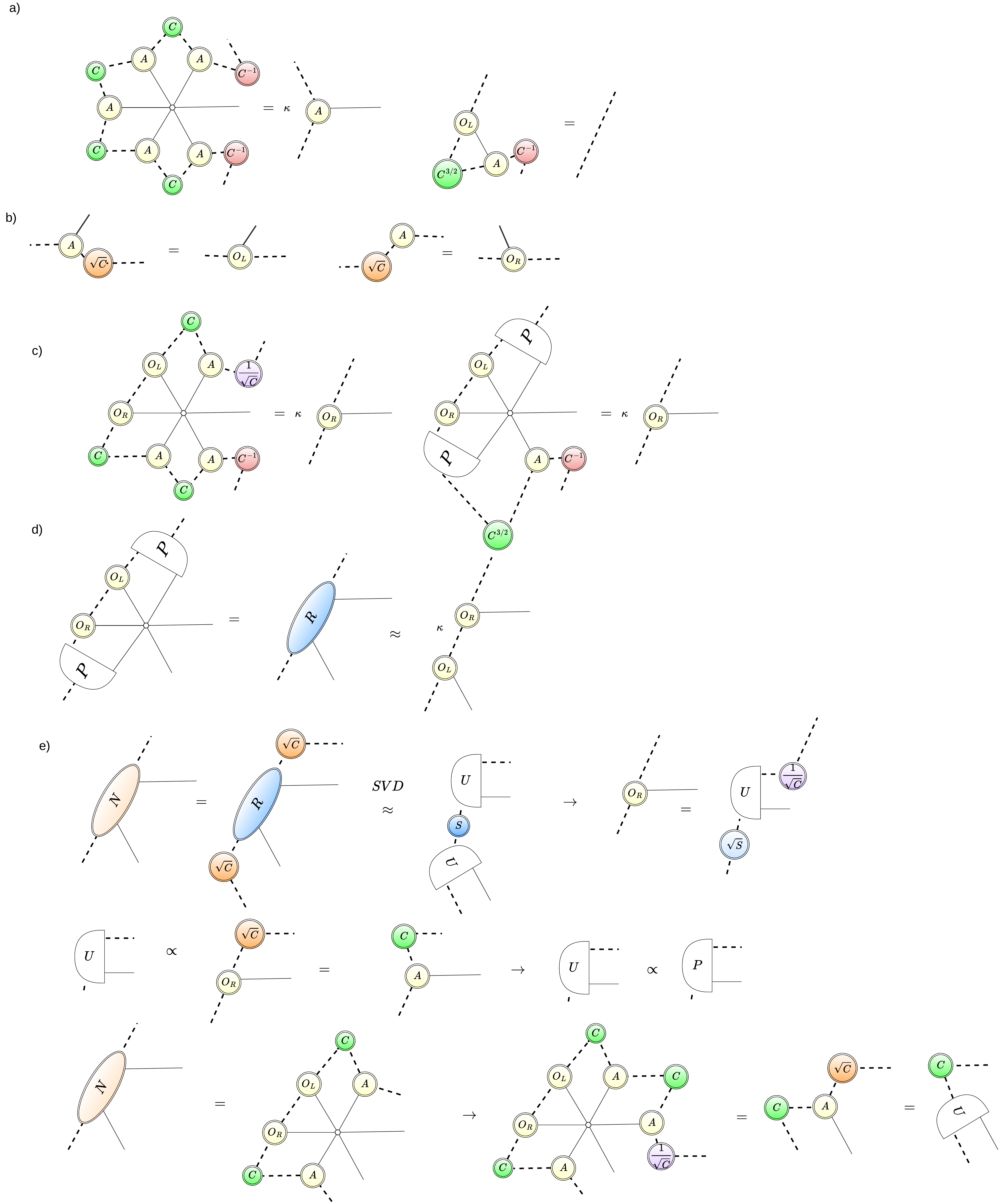} 
  \caption{\label{fig:CTM_triangular_equivalence}%
   Derivation of the CTMRG equations introduced in Ref.~\cite{Lukin2024archimedean} from the new CTM equations. (a) The update equation for the $A$ MPS matrix in the new CTM equations, shown here for easier comparison with the old equations. (b) Definition of the tensors $O_L, O_R$ used in our previous work in terms of the new tensor $A$, and the inverse relation for these new tensors. (c) The update equation for $A$ rewritten in terms of the projectors $P$, the tensors $O_L, O_R$, and the inverse relation for $O_L$. (d) To further compare with the previous formulation, we introduce the matrix $R$ and its approximate factorization; this factorization is fully consistent with equation (c) derived from CTM. (e) In the previous work, the bMPS tensors $O_L, O_R$ were obtained from $R$ as follows: first, the matrix $N$ was formed; this matrix was then SVD-factorized with unitary factor $U$; finally, $O_{R}$ was reconstructed from $U$ and the singular values $S$. Using the relation between $O_R$ and $A$, one can show that $U$ must equal $P$ from the CTM equations, and the CTM expression for $N$ then shows that $U = P$ is indeed the matrix of eigenvectors of $N$. Together, these steps rederive the factorization algorithm of the previous work from the new CTM equations.}
\end{figure}

Let us now briefly discuss how the CTMRG on the triangular lattice presented in this work relates to the triangular-lattice CTMRG from our previous work~\cite{Lukin2024archimedean}. Figure~\ref{fig:CTM_triangular_equivalence}(b) shows the transformation that constructs the bMPS tensors $O_L, O_R$ used in that previous work from the new $A$ message tensor and the corner matrices. In terms of these new tensors, the biorthogonal relations and the diagonalization equation for $C$ are identical to the previous formulation, and the projectors are found in the same way. The main difference lies in the update procedure for the $O_L, O_R$ tensors: in the previous work, this update relied on a novel factorization operation, and our goal here is to show that this factorization can, in fact, be derived from the definition of the projectors in terms of $A, C$ together with the update equation for $A$.

To this end, we first rewrite the update equation for $A$ in the equivalent form shown in Fig.~\ref{fig:CTM_triangular_equivalence}(c), which contains a residual combination $\sqrt{C} A C^{-1}$. Using the definition of $O_L$ together with the biorthogonal relations on the triangular lattice, one can show that this combination is the quasi-inverse of $O_L$ (see the right-hand side of Fig.~\ref{fig:CTM_triangular_equivalence}(a)). In Fig.~\ref{fig:CTM_triangular_equivalence}(d) we then define an auxiliary matrix $R$; from the reformulated update equation in Fig.~\ref{fig:CTM_triangular_equivalence}(c), it follows that $R$ approximately factorizes as the product of $O_L$ and $O_R$ --- precisely the factorization used in our previous work to construct new $O_L, O_R$ tensors. There, this factorization proceeded by forming the auxiliary matrix $N$ and computing its SVD, from which $O_L, O_R$ were then constructed, as shown in Fig.~\ref{fig:CTM_triangular_equivalence}(e).

Let us now show that this SVD-based construction of $O_R$ is equivalent to the message update equation for $A$. First, using the relation between the SVD factor $U$ and $O_R$, and between $O_R$ and $A$, one can show that $U$ must equal $P$ (up to an unfixed gauge on the SVD's factorization index). Then, from the definition of $N$ in terms of $A, O_L, O_R, C$, one can show that the resulting $U$ (equivalently, $P$) indeed diagonalizes the tensor $N$. Thus the $O_R$ tensor can be constructed from $U$, where $U$ diagonalizes $N$ --- precisely the procedure followed in our previous work~\cite{Lukin2024archimedean}.

\subsection{Shastry-Sutherland lattice}

\begin{figure}
\centering
\includegraphics[width= \linewidth]{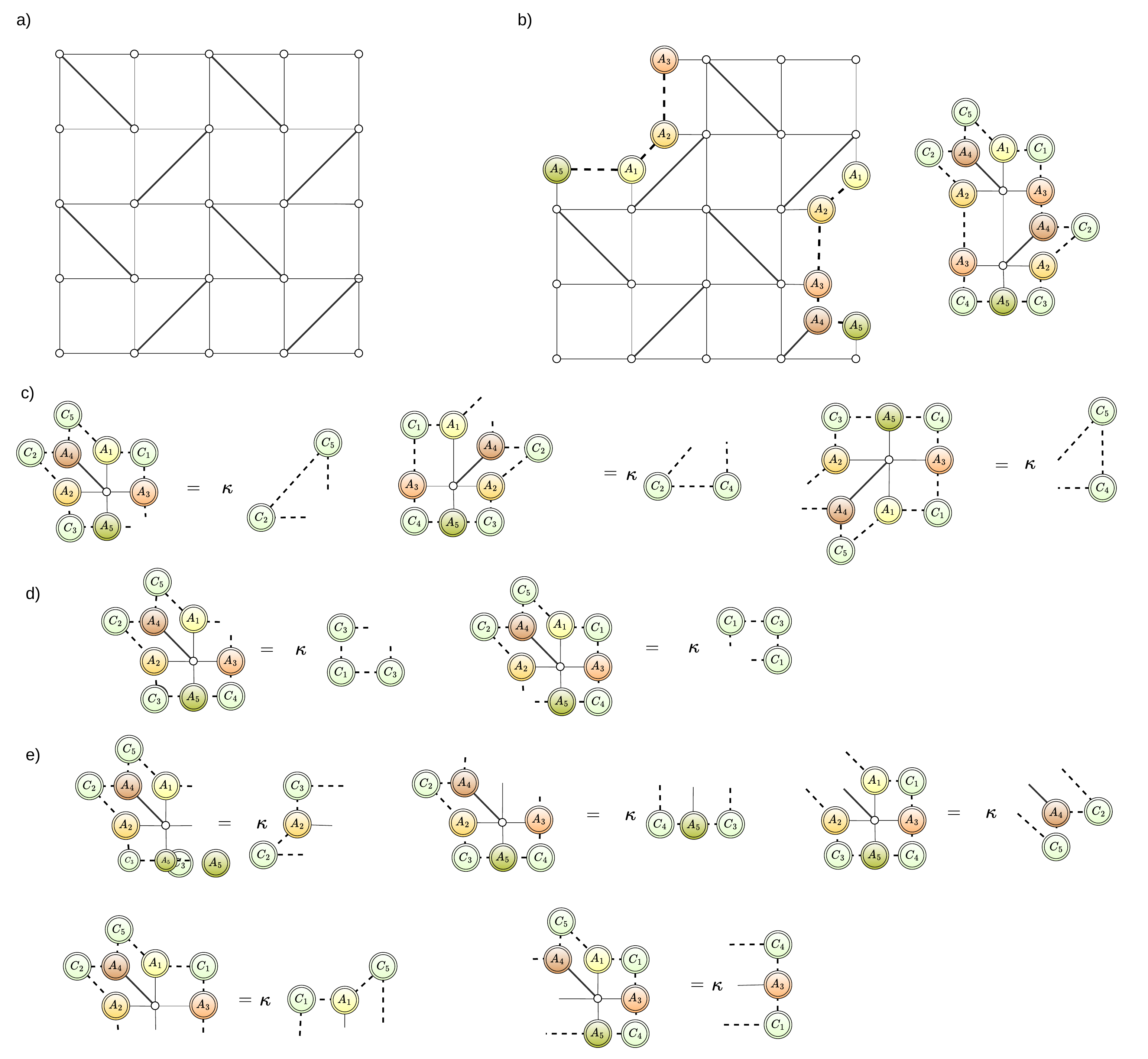} 
  \caption{\label{fig:CTM_shastry_sutherland}%
   Definition of CTM on the Shastry--Sutherland lattice. (a) The Shastry--Sutherland lattice, consisting of valence-$5$ vertices. (b) Some of the MPS environments on the lattice. (c) Corner-matrix equations on the lattice triangles. (d) Corner-matrix equations on the lattice squares. (e) Update equations for the MPS tensors~$A$. }
\end{figure}

\begin{figure}
\centering
\includegraphics[width= \linewidth]{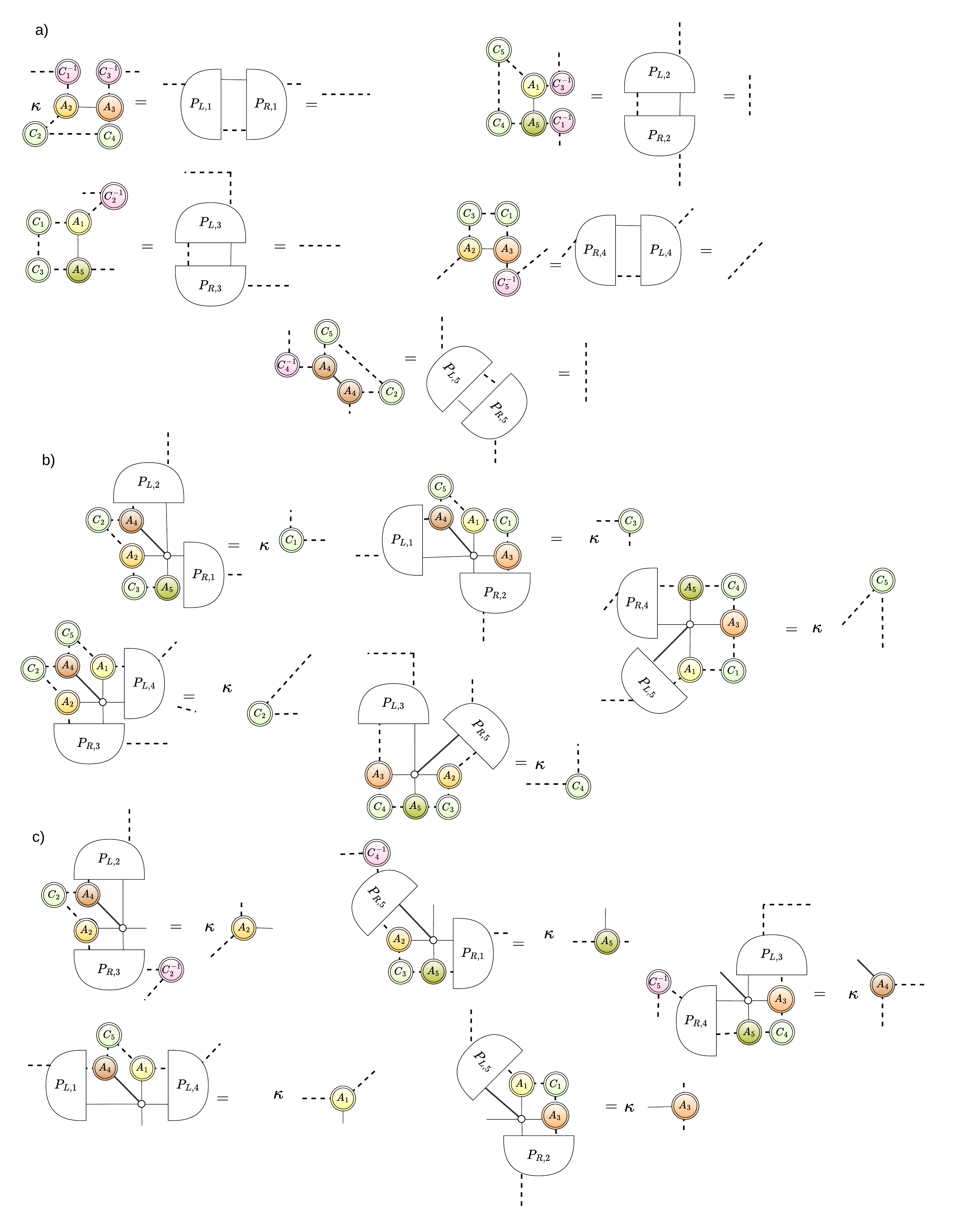} 
  \caption{\label{fig:CTMRG_shastry_sutherland}%
   Definition of CTMRG on the Shastry--Sutherland lattice.  (a) Biorthogonal relations and the corresponding definition of the projectors. (b) The corner-matrix equations in terms of projections of the enlarged corner matrices. (c) Update equations for $A$ in terms of the projectors.}
\end{figure}

In this subsection, we consider the Shastry--Sutherland lattice, a valence-$5$ Archimedean lattice consisting of squares and triangles, as illustrated in Fig.~\ref{fig:CTM_shastry_sutherland}(a). To the best of our knowledge, CTM or CTMRG on this lattice has not been developed before. We concentrate here on simple models defined on this lattice, such as the isotropic Ising and Potts models, and assume translational invariance together with the additional discrete symmetries these models possess.

The typical configurations of the MPS messages on this lattice are shown in Fig.~\ref{fig:CTM_shastry_sutherland}(b). Note that we define five distinct MPS tensors $A$ and five corner matrices. In principle, lattice symmetries might allow this number to be reduced (e.g., by setting $C_{1} = C_{3}$), but we were not confident that such a reduction is compatible with positivity, so we work with the full set of tensors from the start. The CTM equations for the message tensors and the corner matrices are shown in Fig.~\ref{fig:CTM_shastry_sutherland}(c--e). Note that the virtual indices of the message tensors $A$ are not reflection symmetric; nevertheless, it is possible to determine which tensor leg in the diagrams corresponds to which virtual index from the corner matrix attached to that index.

In Fig.~\ref{fig:CTMRG_shastry_sutherland}(a) we further show the biorthogonal equations and the corresponding projectors; note that these projectors are not symmetric. With their help, we define the update of the corner matrices, shown in Fig.~\ref{fig:CTMRG_shastry_sutherland}(b), where the new corner matrices are obtained by projecting the enlarged corner matrix. The projectors can likewise be introduced into the update equations for the MPS messages $A$, as shown in Fig.~\ref{fig:CTMRG_shastry_sutherland}(c). Note that some compensating inverse corner matrices remain in these update equations, owing to the presence of triangles in the lattice.

\subsection{General graphs}

The CTMRG on more general graphs can be constructed systematically from the building blocks presented in the examples above. In most cases, an explicit diagrammatic illustration becomes too cumbersome, owing to the large number of inequivalent corners and messages. Fortunately, the algorithm can be written systematically in terms of the data structure representing the planar graph, and then applied to arbitrary graphs, both finite and periodic infinite. Note that, in contrast to several of the examples above, this more general algorithm does not require any discrete lattice symmetries.

\begin{figure}
\centering
\includegraphics[width= 0.9\linewidth]{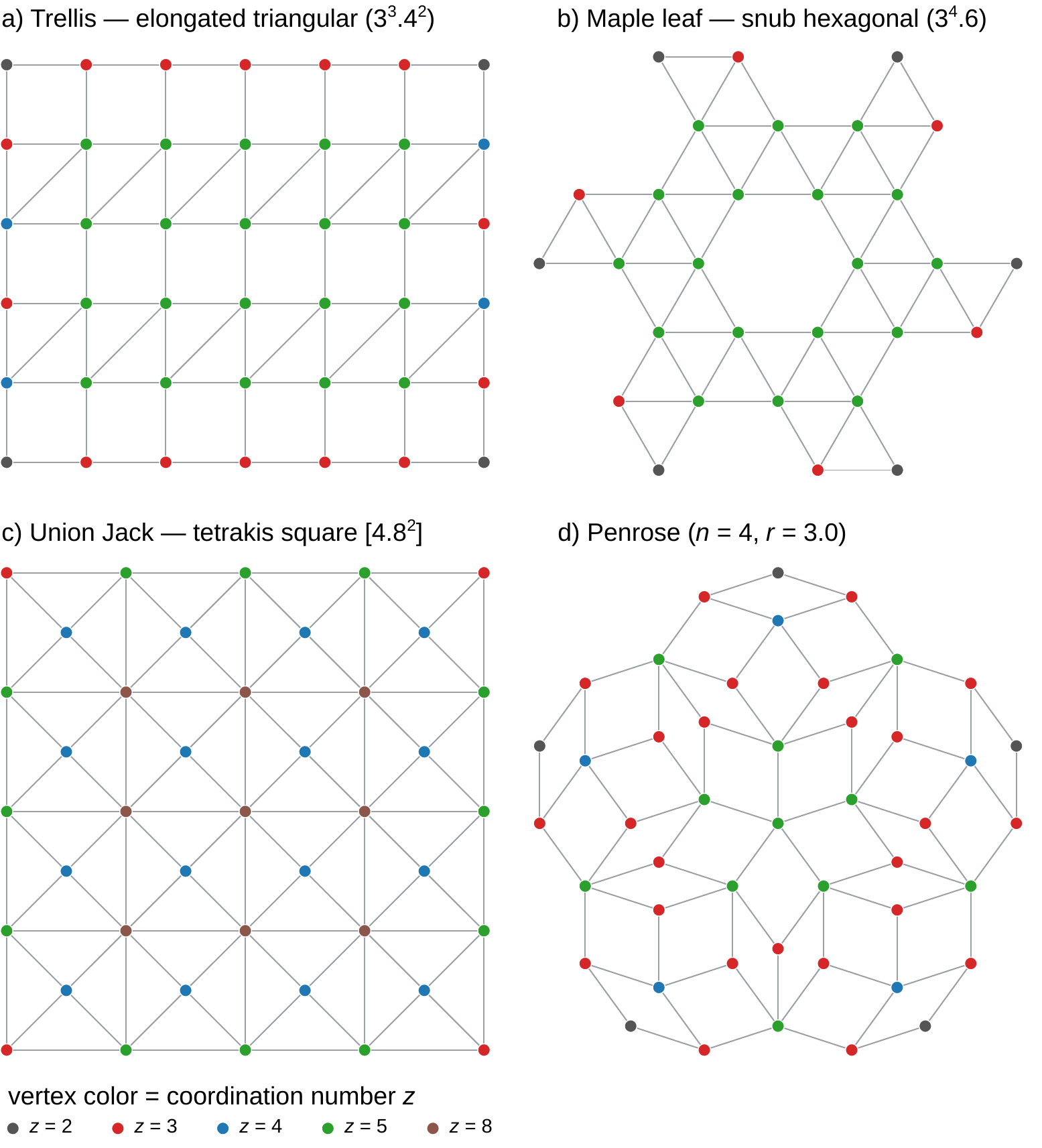} 
  \caption{\label{fig:additional_lattices}%
   Additional lattices and graphs used for CTMRG benchmarks. (a) Trellis. (b) Maple leaf. (c) Union Jack lattice. (d) Finite Penrose tiling. \cite{Penrose1979pentaplexity}}
\end{figure}

In Fig.~\ref{fig:additional_lattices} we show some of the lattices and finite graphs used in our benchmarks of the general algorithm. In particular, we studied the trellis and maple-leaf lattices, which, together with the Shastry--Sutherland lattice, complete the set of Archimedean lattices not studied in our previous work~\cite{Lukin2024archimedean}. The Union Jack lattice is of particular interest, as it is expected to be rather difficult to study with the methods of Ref.~\cite{Lukin2024archimedean}.

Before turning to the results, we want to emphasize several important points arising from the examples above. First, the update equations for the message tensors generally require additional compensating corner matrices, applied after the projectors. Second, if the lattice contains $z=3$ vertices, we must either introduce auxiliary tensors or write the update equations for the message tensors with a corner-matrix inverse placed between the two projectors; in either variant, $z=3$ vertices should always be treated separately from the rest. Third, triangle faces likewise require caution, since triangles are the only faces for which the compensating corner matrices in the tensor update equations are \emph{inverse} corner matrices -- this makes the message update equations less stable for graphs with triangle faces. When solving the equations, we generally first find the projectors and the new projected corner matrices, and then update the message tensors using the \emph{new} projected corner matrices in the compensating part, but the \emph{old} (previous-iteration) corner matrices in the part prior to projector application. To determine which compensating corner matrices should enter the message update equation, it is useful to invoke gauge-invariance logic: a projector's outer index can only be contracted with the index obtained after projecting with the biorthogonal projector.

\section{Results}

\subsection{Benchmark models}

\begin{figure}
\includegraphics[width= 0.8\linewidth]{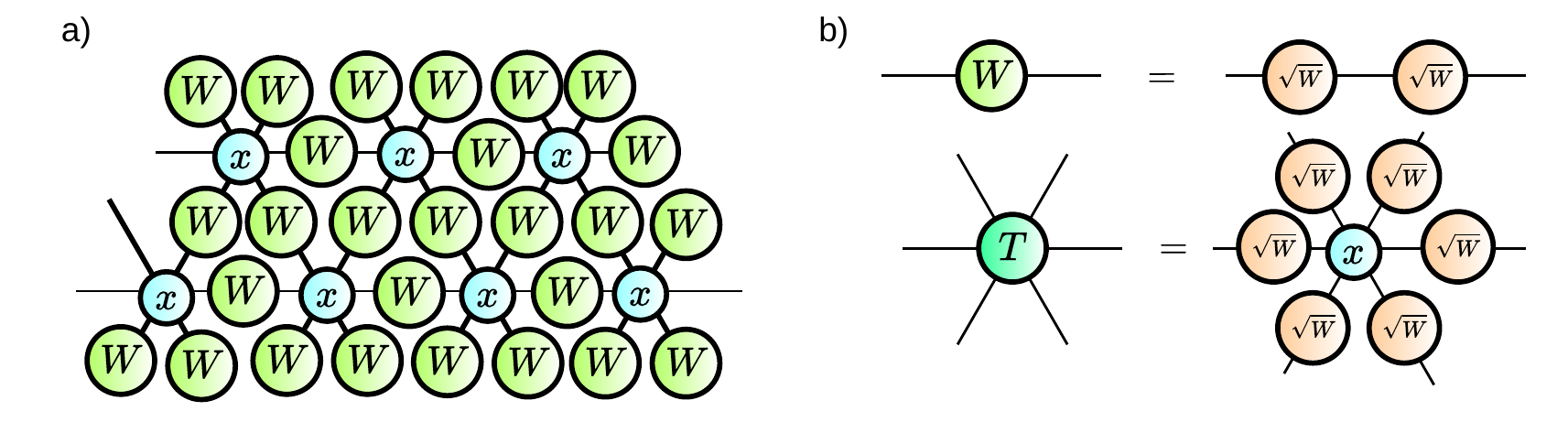} 
 \caption{\label{fig:Ising_TN}%
   Construction of the tensor network for the partition function of the ferromagnetic Ising model with external field. (a) Example of the triangular lattice - place matrices $W$ on the lattice edges and tensors $x$ on the lattice vertices, where $W, x$ are defined in the main text. (b) Symmetrically factorize $W$ and absorb the square root factors in the neighbor tensors $x$, resulting in the tensor network with tensors $T$ on all the sites.    }
\end{figure}

Our main benchmark model was the ferromagnetic Ising model~\cite{Strecka2015review, McCoy1973ising}, with and without an external magnetic field, defined by the Hamiltonian:
\begin{equation}
    H_{Ising} = - \sum_{bonds = \langle v_{1} v_{2}\rangle} \sigma_{v_{1}} \sigma_{v_{2}} - h\sum_{v} \sigma_{v}, 
\end{equation}
with partition function $Z = \sum_{\sigma} \exp[-\beta H]$, where each spin takes values $\sigma = \pm 1$. The construction of the tensor network representation of this partition function is illustrated in Fig.~\ref{fig:Ising_TN}. This construction uses bond matrices $W_{\sigma_{1} \sigma_{2}} = \exp{[\beta \sigma_{1} \sigma_{2}]}$ and vertex tensors 
$$x_{\sigma_{1} \sigma_{2} \cdots \sigma_{z}} = \exp{[\beta h \sigma_{1}}] \delta_{\sigma_{1} \sigma_{2} \cdots \sigma_{z}},$$ where $\delta_{\sigma_{1} \sigma_{2} \cdots \sigma_{z}}=1$ when $\sigma_{1} = \sigma_{2} = \cdots = \sigma_{z}$ and $0$ otherwise.



These models exhibit a second-order phase transition at $h=0$ between the ordered (ferromagnetic) and disordered (paramagnetic) phases. The exact position of this transition is known for many lattices and serves as one of our benchmarks~\cite{Codello2010curie}; the critical exponents at the transition are also known and provide another. Note that this model can be solved by fermionization on general planar graphs, with its free energy obtained from the Kac--Ward determinant, constructed from the free-fermion eigenproblem~\cite{Kac1952kacward, Sherman1960feynman, Burgoyne1963remarks, Chelkak2017revisiting, Chelkak2019icm}. On finite graphs this determinant can usually be computed exactly, while on infinite periodic lattices it can be evaluated via numerical quadrature of the integral over the Brillouin zone. For this reason, the free energy of this model is computable exactly and can be compared with the CTMRG estimate $\kappa = \exp[-F]$. Note that for infinite hyperbolic lattices we have not used the Kac--Ward determinant formula, as it is unclear how the diagonalization of the free-fermion problem should proceed in the infinite-lattice limit.

For some models, such as the Ising model on various hyperbolic lattices, the position of the phase transition and the exact free energy are not known, at least in the infinite-lattice limit. In these cases, we employ the high-temperature expansion of the partition function in terms of generalized lattice loops to compare against our prediction for $\kappa$.

We also analyze the entanglement spectra of the model, defined as $E_{n} = -\log C_{n}$, where $C_{n}$ is the $n$-th diagonal element of the corner matrix $C$~\cite{Peschel2009free, Peschel2012solvable, Jimbo1994algebraic}. Note that for this definition we first equalize the corners around the face. These spectra also allow us to define an entanglement gap $\Delta = E_{2} - E_{1}$. This gap closes at the phase transition (at least for flat lattices) and can be used as a diagnostic of the phase transition. In addition, for the ferromagnetic Ising model in the absence of an external field, the entanglement spectra should naturally organize into multiplets placed at integer multiples of $\Delta$, a consequence of the model's solvability.

\subsection{Finite square lattice}

\begin{figure}
\includegraphics[width= 0.75\linewidth]{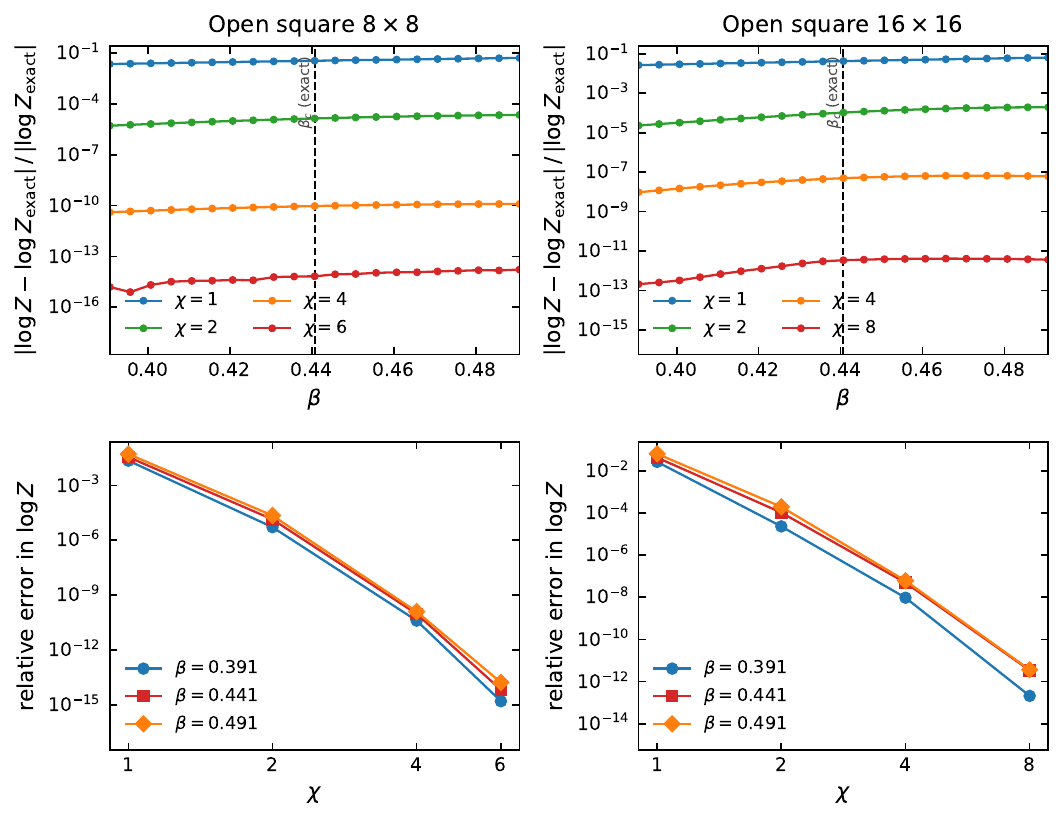} 
 \caption{\label{fig:accuracy_open_square}%
   Relative error of the CTMRG partition function $Z = \prod_{v} \kappa_{v}$ compared with the exact result $Z_{\mathrm{exact}}$, obtained via the Kac--Ward determinant, for the ferromagnetic Ising model on finite open square lattices. Top row: relative error in $\log Z$ versus inverse temperature $\beta$ for several bond dimensions $\chi$, for the $8\times 8$ (left) and $16\times 16$ (right) lattices; the dashed vertical line marks the exact critical inverse temperature~$\beta_{c}$ in thermodynamic limit. Bottom row: relative error in $\log Z$ versus $\chi$ at $\beta_{c}$ and at $\beta_{c}\pm0.05$, for the same two lattice sizes.}
\end{figure}

For small square-lattice graphs with open boundary conditions, we tested that the CTMRG prediction for the partition function, $Z = \prod_{v} \kappa_{v}$, agrees with the exact contraction, computed via the Kac--Ward determinant. We focused on two small graphs, $8\times 8$ and $16 \times 16$. The relative error between the CTMRG and exact partition functions is shown in Fig.~\ref{fig:accuracy_open_square}. Even at inverse temperatures close to the infinite-lattice phase transition, the partition function converges quickly with bond dimension, reaching near-exact agreement with the exact result already at $\chi=8$.

\subsection{Hyperbolic lattices}

\begin{figure}
\includegraphics[width= \linewidth]{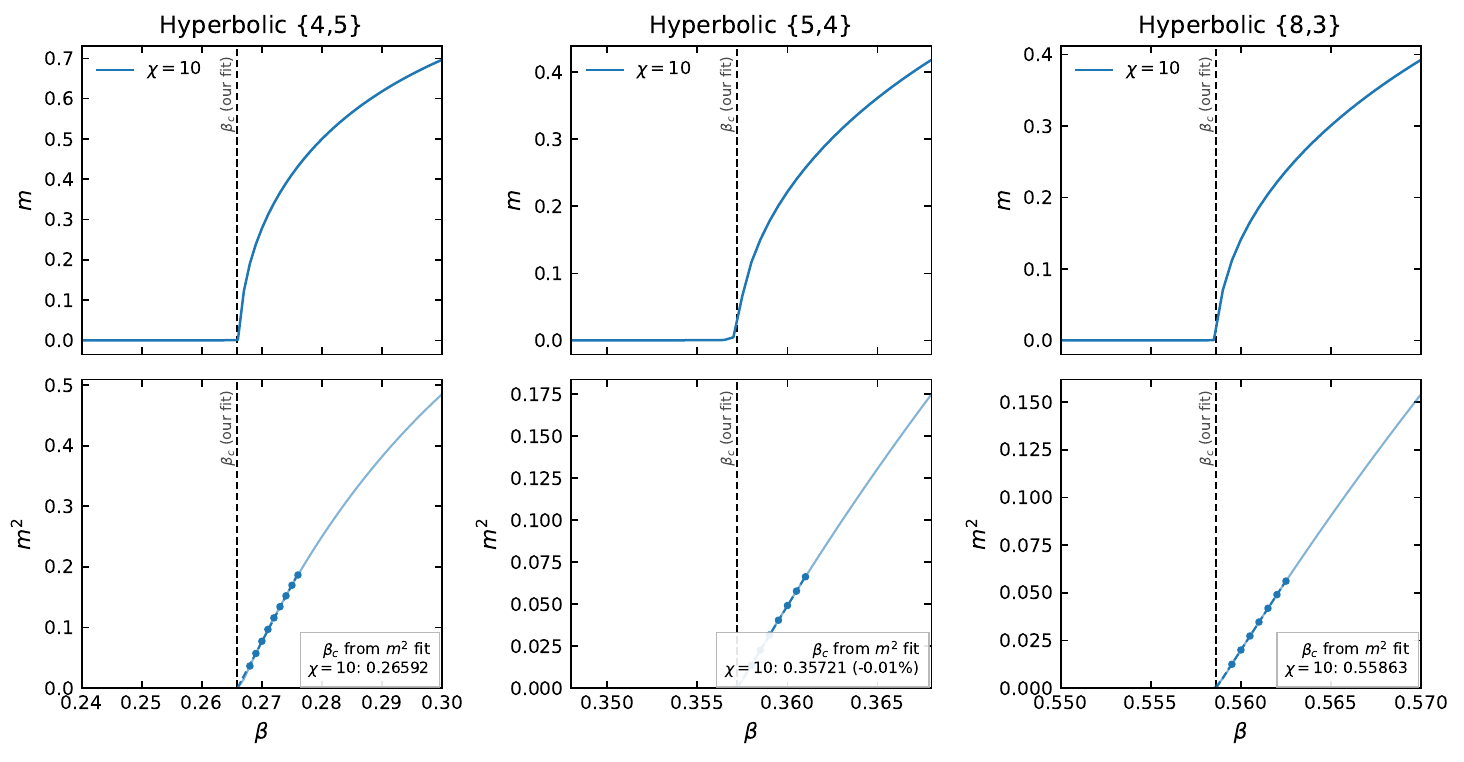} 
 \caption{\label{fig:magnetization_hyperbolic}%
   Magnetization in the ferromagnetic Ising model without an external field, on the hyperbolic $(4,5)$, $(5,4)$, and $(8,3)$ lattices, computed at bond dimension $\chi = 10$. Top row: magnetization $m$ versus inverse temperature $\beta$. Bottom row: squared magnetization $m^{2}$ together with the linear fit used to extract the critical inverse temperature $\beta_{c}$; for the $(5,4)$ lattice, the dashed line marks the published value of $\beta_{c}$, which our fit reproduces to within $0.01\%$.}
\end{figure}

We show the magnetization results for the ferromagnetic Ising model on the hyperbolic $(4,5)$, $(5,4)$, and $(8,3)$ lattices in Fig.~\ref{fig:magnetization_hyperbolic}. We determined the phase transition position from a linear fit of the squared magnetization $m^{2}$; note that hyperbolic phase transitions are known to belong to the mean-field class~\cite{Ueda2007hyperbolic, Genzor2017review}, so the magnetization follows the mean-field exponent $m \propto (\beta - \beta_{c})^{1/2}$, which was used for this fit. Unfortunately, the phase transition position is known only for the $(5, 4)$ lattice~\cite{Ueda2007hyperbolic}, where it matches our prediction to within $0.01\%$. The small remaining discrepancy in the transition temperature is because, in this study, we do not attempt to obtain the most accurate results possible: the results presented here should be regarded as a proof of principle that the proposed algorithm works, rather than an attempt to obtain the most accurate phase-transition data. Note that our calculations for the hyperbolic lattices employ a relatively small CTMRG bond dimension $\chi = 10$, which should be sufficient for mean-field transitions.

\begin{figure}
\includegraphics[width= \linewidth]{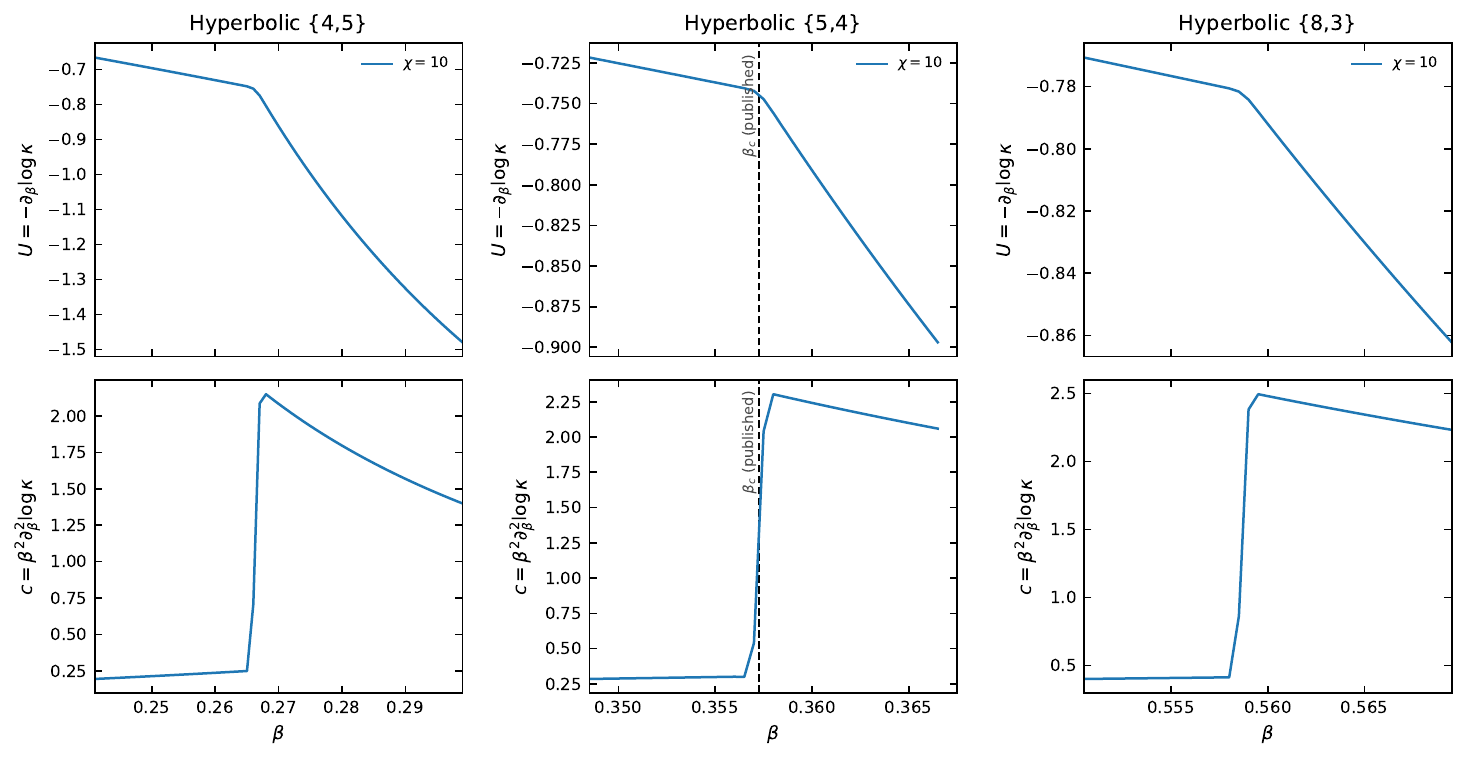} 
  \caption{\label{fig:heat_tr_hyperbolic}%
   Energy $U$ and specific heat $c$ in the ferromagnetic Ising model without an external field, on the hyperbolic $(4,5)$, $(5,4)$, and $(8,3)$ lattices, at bond dimension $\chi = 10$. Top row: energy $U = -\partial_{\beta}\log\kappa$ versus inverse temperature $\beta$. Bottom row: specific heat $c = \beta^{2}\partial_{\beta}^{2}\log\kappa$, showing a clear jump at the phase transition; for the $(5,4)$ lattice, the dashed line marks the published critical inverse temperature $\beta_{c}$.}
\end{figure}

In Fig.~\ref{fig:heat_tr_hyperbolic} we additionally show the energy $U$ and specific heat $c$, obtained from the CTMRG estimate of $\kappa$. Note that the energy $U$ could also be computed directly from a correlation-function measurement with CTMRG, but here we instead use finite-difference differentiation of the $\kappa$ data, as a further check of its validity ($\kappa$ was not computed in previous work~\cite{Lukin2024archimedean}). The specific heat is computed as a second-order finite-difference derivative of $\log \kappa$. The data show a clear jump in the specific heat at the phase transition, the expected behavior for a mean-field phase transition.

\begin{figure}
\includegraphics[width= \linewidth]{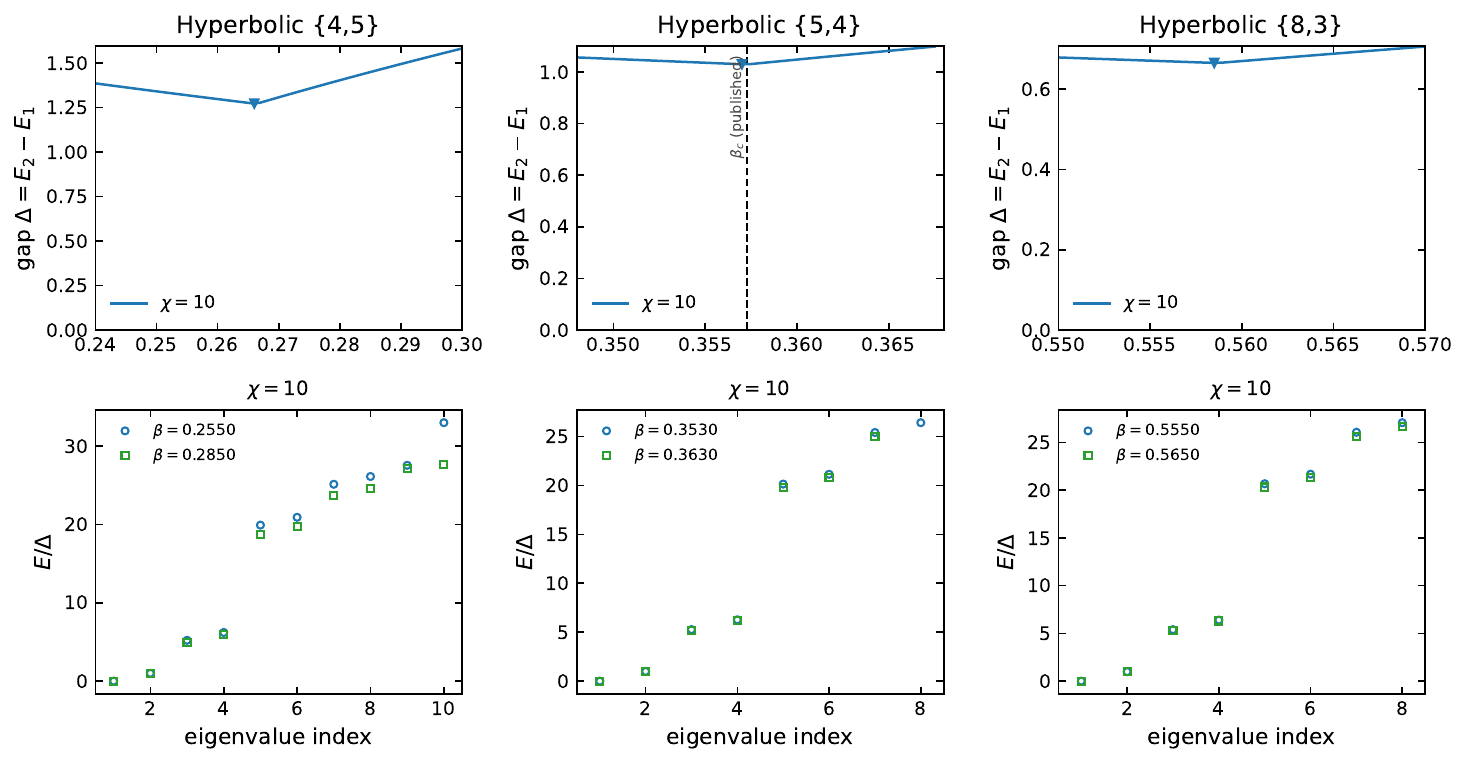}
  \caption{\label{fig:entanglement_hyperbolic}%
   Entanglement gap $\Delta$ and entanglement spectra on both sides of the phase transition, in the ferromagnetic Ising model without an external field, on the hyperbolic $(4,5)$, $(5,4)$, and $(8,3)$ lattices, at bond dimension $\chi = 10$. Top row: entanglement gap $\Delta = E_{2} - E_{1}$ versus inverse temperature $\beta$ (the symbol indicates the minimal value of $\Delta$); for the $(5,4)$ lattice, the dashed line marks the published critical inverse temperature $\beta_{c}$. Bottom row: entanglement spectra $E_{n}/\Delta$ at one inverse temperature below and one above the transition (see legend), showing no change in structure across the transition.}
\end{figure}

Finally, in Fig.~\ref{fig:entanglement_hyperbolic} we show the entanglement gap $\Delta$ and the entanglement spectra on both sides of the phase transition. Our calculations show that the entanglement gap does not close at the phase transition, and that the entanglement spectra do not change structure across the transition either. This is markedly different from the flat lattices considered in the next two subsections, or from the results of our previous work~\cite{Lukin2024archimedean}. The only signature of the phase transition is a minimum in $\Delta$.


\subsection{Honeycomb and triangular lattices}

Let us now discuss the triangular and honeycomb lattices. These lattices are highly symmetric, which allows us to use only one message tensor and one corner matrix for each of them. To avoid simply reproducing our previous results, we apply the new scheme to solve the CTM equations, which involves inverses of the corner matrix $C$, so that the algorithm employed here is somewhat singular. Still, even in this highly symmetric case, we do not encounter any instabilities related to the corner-matrix inverses, of the kind discussed in Appendix~\ref{app:instabilities}. In our calculations, we use moderate bond dimensions $\chi$ up to $40$. Note that precisely at criticality, even larger bond dimensions may be necessary, but $\chi=40$ is sufficient to determine the phase-transition point to high accuracy, using a fit of the data outside the most demanding critical window. For the honeycomb and triangular lattices, both the exact critical temperature and the integral formula for the free energy $F$ are known for the ferromagnetic Ising model, which allows us to make excellent checks of our results.

\begin{figure}
\includegraphics[width= 0.75\linewidth]{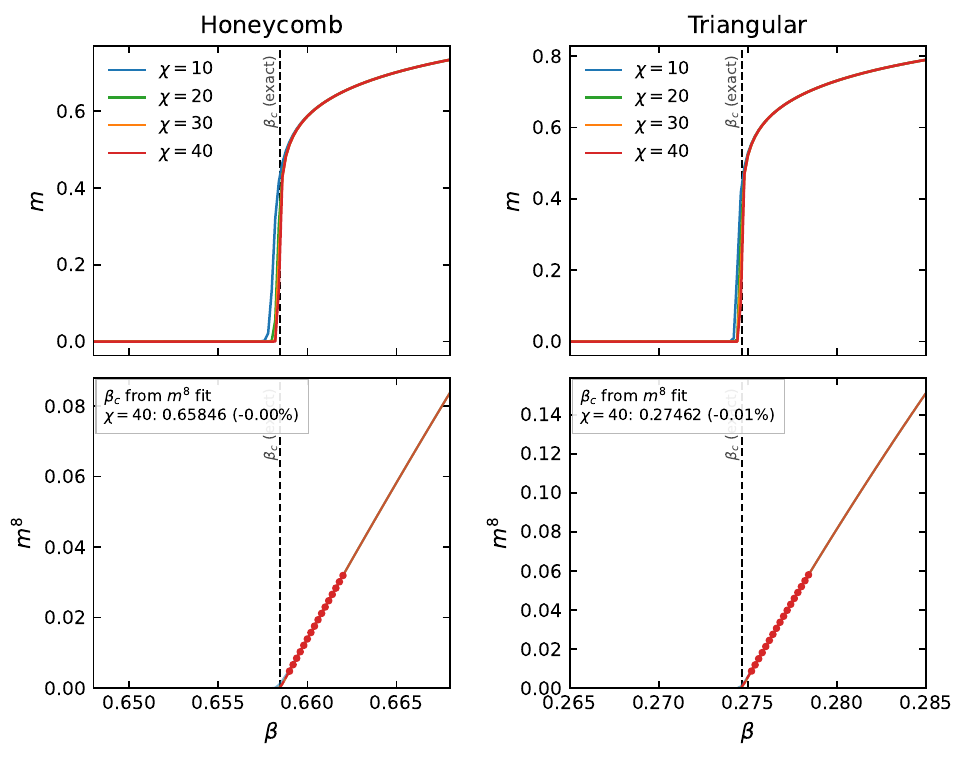} 
  \caption{\label{fig:magnetization_tr_honeycomb}%
   Magnetization in the ferromagnetic Ising model without an external field, on the honeycomb and triangular lattices, at bond dimensions $\chi \in \{10, 20, 30, 40\}$. Top row: magnetization $m$ versus inverse temperature $\beta$, converging to the exact critical point (dashed line) with increasing $\chi$. Bottom row: $m^{8}$ together with the linear fit used to extract $\beta_{c}$, which reproduces the exact value to within $0.02\%$.}
\end{figure}

In Fig.~\ref{fig:magnetization_tr_honeycomb} we show the magnetization together with the position of the exact transition point. The critical behavior of the magnetization visibly converges to the exact result with increasing bond dimension $\chi$. It is known that the phase transition of the ferromagnetic Ising model on planar flat lattices is described by the Ising CFT~\cite{Belavin1984bpz, Chelkak2012universality}, with the spontaneous magnetization near the phase transition scaling as $m \propto (\beta - \beta_{c})^{1/8}$. For this reason, we can obtain the phase transition position from a linear fit of $m^{8}$. The critical inverse temperatures $\beta_{c}$ obtained in this way lie within $0.01\%$ of the exact result.

\begin{figure}
\includegraphics[width= 0.75\linewidth]{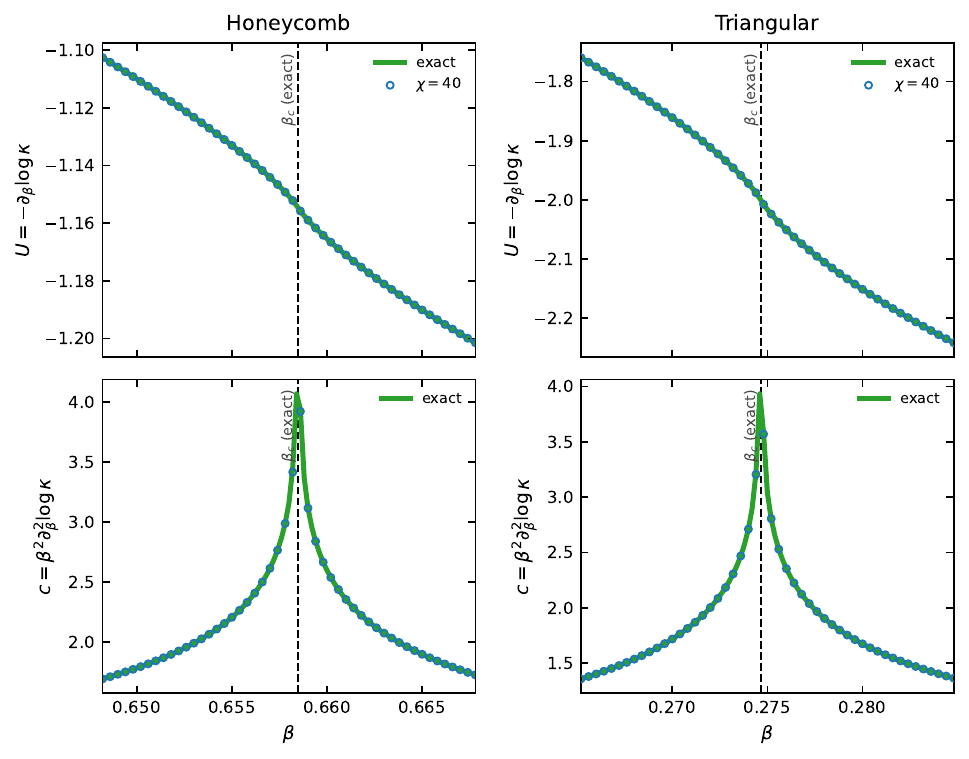} 
  \caption{\label{fig:heat_tr_honeycomb}%
   Energy $U$ and specific heat $c$ in the ferromagnetic Ising model without an external field, on the honeycomb and triangular lattices, at bond dimension $\chi = 40$, compared with the model's exact solution. Top row: energy $U = -\partial_{\beta}\log\kappa$ versus inverse temperature $\beta$. Bottom row: specific heat $c = \beta^{2}\partial_{\beta}^{2}\log\kappa$, reproducing the logarithmic divergence at the phase transition (dashed line marks the exact critical point).}
\end{figure}

In Fig.~\ref{fig:heat_tr_honeycomb} we show the energy $U$ and specific heat $c$ near the phase transition at $\chi=40$. The CTMRG results were obtained using finite-difference derivatives of the $\kappa$ data. The specific heat reproduces the logarithmic divergence at the phase transition, although the precise divergence is masked by the finite grid and finite bond dimension. The data for $\kappa$ are sufficiently regular for these finite-difference derivatives to be reliable, as confirmed by comparison with the exact solution.

\begin{figure}
\includegraphics[width= 0.75\linewidth]{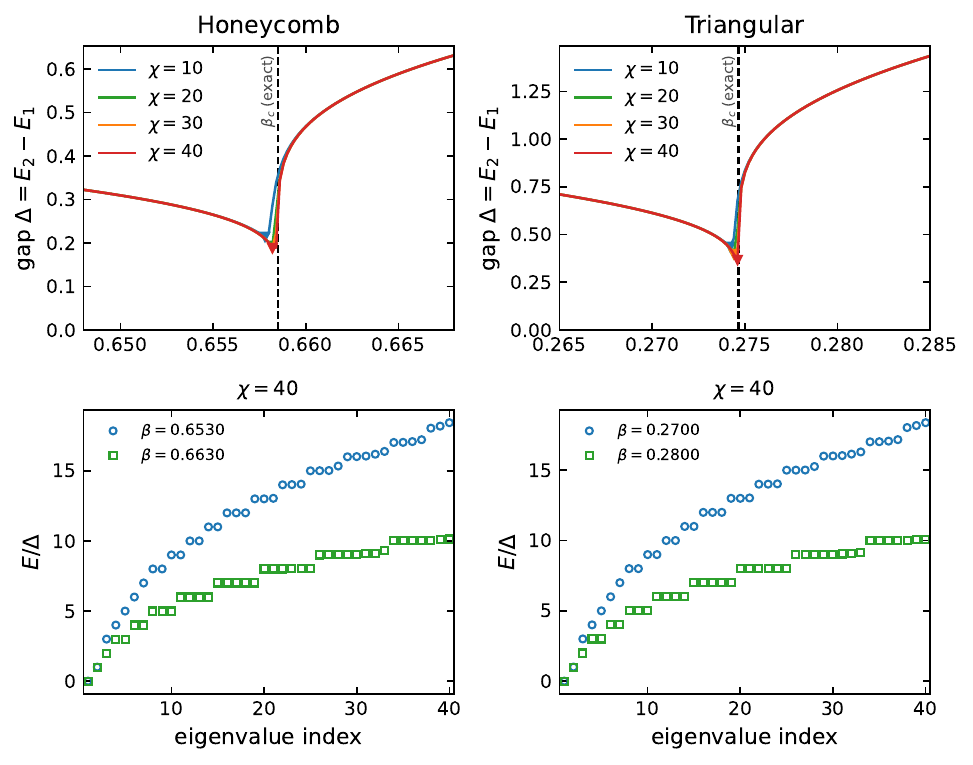} 
  \caption{\label{fig:entanglement_tr_honeycomb}%
   Entanglement gap $\Delta$ and entanglement spectra on both sides of the phase transition, in the ferromagnetic Ising model without an external field, on the honeycomb and triangular lattices. Top row: entanglement gap $\Delta = E_{2} - E_{1}$ versus inverse temperature $\beta$, at bond dimensions $\chi \in \{10, 20, 30, 40\}$; the dashed line marks the exact critical inverse temperature $\beta_{c}$. Bottom row: entanglement spectra $E_{n}/\Delta$ at $\chi = 40$, at one inverse temperature below and one above the transition (see legend), showing the same characteristic degeneracies at integer multiples of $\Delta$ on both sides.}
\end{figure}

In Fig.~\ref{fig:entanglement_tr_honeycomb} we additionally show the entanglement gap and entanglement spectra across the transition. After proper normalization, the entanglement spectra on both lattices show the same characteristic degeneracies, which change discontinuously across the phase transition. Note additionally that the spectra are integer multiples of $\Delta$, a characteristic property of integrable models, which can be used as an additional check of the numerical accuracy of the algorithm.

\begin{figure}
\includegraphics[width= 0.9\linewidth]{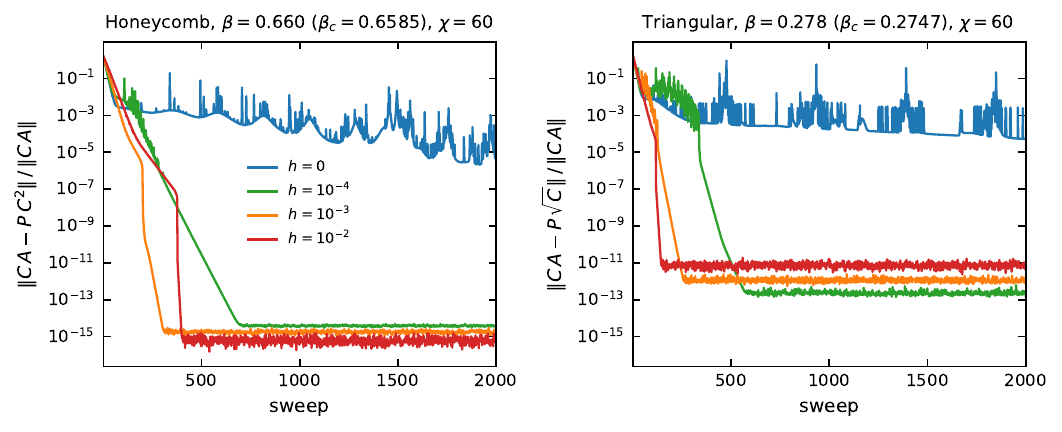} 
  \caption{\label{fig:ap_relations}%
   From the CTM relations it follows that the message tensors $A$ should be related to the projectors $P$ and to the corner matrix~$C$ through the biorthogonal relation. We show that this relation is indeed satisfied, to a precision limited only by machine accuracy, once a nonzero external field $h$ is present in the ferromagnetic Ising model. Each panel shows the relative residual of this relation as a function of the number of CTMRG sweeps, for the honeycomb (left)
   and triangular (right)
   lattices, at several values of $h$. At $h=0$ the residual fails to decay, consistent with the residual gauge freedom discussed in the main text. Note that the relation between the $A, C$ and $P$ should hold only in the fully fixed gauge. }
\end{figure}

Finally, in Fig.~\ref{fig:ap_relations} we show the relation between the tensors $A$ and the projectors $P$ along the CTMRG trajectory. At a nonzero external field $h$, the projectors and the tensors $A$ are indeed related as expected from the biorthogonal relation and the definition of the projectors. At $h=0$, however, the system is integrable and the corner-matrix spectrum acquires large additional degeneracies. These degeneracies produce a massively enlarged residual gauge freedom that is not fixed by the procedure in Appendix~\ref{app:residual_gauge}; as a result, the relation between the projectors and the $A$ tensors fails to converge unless a more elaborate gauge-fixing scheme is used -- it converges only once such a gauge is fixed (note that the relation between $A, C$ and $P$ holds only if the gauge on the final index of $P$ is the same as the gauge on the initial index, that is if the gauge is fully fixed between CTMRG steps). 

\subsection{Shastry-Sutherland, Trellis, Maple-Leaf, Union Jack lattices}

In this section, we summarize our results for the ferromagnetic Ising model on the Shastry-Sutherland, trellis, maple-leaf, and Union Jack lattices. All of these lattices have triangular faces, and triangular faces generally require an inverse corner matrix $C^{-1}$, applied after the projectors, in the message-tensor update equations. If the corner matrix $C$ is poorly conditioned, this inverse becomes the leading source of potential algorithmic instability, as discussed in Appendix~\ref{app:instabilities}. This instability usually appears at large bond dimensions. For this reason, the results in this subsection are obtained not at a fixed bond dimension, but rather with a fixed cutoff on the truncated eigenvalues, so that the smallest elements of $C$ cannot fall below a fixed threshold and destabilize the algorithm. In practice, this dynamically chosen bond dimension is largest at the phase transition (we specify this value in the corresponding plots). Note also that, because the cutoff is determined by the corner spectra, it results in slightly different bond dimensions on different types of faces (for example, on the hexagons and triangles of the maple-leaf lattice). For this reason, we report not one but two bond dimensions for the trellis, maple-leaf, and Shastry-Sutherland lattices. Note that the exact transition temperatures for the ferromagnetic Ising model on these lattices are known, and that the partition function on these lattices can be computed by numerical integration of the Kac--Ward determinant.

\begin{figure}
\includegraphics[width= \linewidth]{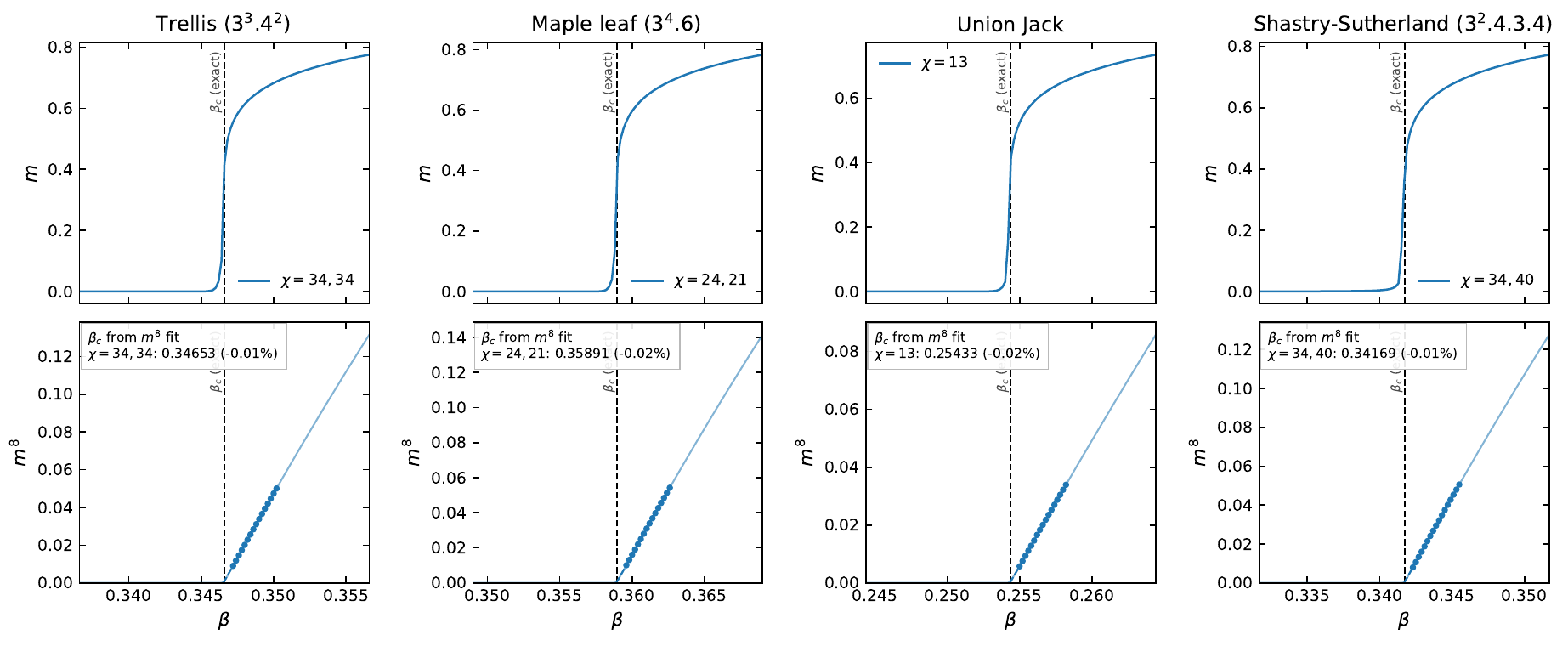} 
  \caption{\label{fig:magnetization_archimedian}%
   Magnetization in the ferromagnetic Ising model without an external field, on the trellis, maple-leaf, Union Jack, and Shastry-Sutherland lattices (vertex configurations shown above each column; bond dimensions $\chi$ on the inequivalent face types shown in each panel's legend). Top row: magnetization $m$ versus inverse temperature $\beta$; the dashed line marks the exact  critical inverse temperature $\beta_{c}$. Bottom row: $m^{8}$ together with the linear fit used to extract $\beta_{c}$, which reproduces the reference value to within $0.02\%$ in all four cases.}
\end{figure}

In Fig.~\ref{fig:magnetization_archimedian} we show the magnetization near the phase transition for all four lattices. As for the honeycomb and triangular lattices, the critical inverse temperature was estimated from a linear fit of $m^{8}$, and was found to lie within $0.02\%$ of the exact (or published) value.

\begin{figure}
\includegraphics[width= \linewidth]{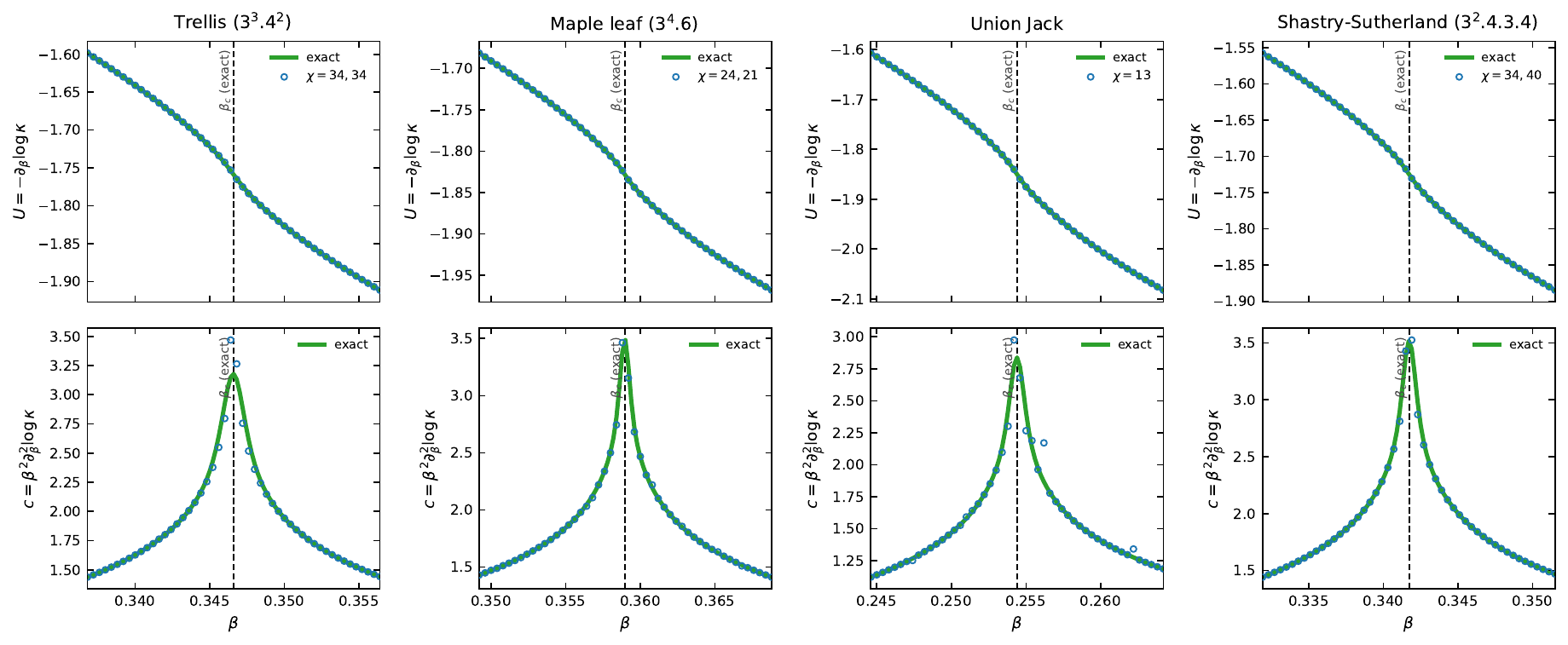} 
  \caption{\label{fig:heat_archimeadian}%
   Energy $U$ and specific heat $c$ in the ferromagnetic Ising model without an external field, on the trellis, maple-leaf, Union Jack, and Shastry-Sutherland lattices, compared with the exact result obtained by integrating the Kac--Ward determinant. Top row: energy $U = -\partial_{\beta}\log\kappa$ versus inverse temperature $\beta$. Bottom row: specific heat $c = \beta^{2}\partial_{\beta}^{2}\log\kappa$; because the specific heat is obtained from a numerical derivative, its prediction can slightly deviate from the exact result near the phase transition, as seen most clearly for the Union Jack lattice.}
\end{figure}

In Fig.~\ref{fig:heat_archimeadian} we show the energy $U$ and specific heat $c$ on these four lattices, and compare our results with the exact values obtained by integrating the Kac--Ward determinant over the Brillouin zone. The results clearly show agreement in both the free energy and the specific heat, with the sole exception of a few specific-heat points for the Union Jack lattice near the phase transition.

\begin{figure}
\includegraphics[width=\linewidth]{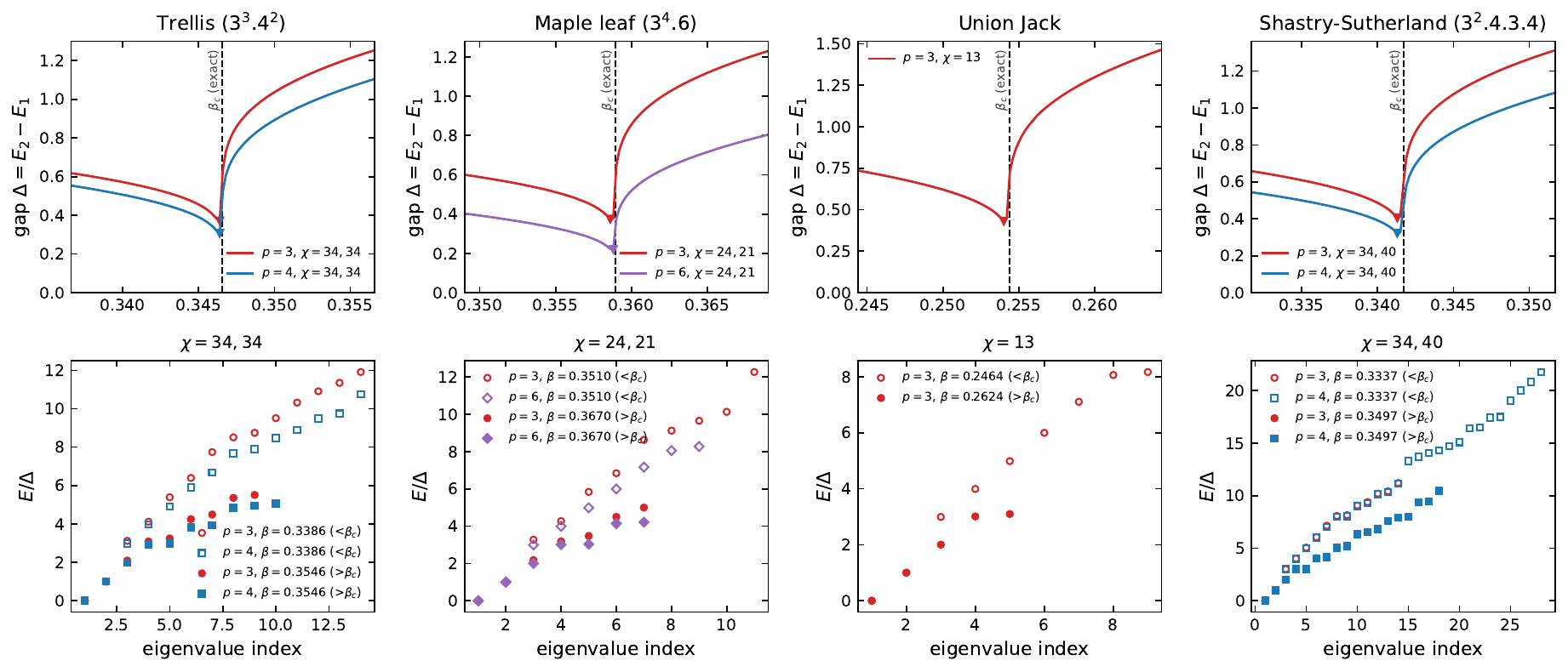} 
  \caption{\label{fig:entanglement_archimedian}%
   Entanglement gap $\Delta$ and entanglement spectra on both sides of the phase transition, in the ferromagnetic Ising model without an external field, on the trellis, maple-leaf, Union Jack, and Shastry-Sutherland lattices. Top row: entanglement gap $\Delta = E_{2} - E_{1}$ versus inverse temperature $\beta$, shown separately for each inequivalent face type ($p$ = polygon size; see legend); the dashed line marks the reference critical inverse temperature $\beta_{c}$. Bottom row: entanglement spectra $E_{n}/\Delta$ at one inverse temperature below and one above the transition, again shown separately for each face type.}
\end{figure}

Finally, in Fig.~\ref{fig:entanglement_archimedian} we show the entanglement gap $\Delta$ and the entanglement spectra on these lattices. Unfortunately, the entanglement spectra on these lattices show, to a much lesser degree, the expected structure of degenerate multiplets at integer multiples of $\Delta$ that should remain stable throughout each phase. Some residual structure can still be noticed among the low-lying levels, but it is essentially lost for the higher ones. This is a general sign of the greater numerical instability of these non-symmetric, matrix-inverse-based results. The Union Jack lattice appeared to be especially hard to stabilize in this respect, and is probably the most unstable of the infinite lattices we worked with, either in this study or in Ref.~\cite{Lukin2024archimedean}.

\subsection{Penrose tilings and finite hyperbolic graphs}\label{sec:results_penrose}

\begin{figure}
\includegraphics[width= \linewidth]{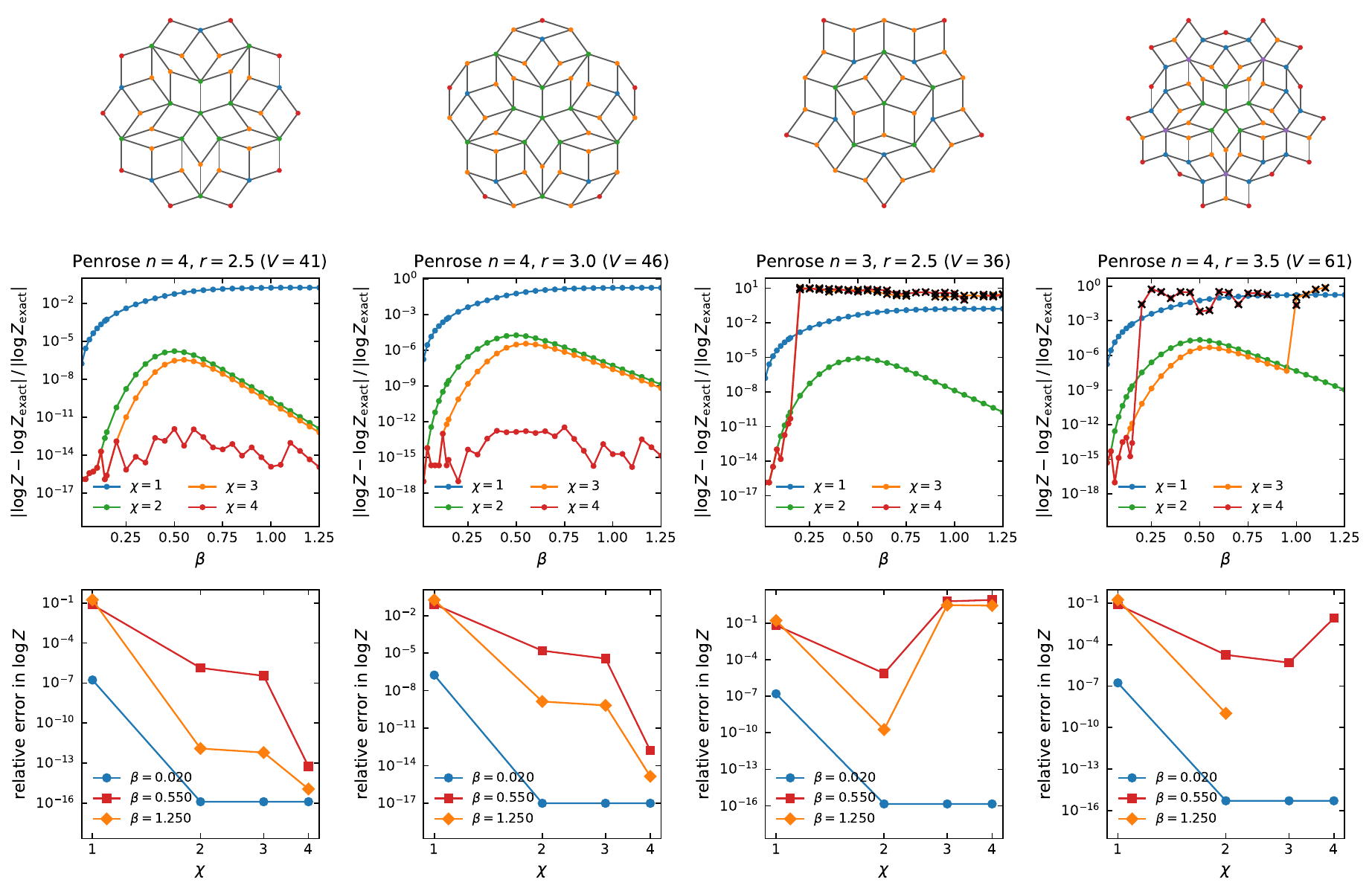} 
 \caption{\label{fig:penrose_tilings}%
   Comparison of the CTMRG partition function with the exact result for four Penrose tiling graphs (top row: the graphs themselves, labeled by the number of substitutions $n$, truncation radius $r$, and vertex count $V$). Middle row: relative error in $\log Z$ versus inverse temperature $\beta$, for bond dimensions $\chi = 1$ (belief propagation) through $\chi = 4$; crosses mark points where the algorithm failed to converge. Bottom row: relative error in $\log Z$ versus $\chi$, at three representative values of $\beta$.}
\end{figure}

In this subsection we report our results for Penrose tilings~\cite{Penrose1979pentaplexity}, where we encountered the second type of numerical-convergence issue discussed in Appendix~\ref{app:instabilities}. The graphs were generated using substitution rules for Robinson triangles~\cite{Baake2013aperiodic}, with $n$ substitutions starting from the ``sun'' seed. After the initial generation, we also introduced an additional truncation parameter $r$, which removed rhombs whose centers lie more than $r$ rhomb side-lengths from the tiling center. As in the case of the open square lattice, we compare the CTMRG prediction for the partition function with the exact result at different bond dimensions $\chi$. The results are shown in Fig.~\ref{fig:penrose_tilings}. While at bond dimension $\chi=2$ the algorithm converges and substantially improves on the results of belief propagation ($\chi=1$), at higher bond dimensions some of the graphs fail to converge at any $\beta$. The convergence at $\chi=2$ indicates that the algorithm itself is correct on these lattices, but that convergence is destabilized at higher bond dimensions.

\begin{figure}
\includegraphics[width= \linewidth]{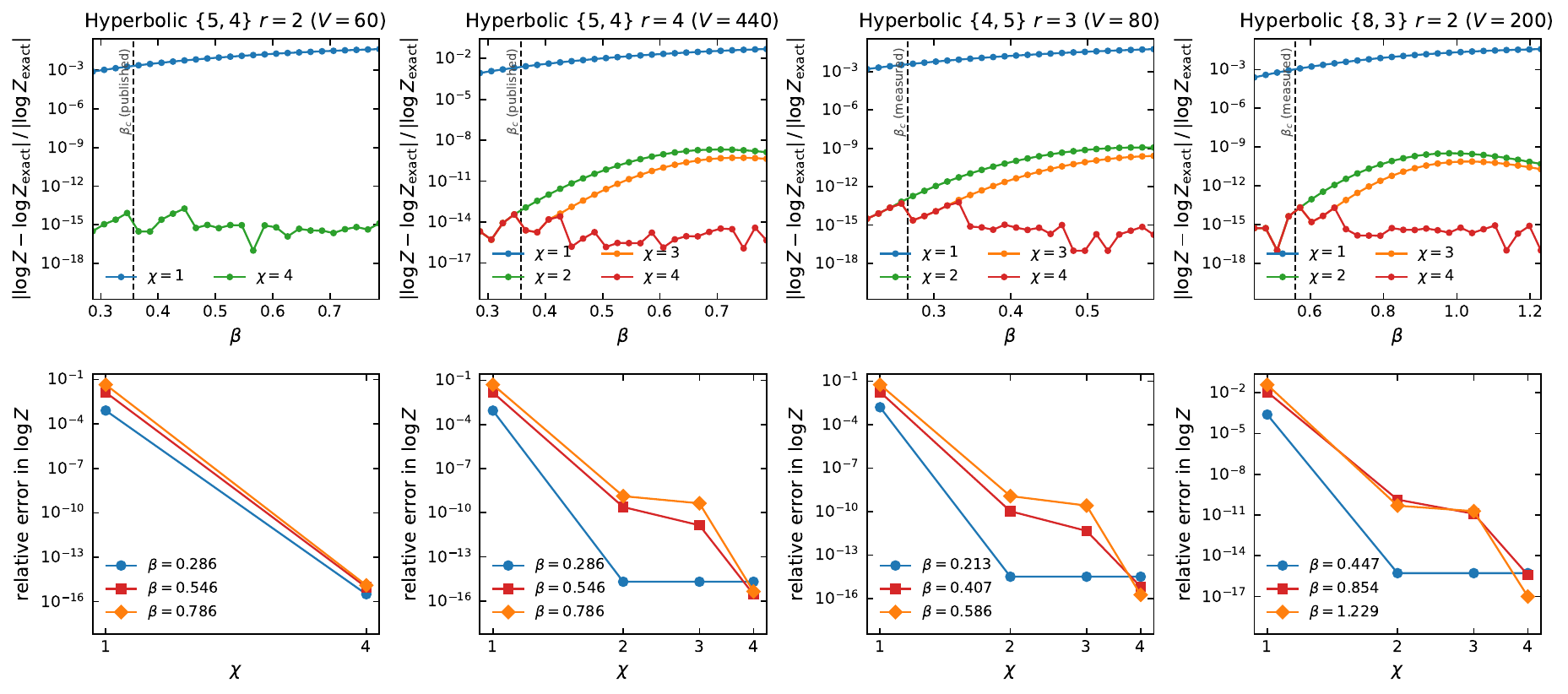} 
 \caption{\label{fig:finite_hyperbolic}%
   Comparison of the CTMRG partition function with the exact result, obtained via the Kac--Ward determinant, for four finite hyperbolic graphs (labeled by lattice type, truncation radius $r$, and vertex count $V$; the dashed line marks the published or measured critical inverse temperature $\beta_{c}$, as indicated). Top row: relative error in $\log Z$ versus inverse temperature $\beta$, for bond dimensions $\chi = 1$ through $\chi = 4$. Bottom row: relative error in $\log Z$ versus $\chi$, at three representative values of $\beta$.}
\end{figure}
Aside from Penrose tilings, we also studied the finite hyperbolic graphs of $(4,5)$, $(5,4)$, and $(8,3)$ lattices. These graphs are much more numerically stable, and we did not encounter numerical instabilities with them. In Fig.~\ref{fig:finite_hyperbolic} we compare the CTMRG partition function with the exact value; as before, we see exponential convergence of the result with bond dimension~$\chi$. These finite hyperbolic graphs of the lattices were generated in the Poincaré disk by iterated reflection of a central regular polygon across its edges (geodesic inversions), keeping all faces within $r$ reflection generations of the center~\cite{Dunham1981hyperbolic, Anderson2005hyperbolic}. Coincident vertices were merged using the hyperbolic distance, and the graph was embedded in the plane with straight chords between vertices for the Kac--Ward computation.

\begin{figure}
\includegraphics[width= \linewidth]{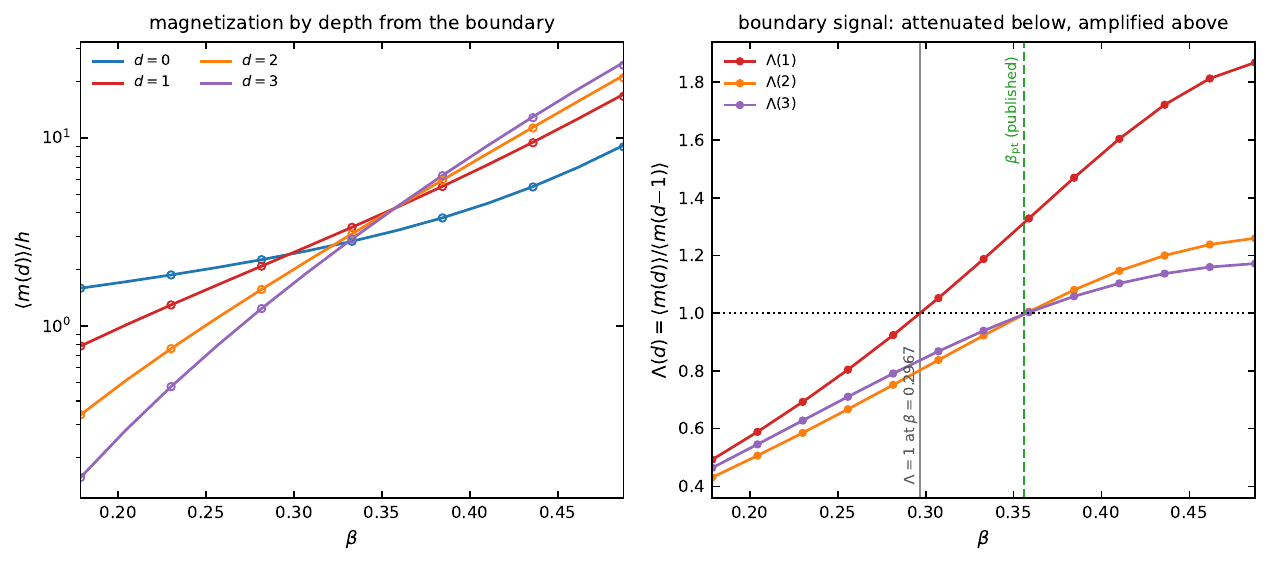} 
 \caption{\label{fig:boundary_order_hyperbolic}%
   Response to a small boundary magnetic field $h$ in the finite hyperbolic $(5, 4)$ lattice. Left: induced magnetization $\langle m(d)\rangle / h$ versus inverse temperature $\beta$, at distances $d = 0, 1, 2, 3$ from the boundary. Right: the ratio $\Lambda(d) = \langle m(d)\rangle / \langle m(d-1)\rangle$ versus $\beta$, for $d = 1, 2, 3$; the dashed green line marks the published critical inverse temperature $\beta_{c}$. The gray line marks where $\Lambda(1) = 1$; note that $\Lambda(2)$ and $\Lambda(3)$ instead cross $1$ close to $\beta_{c}$, while $\Lambda(1)$ crosses earlier.}
\end{figure}

It is also interesting to confirm the phase-transition temperature obtained previously with infinite-lattice CTMRG. To this end, we study the following setup: we introduce a small magnetic field $h=10^{-3}$ on the boundary sites and study how it induces magnetization in the bulk, as a function of the (graph) distance $d$ from the boundary. We used the $r=4$ hyperbolic graph of the $(5, 4)$ lattice, with $440$ sites and four distances from the boundary ($d=0, 1, 2, 3$). If the magnetization decays with $d$, the bulk has no tendency toward ferromagnetic ordering; if it instead grows with $d$, even a small boundary field orders the bulk ferromagnetically. In Fig.~\ref{fig:boundary_order_hyperbolic} we show both the magnetization itself and the more revealing ratio $\Lambda(d) = m(d)/m(d-1)$, introduced in the recent Ref.~\cite{jaroszewicz2026temperaturescalesisingmodel} (this ratio exceeds $1$ if the bulk magnetization grows with $d$). We find that both $\Lambda(2)$ and $\Lambda(3)$ cross $1$ at nearly the same $\beta$, close to the value where infinite-lattice CTMRG predicted the phase transition.

\section{Conclusions}

In this study, we proposed and tested an algorithm for contracting tensor networks on planar graphs, based on a generalization of CTMRG. Compared to conventional CTMRG, this approach introduces some additional auxiliary structures, such as inverse corner matrices and relations between the projectors and the CTM tensors, and it allows for a systematic formulation on general planar graphs.

In our tests, we encountered two potential sources of convergence failure for the algorithm. We identified metrics that allow these instabilities to be detected, but we are not currently aware of a systematic way to cure these convergence problems, although the main text presents some potential remedies. We should also note that, while we studied several additional models beyond those presented in this paper, we cannot be certain that the algorithm is free of other, as yet unidentified, instabilities.

There are several directions for future research. The most important is a generalization of CTMRG to general 3d graphs~\cite{Nishino1998ctmrg3d, Baxter1984cubic} (or, more generally, to 3d cell complexes). There are also several potential ways to solve the underlying CTM equations without relying on the CTMRG scheme presented in this work. For example, the CTM equations can be solved with Variational Uniform Matrix Product States (VUMPS)~\cite{ZaunerStauber2018vumps, Vanderstraeten2019tangent, Nietner2020unitcell}, and it may be possible to generalize this approach to other lattices. Yet another direction is a variational formulation of the CTM equations~\cite{Liu2022vctmrg}. Given the formulation of the partition function in terms of CTM corners and tensors, it may be possible to study in what sense this partition function is stationary with respect to the CTM tensors, and thereby obtain a generalized variational principle. Another recent research direction employs QR decomposition to obtain the projectors from the CTM tensors and corners~\cite{Zhang2026qrctmrg, Yang2025qrhoneycomb}. This construction is very natural in light of the relation between CTM tensors and projectors established in this work, and it should be generalizable to other symmetric lattices, such as kagome or triangular, beyond the previous constructions for the square and honeycomb lattices. Yet another potential generalization to planar graphs is the fixed-point corner-matrix (FPCM) approach of Ref.~\cite{Fishman2018fpcm}. Finally, the most straightforward generalization of the approach presented here is its dualization to CTMRG on the interaction-round-a-face (IRF) model defined on the dual graph.

Beyond CTM-related approaches, there may be room for further adaptation of ideas from the belief-propagation literature. Recent work has used cluster expansions with BP environments to systematically correct BP calculations; it may be possible to carry out analogous cluster expansions with CTM environments. It would be interesting to know whether there are regimes in which a beneficial trade-off can be achieved between the CTM bond dimension~$\chi$ and the maximal cluster size.

Yet another potential application is the derivation of series expansions from the CTM equations~\cite{Baxter1979series, Baxter1999hardsquares, Chan2012ctmrgseries}; the CTM equations on other lattices could potentially be used to derive series expansions for other models.

Finally, we want to mention that, while preparing this manuscript, we became aware of Ref.~\cite{Woolls2026mpbp}, which has some overlap with our current study. In particular, the authors of that paper also introduce a choice of projectors based on eigenvalue decomposition, and show how to apply CTMRG to the finite square lattice with open boundary conditions.

\section*{Acknowledgments}

Claude Code was used for code development and documentation. All code was human-reviewed, and the authors hold full responsibility for the correctness of the results.

\paragraph{Funding information}
The authors acknowledge support by the National Research Foundation of Ukraine, project No.~2023.03/0073 and the Simons Foundation in the framework of the call ``Centers of Excellence in Ukraine'', grant No.~SFI-PD-COE-00032628.

\paragraph{Data availability} All the numerical data present in the paper will be available from the authors upon reasonable request. The code will be available in the public online repository before final publication.

\begin{appendix}
\section{Discussion of the residual gauge choices in the projectors}\label{app:residual_gauge}

Our choice of projectors stems from the diagonalization of the product of enlarged corner matrices over the lattice face. There are several important points about this diagonalization problem, which should be carefully addressed. First, note that the eigenvalues of the eigendecomposition problems for different bonds of the same face are the same, and the rank is also the same for all bonds. Hence, if there are no degenerate eigenvalues, the labeling of projectors is unambiguous (they can be labeled by the shared eigenvalues) across all bonds of the face, and the truncated bond dimension can also be chosen to be the same on all bonds of the face. The eigenvalues of the product of enlarged corner matrices are not guaranteed to be real in the absence of additional reflection symmetries, which may force these matrices to be Hermitian. Since in this paper we deal with real-valued tensor networks, this means that eigenvalues may come in complex-conjugate pairs $\lambda, \bar{\lambda}$; the corresponding left and right eigenvectors are also complex-conjugate. As we generally do not want complex projectors and message tensors for a real-valued tensor network, we treat complex-conjugate pairs as follows: (1) always choose the truncated bond dimension $\chi_{f}$ on face $f$ such that complex-conjugate pairs of eigenvalues are not separated by the truncation; (2) form the projectors from the real and imaginary parts of the complex-conjugate eigenvectors, so that the projectors themselves are real. Note that additional normalization factors are then necessary so that biorthogonality still holds. Note also that projecting the enlarged corner matrix with these real projectors yields a non-diagonal corner matrix, with $2\times2$ blocks corresponding to the complex-conjugate eigenvalue pairs. In this case, we also chose to truncate the bonds by the eigenvalue modulus $|\lambda|$ (an alternative choice would be to truncate by $|\mathrm{Re}\,\lambda|$). Another issue is the potential degenerate multiplets of $|\lambda|$, which may appear in systems with additional symmetries; here too, we chose to truncate without splitting the multiplet.

The next important issue is the choice of residual gauge on the projectors. Note that the eigenvectors inside the projectors already correspond to fixed matrix eigenvalues, so we generally do not have a full $GL(\chi)$ freedom to rotate them (such a rotation would completely break the block-diagonalization of the truncated corner matrices). We do, however, have a residual $(\mathbb{R}\setminus\{0\})^{\chi}$ gauge, which rescales a left eigenvector $v_{L} \to t v_{L}$ while simultaneously rescaling the corresponding right eigenvector in the opposite direction, $v_{R} \to v_{R}/t$ (this combined rescaling preserves biorthogonality). Note that eigenvectors corresponding to the same complex-conjugate pair of eigenvalues must be rescaled by the same factor.

There are several different ways to fix this residual gauge. The first is to try to equalize the truncated corner matrices $C_{k}$ around the face. Note that the truncated corner matrices are forced into block-diagonal form after projection (with blocks corresponding to complex-conjugate pairs), but in general the different $C_{k}$ are not identical unless some additional rotational symmetry forces them to be. The factors $t_{b}$ on different bonds of the face can then be chosen to equalize these corner matrices. Still, some additional freedom remains even after equalization, since we can simultaneously rescale, by the same factor $t$, the vectors $v_{L,b}$ on different bonds $b$ that correspond to the same eigenvalue. This residual symmetry leaves the corner matrices on every face unchanged, but can affect the message tensors: they transform as $A_{lr}^{p} \to t^{\pm1}_{1,l} t^{\pm1}_{2, r} A_{lr}^{p}$, where the sign of the exponent depends on convention, $t_{1}$ corresponds to the face to the left of the message edge, and $t_{2}$ to the face to its right. We can then try to choose $t_{k}$ on the different faces so that the distance between the message tensors $A$ at successive iterations is small (i.e., so that the message tensors converge element-wise). To do this, we first parametrize $t_{k} = \exp(u_{k})$, where $u_{k}$ may have an imaginary part equal to $i\pi$. The equation $A_{old, lr}^{p} = t^{\pm1}_{1,l} t^{\pm1}_{2, r} A_{new, lr}^{p}$ can then be rewritten, for each $u_{k}$, as a linear equation, and all the $u_{k}$ can be solved for simultaneously by least squares. In practice, we usually solved this least-squares problem for a single $A$ tensor first, and then proceeded sequentially through the remaining $A$ tensors with the already-determined $u_{k}$ held fixed. This procedure allowed us to converge the CTM tensors element-wise. Note that, in principle, the least-squares problem should be weighted, so that small elements of $A$ are treated with lower precision, but in most cases even an unweighted least-squares fit was sufficient to restore the gauge.

In principle, this element-wise convergence allows us to use additional techniques from BP message passing, such as damping. While we tried converging CTMRG with this additional damping of the message tensors, we found that in most cases it only slowed down numerical convergence.

\section{The role of the inverse matrices}\label{app:inverse_matr}

In Ref.~\cite{2016_Vanderstraeten_gradients}, a new type of corner matrix was introduced. These corner matrices occupy the same position between MPS environments as our corner matrices $C^{-1}$, which we showed to be inverses of the more conventional CTMRG corner matrices $C$. Our goal here is to show that our inverse corner matrices $C^{-1}$ can indeed play the role of the matrices from Ref.~\cite{2016_Vanderstraeten_gradients}.

\begin{figure}
\includegraphics[width= \linewidth]{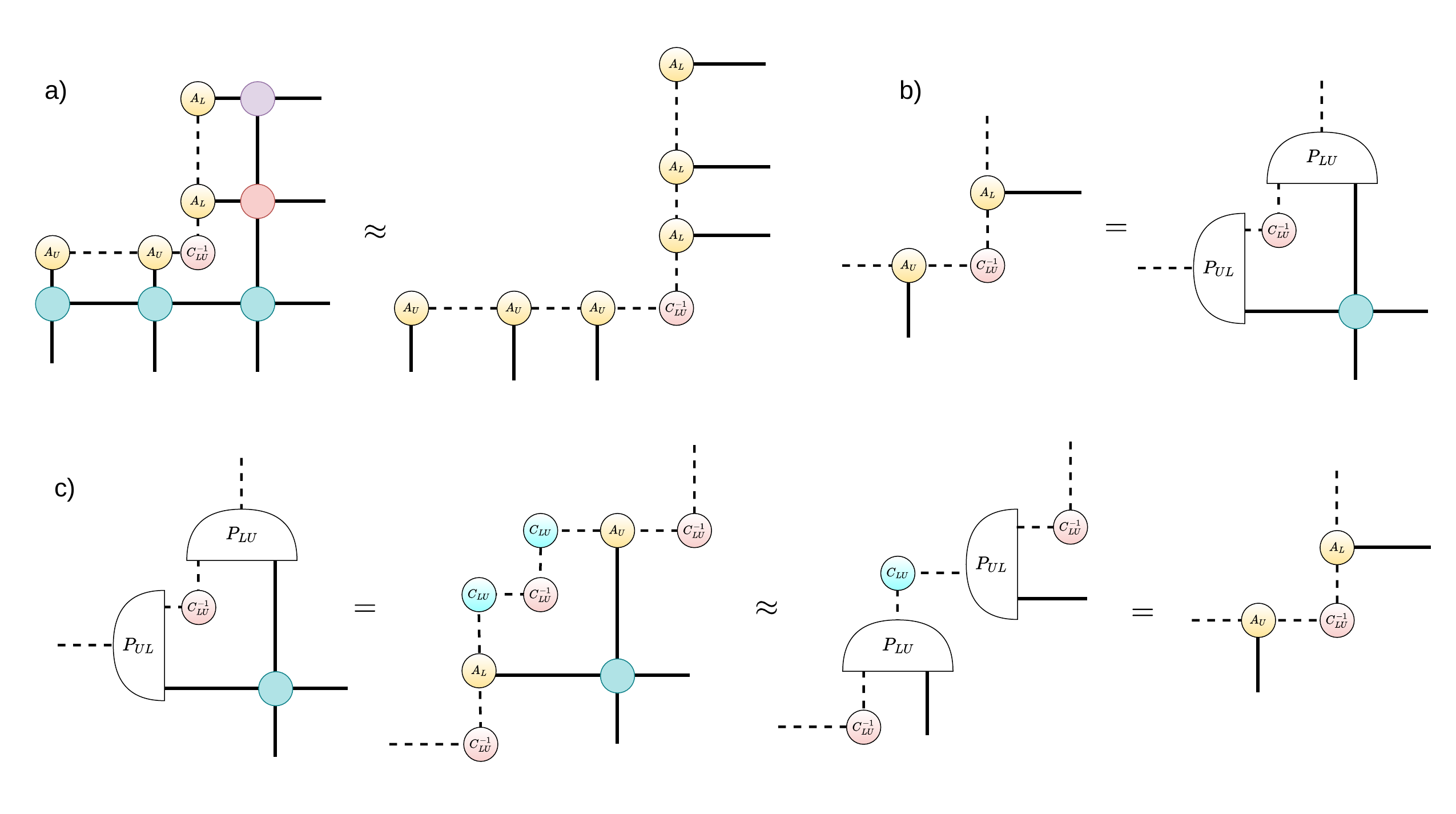} 
  \caption{\label{fig:CTM_inverse_corners}%
   Refs.~\cite{2016_Vanderstraeten_gradients, 2026_vanderstraeten_corner} introduced modified corner matrices, which satisfy the equation in (a). These corner matrices play approximately the same role as our inverse matrices $C^{-1}$, and we show below that $C^{-1}$ indeed approximately obeys the equation in (a). To handle the large MPS environment in (a), we introduce projectors~$P$ to absorb the row and column tensors into $A$. As a result, we obtain a local equation for $C^{-1}$, shown in (b). We can then use the expression for $P$ in terms of $C$ and $A$ to simplify the right-hand side of (b), as shown in (c). In (c) we additionally use the fact that the projectors diagonalize the enlarged corner matrices, writing the eigendecomposition of the enlarged corner matrix in terms of these projectors. Finally, we again express the projectors in terms of $A$ and $C$, arriving at the left-hand side of the equation in (b).}
\end{figure}
The new matrices from Ref.~\cite{2016_Vanderstraeten_gradients} were defined via the equation shown in Fig.~\ref{fig:CTM_inverse_corners}(a). We now show that this equation approximately holds for our inverse corner matrices $C^{-1}$. Note that the equation in Fig.~\ref{fig:CTM_inverse_corners}(a) is a global equation involving a large environment. In the original study, this equation was solved variationally by minimizing the distance between its left- and right-hand sides; this minimization leads to a linear equation for $C^{-1}$. While this approach reduces to our CTM equations for $C^{-1}$ when the bulk tensor network is symmetric under reflections ($A_{U} = A_{D}$, $A_{L} = A_{R}$), in the general case this variational approach does not lead to our CTMRG equations. 

Nonetheless, we can show that our $C^{-1}$ matrices approximately obey the equation in Fig.~\ref{fig:CTM_inverse_corners}(a) even in the fully non-symmetric case. First, we insert biorthogonal projectors on the products of the bulk-tensor-network row and column indices and the MPS message-tensor virtual indices. After inserting these projectors, we obtain a local equation for $C^{-1}$, shown in Fig.~\ref{fig:CTM_inverse_corners}(b). The right-hand side of this equation is further transformed in Fig.~\ref{fig:CTM_inverse_corners}(c): first, we use the expression for the projectors $P$ in terms of $C$ and $A$, which allows us to cancel one $C$ and one $C^{-1}$ matrix. The resulting tensor network is precisely the enlarged corner matrix (with two additional $C^{-1}$ factors on its outer virtual indices). But the enlarged corner matrices are diagonalized by the projectors $P$, which allows us to again express the enlarged corner matrix in terms of the projectors. Finally, expressing the projectors once more in terms of $A$ and $C$, we arrive at the left-hand side of the equation in Fig.~\ref{fig:CTM_inverse_corners}(b). 

This proves that our $C^{-1}$ matrices approximately follow the equation in Fig.~\ref{fig:CTM_inverse_corners}(a), and can therefore be considered equivalent to the corner matrices defined in Ref.~\cite{2016_Vanderstraeten_gradients}. Moreover, this construction can naturally be generalized to other lattices. Recently, corner matrices of this type were also used in Ref.~\cite{2026_vanderstraeten_corner} to compute the corner entanglement of PEPS states on the square lattice; using the CTMRG approach from our study, this corner-entanglement calculation could likely be generalized to other lattices.

\section{Numerical instabilities}\label{app:instabilities}

In our experiments, we encountered two types of numerical instability that can hinder CTMRG convergence in certain cases. The first is connected to the presence of $C$-matrix inverses in the update equations, when the underlying graph $G$ has either $z=3$ vertices or triangular faces. During the algorithm's convergence, it is important to monitor the condition number of the corner matrices $C$ that are inverted; if the condition number becomes as large as $10^{9}$, the algorithm can become unstable. This instability can be mitigated with a special gauge that equalizes the corner matrices around a face. This gauge redistributes the conditioning among the corners on the same face, and if only one of them is very ill-conditioned, such redistribution can substantially decrease the condition number. If the redistribution does not help and all the corners around the face remain ill-conditioned, the only solution we have found so far is to reduce the truncated bond dimension on the face with ill-conditioned corners.

It is probably possible to derive more stable CTMRG schemes that do not rely on corner-matrix inverses in their definitions. In particular, for the symmetric honeycomb lattice, we have shown that introducing auxiliary $L$ tensors allows us to obtain CTMRG equations without matrix inverses, even for a lattice with $z=3$ vertices. Our previous attempts to build CTMRG on the triangular lattice also resulted in algorithms without explicit matrix inverses. The problem is that these constructions were unsystematic, and we do not yet have a systematic method for constructing the general-graph algorithm without these matrix inverses.

A second numerical instability we found was the ill-conditioning of the biorthogonal projectors, $\|P_{L}\| \|P_{R}\|$. Recall that these projectors are obtained as the left and right eigenvectors of the product of enlarged corner matrices over a graph face. In the rotationally symmetric cases, these projectors are orthogonal matrices with perfect conditioning. In more general cases, however --- especially for the Penrose tilings (see Sec.~\ref{sec:results_penrose}) --- we found that the condition number of the projectors can be as high as $10^{7}-10^{9}$. This usually happens when the left and right eigenvectors of the matrix become nearly orthogonal, so that the additional normalization, which is necessary for biorthogonality, leads to a rapid growth of the vector norms. 

Such large condition numbers can produce a peculiar instability in which the biorthogonal relations of the message tensors diverge while the corner matrices themselves converge. The biorthogonal relations normally converge rather quickly, and we used them as one of our metrics for algorithm convergence; they are precisely the conditions for the orthogonality of the projectors. Therefore, when the projectors' condition numbers are large, restoring biorthogonality generally requires inverting some very ill-conditioned matrices. The biorthogonality relation also enters the normalization of the message tensors, so when it fails to converge, the message tensors become increasingly difficult to normalize and their norms become very large 
as well, further compounding the instability. 

At present, the only solution we have found for this instability is likewise limiting the maximal bond dimension on faces with ill-conditioned projectors. We have not encountered this problem for infinite lattices, but we did encounter it for the Penrose tilings in Sec.~\ref{sec:results_penrose}; regular lattices with open boundary conditions also do not appear to suffer from it.

\section{Potts model}\label{app: Potts}
\begin{figure}
\includegraphics[width= \linewidth]{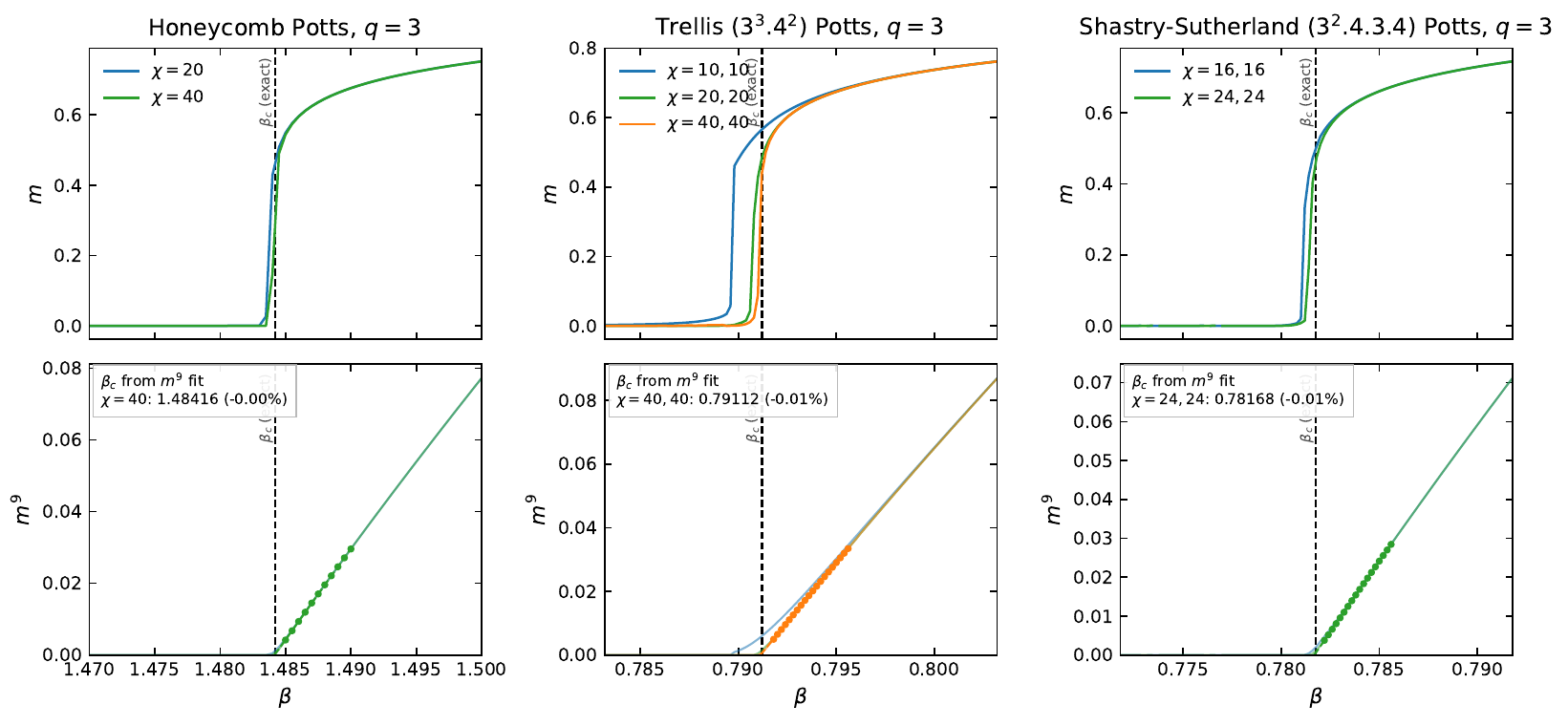} 
  \caption{\label{fig:magnetization_potts}%
   Magnetization in the $q=3$ Potts model on the honeycomb, trellis, and Shastry-Sutherland lattices (bond dimensions $\chi$ on the inequivalent face types shown in each panel's legend). Top row: magnetization $m$ versus inverse temperature $\beta$; the dashed line marks the exact critical inverse temperature~$\beta_{c}$. Bottom row: $m^{9}$ together with the linear fit used to extract $\beta_{c}$, which reproduces the reference value to within $0.02\%$ in all cases.}
\end{figure}

In this appendix, we additionally apply the proposed algorithm to the $q=3$ Potts model. The $q=3$ Potts model is defined by the Hamiltonian:
\begin{equation}
    H_{Potts} = - \sum_{bonds=\langle v_{1} v_{2} \rangle} \delta_{\sigma_{v_{1}}\sigma_{v_{2}}},
\end{equation}
where $\sigma_{v_{1}}=1,2,3$. The tensor network for this model partition function is constructed similarly to the Ising model, with the difference being $W_{\sigma_{1} \sigma_{2}} = \exp{[\beta \delta_{\sigma_{1} \sigma_{2}}]}$, which is now a $3\times3$ matrix. Hence, the corresponding tensor network has a bond dimension of $D=3$, and the model is a good test to ensure that the algorithm works with $D>2$. The model exhibits a ferromagnetic phase transition, with known critical exponents; in particular, magnetization $m \propto (\beta-\beta_{c})^{1/9}$. There is no analytic expression for the model partition function, but the critical temperature is known for many lattices~\cite{feldmann1998potts, jacobsen2014graphpolynomials}. We studied the model on honeycomb, trellis, and Shastry-Sutherland lattices and computed the magnetizations and critical temperatures. The results are shown in Fig.~\ref{fig:magnetization_potts} and show clear agreement with the known critical temperatures and magnetization critical exponent. 

\section{High-temperature series for hyperbolic lattices}

For infinite hyperbolic lattices we generally do not have an exact expression for the partition function of the Ising model. Instead, we can use high-temperature expansions and compare the partition function from CTMRG with this expansion. The leading order terms for $\log\kappa$ of the Ising model on a $(p, q)$ hyperbolic lattice are the following:
\begin{equation}\label{eq:log_k}
    \log\kappa = \log{2} + (q/2) \log[\cosh(\beta)] + (q/p)\tanh(\beta)^{p} + (q/2)\tanh(\beta)^{2p-2} + O(\tanh(\beta)^{2p}).
\end{equation}
Here, the $\log{2}$ term corresponds to the sum over two $\sigma$ values on every site, $\log{\cosh{\beta}}$ is the leading term of the expansion, $\tanh(\beta)^{p}$ corresponds to the first loop correction counting minimal $p$-gons in the lattice, and $\tanh(\beta)^{2p-2}$ counts the next loop correction in the form of two loops sharing the edge. 

\begin{figure}
\includegraphics[width= \linewidth]{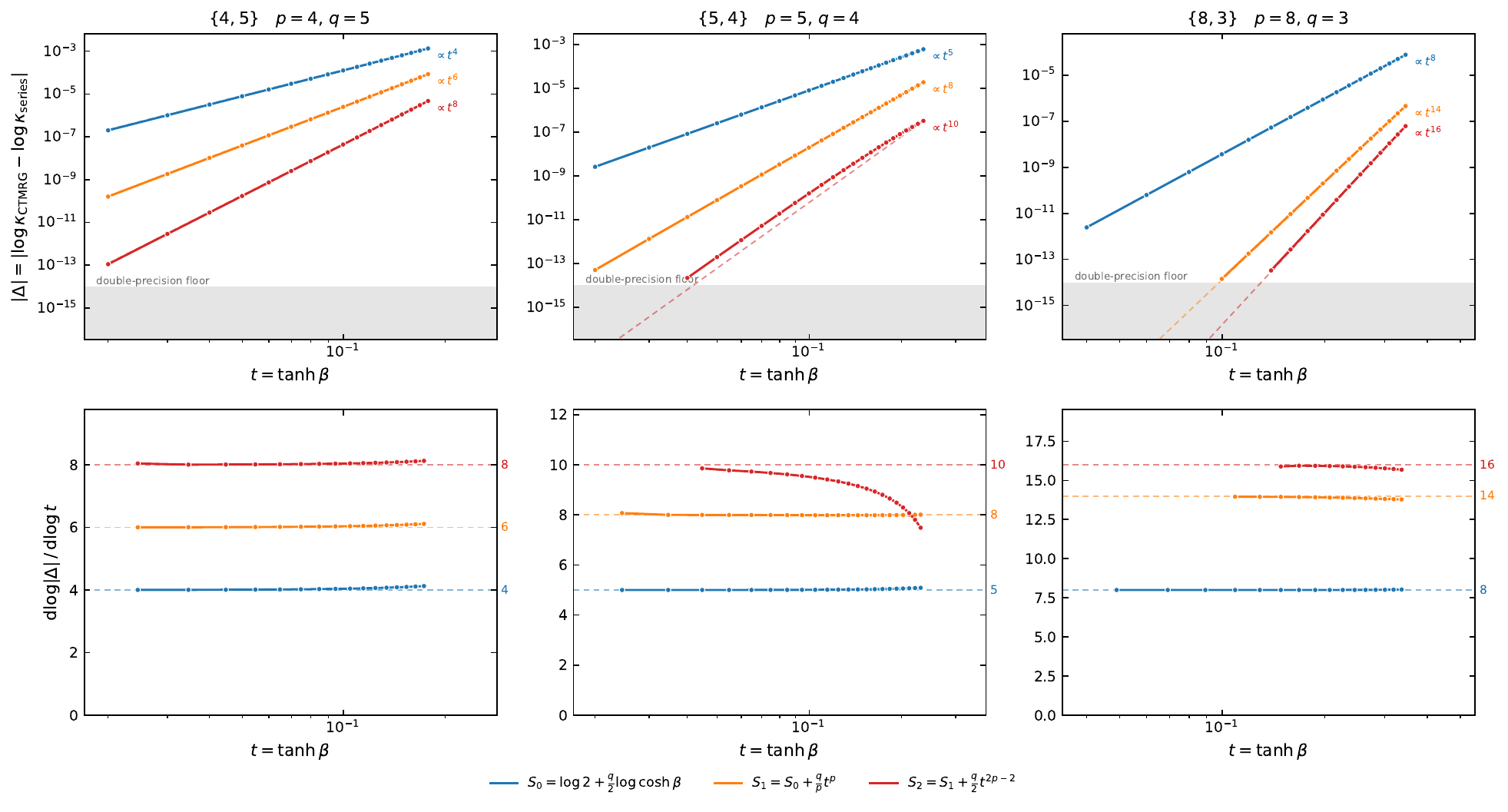} 
  \caption{\label{fig:hts}%
   The difference $\Delta$ in free energies per site $\log{\kappa}$ obtained from CTMRG and from different orders of high-temperature expansion against $t=\tanh{\beta}$ for $(4, 5), (5, 4), (8,3)$ hyperbolic lattices. In the bottom row of plots we also show its logarithmic derivative, which clearly shows the correct slopes of the residuals predicted by the series expansions.  }
\end{figure}

To verify our results against this formula, we evaluated Eq.~\eqref{eq:log_k} numerically with CTMRG partition functions on $(4,5), (5,4), (8,3)$ hyperbolic lattices at the high-temperature limit and compared our calculations with the series expansions. In particular, in Fig.~\ref{fig:hts} we show a log-log plot of the difference between the CTMRG $\log \kappa$ and the finite-order truncation of the series. We see that the difference behaves as the next non-vanishing term in the series. The CTMRG prediction agrees very well with high-temperature expansions, with the latter being rather sensitive to the lattice structure. This confirms that CTMRG correctly takes into account the hyperbolic lattice geometry. 

\end{appendix}

\bibliography{refs.bib}

\end{document}